\documentclass[preprint,superscriptaddress,nofootinbib]{revtex4}
     
     \usepackage{amssymb,amsmath,graphicx,subfigure,color}
     \usepackage{rotating,multirow,dcolumn}
     \usepackage[colorlinks]{hyperref}
     \allowdisplaybreaks[4]
     \begin{document}

     \title{The QED nonfactorizable correction to the
            semileptonic charmed three-body $B$ decays}
     \author{Yueling Yang}
%%   \email[corresponding author, ]{yangyueling@htu.edu.cn}
     \affiliation{Institute of Particle and Nuclear Physics,
                 Henan Normal University, Xinxiang 453007, China}
     \author{Liting Wang}
     \affiliation{Institute of Particle and Nuclear Physics,
                 Henan Normal University, Xinxiang 453007, China}
     \author{Jiazhi Li}
     \affiliation{Institute of Particle and Nuclear Physics,
                 Henan Normal University, Xinxiang 453007, China}
     \author{Qin Chang}
%%   \email[corresponding author, ]{changqin@htu.edu.cn}
     \affiliation{Institute of Particle and Nuclear Physics,
                 Henan Normal University, Xinxiang 453007, China}
     \author{Junfeng Sun}
%%   \email[corresponding author, ]{sunjunfeng@htu.edu.cn}
     \affiliation{Institute of Particle and Nuclear Physics,
                 Henan Normal University, Xinxiang 453007, China}

     \begin{abstract}
     In this paper, the nonfactorizable photonic corrections to
     the effective four-fermion vertex for the
     $\overline{B}$ ${\to}$ $D^{({\ast})}$ $+$ ${\ell}^{-}$
     $+$ $\bar{\nu}_{\ell}$ decays are considered at the
     quark level within SM, where ${\ell}$ $=$ $e$, ${\mu}$
     and ${\tau}$.
     It is found that the QED corrections are closely related
     with the mass of the charged lepton.
     The QED contributions will enhance branching ratios, which
     results in the slight reduction of ratios $R(D^{(\ast)})$
     and the relationship $R(D^{(\ast)})_{e}$ $<$
     $R(D^{(\ast)})_{\mu}$.

     \href{https://doi.org/10.1140/epjc/s10052-024-13647-z}
          {Eur. Phys. J. C 84, 1282 (2024).}
     \href{https://doi.org/10.1140/epjc/s10052-024-13647-z}
          {https://doi.org/10.1140/epjc/s10052-024-13647-z}

     \href{http://arxiv.org/abs/2410.06087}
          {arXiv:2410.06087.}
     \end{abstract}
     \maketitle

     \section{Introduction}
     \label{sec01}
     Within the standard model (SM) of particle physics,
     the exclusive semileptonic charmed three-body $B$ decays,
     $\overline{B}$ ${\to}$ $D^{({\ast})}$ $+$ ${\ell}^{-}$ $+$
     $\bar{\nu}_{\ell}$, are induced by the sequential decay
     $b$ ${\to}$ $c$ $+$ $W^{{\ast}-}$ and $W^{{\ast}-}$ ${\to}$
     ${\ell}^{-}$ $+$ $\bar{\nu}_{\ell}$ at the quark level,
     and therefore they are closely connected with the
     determination of the Cabibbo-Kobayashi-Maskawa (CKM) matrix
     element ${\vert} V_{cb} {\vert}$ describing the flavor-change
     transition $b$ ${\to}$ $c$ and couplings between the
     charged $W$ boson and leptons.
     Despite the importance and significance of these semileptonic
     $B$ decays, it is striking that they are at present obscured
     by two clouds.
     One is that the discrepancy between the up-to-date values of
     ${\vert} V_{cb} {\vert}$ obtained from the inclusive and
     exclusive semileptonic $B$ decay is of approximately $3.0\,{\sigma}$
     \cite{PhysRevD.110.030001}.
     The other is that deviation of the $R(D)$-$R(D^{\ast})$
     correlation distribution between the averaged measurements and SM
     predictions exceeds $3.0\,{\sigma}$ \cite{hflav}, where the
     ratios of branching fractions,
     %------------------------------------
     \begin{eqnarray}
     R(D) & {\equiv} &
     \frac{ {\cal B}(\overline{B}{\to}D{\tau}^{-}\bar{\nu}_{\tau}) }
          { {\cal B}(\overline{B}{\to}D{\ell}^{-}\bar{\nu}_{\ell}) }
     \label{rd-definition}, \\
     R(D^{\ast}) & {\equiv} &
     \frac{ {\cal B}(\overline{B}{\to}D^{\ast}{\tau}^{-}\bar{\nu}_{\tau}) }
          { {\cal B}(\overline{B}{\to}D^{\ast}{\ell}^{-}\bar{\nu}_{\ell}) }
     \label{rdstar-definition},
     \end{eqnarray}
     %------------------------------------
     with ${\ell}$ $=$ $e$ and ${\mu}$.
     With the advent of high precision era of the heavy-flavor physics
     research, these semileptonic charmed $B$ decays have attracted
     considerable attention in scrutinizing SM and searching for
     possible new physics (NP) beyond SM.

     Phenomenologically, according to the conventions of Ref.~\cite{zpc46.p93},
     the differential decay rate distribution for the $\overline{B}$
     ${\to}$ $D^{({\ast})}$ $+$ ${\ell}^{-}$ $+$
     $\bar{\nu}_{\ell}$ decays
     is generally written as,
     %------------------------------------
     \begin{eqnarray}
     \frac{ d\,{\Gamma} }{ d\,q^{2} \ d\,{\cos}{\theta} }
     & = & {\vert} {\eta}_{\rm EW} {\vert}^{2}\,
     \frac{ G_{F}^{2} \, {\vert} V_{cb} {\vert}^{2} \,
            {\vert} {\bf p} {\vert} \, q^{2} }
          { 256 \, {\pi}^{3} \, m_{B}^{2} } \,
     \Big( 1 - \frac{ m_{\ell}^{2} }{ q^{2} } \Big)^{2}
     \nonumber \\ &   &
     \big\{ \big[ H_{U} \, ( 1 + {\cos}^{2}{\theta} )
         + 2 \, H_{L} \, {\sin}^{2}{\theta}
         + 2 \, H_{P} \, {\cos}{\theta} \big]
     \nonumber \\ &  & \!\!\!\! \!\!\!\! \!\!\!
     + \frac{ m_{\ell}^{2} }{ q^{2} } \, \big[ 2 \, H_{S}
          + 2 \, H_{L} \, {\cos}^{2}{\theta}
          + 4 \, H_{SL} \, {\cos}{\theta}
          + H_{U} \, {\sin}^{2}{\theta} \big] \big\}
     \label{eq:differential-width-dq2-dcos},
     \end{eqnarray}
     %------------------------------------
     where
     \begin{itemize}
     \item  The momentum of $D^{({\ast})}$ meson in the rest frame of $\overline{B}$ meson,
            %------------------------------------
            \begin{equation}
           {\vert} {\bf p} {\vert} \, = \,
            \frac{ \sqrt{ {\lambda}(m_{B}^{2},m^{2}_{D^{({\ast})}},q^{2}) } }{ 2\,m_{B} }
            \label{eq:momentum-of-D-meson},
            \end{equation}
            %------------------------------------
            $m_{B}$, $m_{D^{({\ast})}}$ and $m_{\ell}$ are respectively the mass
            of the initial $\overline{B}$ meson, the recoiled $D^{({\ast})}$ meson and
            the charged lepton ${\ell}$,  and the kinematical function
            %------------------------------------
            \begin{equation}
            {\lambda}(x,y,z) \, = \, x^{2}+y^{2}+z^{2}-2\,x\,y-2\,y\,z-2\,z\,x
            \label{eq:kallen-function}.
            \end{equation}
            %------------------------------------
     \item  $q$ $=$ $p_{B}$ $-$ $p_{D^{({\ast})}}$
                $=$ $p_{\ell}$ $+$ $p_{\bar{\nu}}$
            is the momentum transfer,
            $p_{B}$, $p_{D^{({\ast})}}$,
            $p_{\ell}$ and $p_{\bar{\nu}}$ respectively denote the
            four-momentum of % the off-shell $W^{{\ast}-}$ boson,
            the $\overline{B}$ meson,
            the $D^{({\ast})}$ meson,
            the charged lepton ${\ell}^{-}$ and
            the anti-neutrino $\bar{\nu}_{\ell}$.
            The bounds on $q^{2}$ are given by
            %------------------------------------
            \begin{equation}
            m_{\ell}^{2} \, \le \, q^{2} \, \le \, (m_{B}-m_{D^{({\ast})}})^{2}
            \label{eq:q2-min-max},
            \end{equation}
            %------------------------------------
            and the lower and upper bounds correspond to the
            maximum momentum ${\vert} {\bf p} {\vert}_{\rm max}$ $=$
            ${\lambda}^{1/2}(m_{B}^{2},m^{2}_{D^{({\ast})}},m_{\ell}^{2})/2\,m_{B}$
            and the minimum momentum ${\vert} {\bf p} {\vert}_{\rm min}$ $=$ $0$, respectively.
     \item  ${\theta}$ denotes the polar angular between the
            $D^{({\ast})}$ meson and the lepton ${\ell}^{-}$.
            The case of ${\cos}{\theta}$ $<$ $0$ means that the movement
            of $D^{({\ast})}$ and ${\ell}^{-}$ is in opposite direction.
     \item  The Fermi weak-interaction coupling constant $G_{F}$
            ${\approx}$ $1.166\,{\times}\,10^{-5}\,{\rm GeV}^{-2}$
            \cite{PhysRevD.110.030001}.
     \item  The distribution Eq.(\ref{eq:differential-width-dq2-dcos})
            has been separated into the spin no-flip and flip contributions.
            The flip contributions brings in the characteristic flip
            factor $m_{\ell}^{2}/q^{2}$, which will vanish in the
            zero lepton mass limit.
            The lepton mass effects should be considered
            when one analyzes semi-tauonic $B$ decays.
            What's more, the lepton mass effects allow one to probe
            the scalar hadronic contributions $H_{S}$ which is not
            accessible in the zero lepton mass limit.
            The hadronic contributions,
            %------------------------------------
            \begin{eqnarray}
            H_{U} & = & {\vert} H_{+} {\vert}^{2}  + {\vert} H_{-} {\vert}^{2}
            \label{eq:unpolarized-transverse-component}, \\
            H_{P} & = & {\vert} H_{+} {\vert}^{2}  - {\vert} H_{-} {\vert}^{2}
            \label{eq:parity-odd-component}, \\
            H_{L} & = & {\vert} H_{0} {\vert}^{2}
            \label{eq:longitudinal-component}, \\
            H_{S} & = & {\vert} H_{t} {\vert}^{2}
            \label{eq:scalar-component}, \\
            H_{SL} & = & {\rm Re}(H_{t}\,H_{0}^{\ast})
            \label{eq:scalar-longitudinal-interference-component},
            \end{eqnarray}
            %------------------------------------
            denote respectively the unpolarized-transverse,
            parity-odd, longitudinal, scalar, scalar-longitudinal
            interference components of the hadronic amplitudes.
            The subscript of the helicity amplitudes $H_{\lambda}$,
            ${\lambda}$ $=$ $+$, $-$, $0$ and $t$, corresponds to
            the helicity projections of the spin of the virtual
            $W^{\ast}$ particle in the $B$ meson rest frame.
            The analytical expressions of $H_{\lambda}$ are the
            functions of the $B$ ${\to}$ $D^{({\ast})}$ transition
            form factors and $q^{2}$ rather than
            the angular variable ${\theta}$.
            The $B$ ${\to}$ $D^{({\ast})}$ transition
            form factors and helicity amplitudes $H_{\lambda}$
            are listed in Appendix \ref{app01}, \ref{app02},
            \ref{app03}, \ref{app04}.
     \item  The factor ${\eta}_{\rm EW}$ accounts for the
            structure-independent electroweak
            corrections \cite{npb.196.p83}, and even sometimes includes
            a long-distance electromagnetic radiation effect for neutral
            $B$ meson decays,
            ${\vert} {\eta}_{\rm EW} {\vert}^{2}\, (1+{\alpha}_{\rm em}\,{\pi})$
            \cite{PhysRevD.89.114504,PhysRevD.109.094515,EPJC.82.1141}.
            This factor is often considered to be universal in many
            literature, such as
            Refs.~\cite{PhysRevD.89.114504,PhysRevD.92.034506,PhysRevD.92.054510,
                  PhysRevD.94.094008,PhysRevD.95.115008,
                  PhysRevD.100.052007,PhysRevD.101.074513,PhysRevD.107.052008,
                  PhysRevD.109.094515,JHEP.2017.11.061,EPJC.82.1141}
            and its values is commonly taken as
            ${\eta}_{\rm EW}$ ${\approx}$ $1.0066$.
     \end{itemize}
     %------------------------------------

     By performing integration over ${\cos}{\theta}$ in
     Eq.(\ref{eq:differential-width-dq2-dcos}),
     we can obtain,
     %------------------------------------
     \begin{eqnarray}
     \frac{ d\,{\Gamma} }{ d\,q^{2} }
     & = & {\vert} {\eta}_{\rm EW} {\vert}^{2}\,
     \frac{ G_{F}^{2} \, {\vert} V_{cb} {\vert}^{2} \,
            {\vert} {\bf p} {\vert} \, q^{2} }
          { 96 \, {\pi}^{3} \, m_{B}^{2} } \,
     \Big( 1 - \frac{ m_{\ell}^{2} }{ q^{2} } \Big)^{2}
     \nonumber \\ &   &
     \big\{ ( H_{U}+H_{L} ) \, \big( 1 + \frac{ m_{\ell}^{2} }{ 2\, q^{2} } \big)
         + \frac{ 3 \, m_{\ell}^{2} }{ 2\, q^{2} } \, H_{S} \big\}
     \label{eq:differential-width-dq2-01} \\ & = &
    {\vert} {\eta}_{\rm EW} {\vert}^{2}\,
     \frac{ G_{F}^{2} \, {\vert} V_{cb} {\vert}^{2} \,
            {\vert} {\bf p} {\vert} \, q^{2} }
          { 96 \, {\pi}^{3} \, m_{B}^{2} } \,
     \Big( 1 - \frac{ m_{\ell}^{2} }{ q^{2} } \Big)^{2}
     \nonumber \\ &   &
     \big\{ ( {\vert} H_{+} {\vert}^{2} +
              {\vert} H_{-} {\vert}^{2} +
              {\vert} H_{0} {\vert}^{2} ) \,
     \big( 1 + \frac{ m_{\ell}^{2} }{ 2\, q^{2} } \big)
         + \frac{ 3 \, m_{\ell}^{2} }{ 2\, q^{2} } \, {\vert} H_{t} {\vert}^{2} \big\}
     \label{eq:differential-width-dq2-02}.
     \end{eqnarray}
     %------------------------------------
     In comparison, it is easily seen that the parity-odd and
     scalar-longitudinal interference components, $H_{P}$ and $H_{SL}$,
     which is directly proportional to ${\cos}{\theta}$,
     contribute nothing to the partial decay width.

     Theoretically, a lot of efforts have been made to try to tackle
     and understand the lepton flavor universality (LFU) problem
     arising from the $R(D)$-$R(D^{\ast})$ correlation distribution
     in semileptonic charmed $B$ decays, which can basically be
     summarized into two aspects, namely, the $B$ ${\to}$
     $D^{({\ast})}$ form factors and the couplings between $W$
     bosons and leptons, corresponding to the lower and upper
     circles in Fig.~\ref{fig01}, respectively.
     (1)
     The form factors have the direct and
     significant influence on the helicity amplitudes, the
     differential rate distribution, and partial decay width.
     In general, the form factors are the functions of successive
     variable $q^{2}$.
     It is usually believed that the $B$ ${\to}$ $D^{({\ast})}$
     form factors could be described by the Isgur-Wise function
     with the heavy quark effective theory.
     Especially at the $q_{\rm max}^{2}$ domination, the recoil
     velocity of the $D^{({\ast})}$ in the $B$ rest frame is zero,
     and the Isgur-Wise function is normalized to one in
     the heavy quark limits.
     For the case of heavy quark with finite mass,
     form factors near the $q_{\rm max}^{2}$ domination
     can be obtained with the lattice calculation.
     In addition, it is usually assumed that the form factors at
     the $q^{2}$ ${\sim}$ $0$ regions are calculable for the hard
     gluon-exchange contributions with the perturbative theory,
     and/or determinable with the light-cone sum rules or light
     front quark models by an additional $q^{2}$ extrapolation
     from the space-like to time-like regions,
     such as Refs. \cite{PhysRevD.108.L071504,JHEP.2022.05.024,
     PhysRevD.101.074035,JHEP.2019.12.102,EPJC.79.422,PhysRevD.98.114018}.
     The knowledge of the shapes of the form factors versus $q^{2}$
     will provide crucial compatibility checks between theoretical
     expectations with experimental measurements,
     and improve the precision of the determinations of the
     CKM element ${\vert} V_{cb} {\vert}$.
     The shapes of the form factors are in principle obtained
     by the combined fit of the predictions at small $q^{2}$
     regions and the lattice calculation at large $q^{2}$ regions.
     Unfortunately, the values of form factors across the entire
     region $q^{2}$ ${\in}$ $(0,\,q_{\rm max}^{2}]$,
     for example, in the vicinity of $q^{2}$ ${\approx}$
     $m_{\tau}^{2}$, can not be
     effectively evaluated with a desirable theoretical method
     for the moment.
     The discrepancy between the distribution shapes of form factors
     from theoretical calculation and that from experimental
     measurement might be the key to the LFU problem.
     Here, we will use the form factors of lattice calculation
     \cite{PhysRevD.92.034506,PhysRevD.109.074503}
     recommended by the Heavy Flavor Averaging Group
     Collaboration (HFLAV) in order to compared with the HFLAV
     results on semileptonic $B$ decays.
     (2)
     In the meantime the LFU problem fuel inquisition and/or
     speculation about the gauge couplings between $W$ bosons
     and leptons within SM.
     The characteristic couplings corresponding to specific lepton,
     and even new operators arising from characteristic particles,
     are introduced in many NP models.
     An essential precondition for NP is the more and more precise
     check on SM.
     It is common sense that even without NP, the higher order
     corrections can also modify the effective interaction vertex
     couplings  within SM.
     In this paper, we will investigate the nonfactorizable
     contributions arising from the QED radiative corrections
     to the vertex couplings within SM, which correspond to the
     factor ${\eta}_{\rm EW}$.
     We find that the nonfactorizable contributions from the QED
     radiative corrections are function of variables $q^{2}$,
     ${\cos}{\theta}$ and the lepton mass,
     so the factor ${\eta}_{\rm EW}$ should be lepton-flavor
     dependent, and the integration over ${\cos}{\theta}$ in
     Eq.(\ref{eq:differential-width-dq2-dcos}) become nontrivial.

     The remainder of this paper is organized as follows.
     The theoretical frame for the $\overline{B}$ ${\to}$
     $D^{({\ast})}$ $+$ ${\ell}^{-}$ $+$ $\bar{\nu}_{\ell}$
     decays and the QED vertex radiative corrections
     within SM are presented in Section \ref{sec02}.
     The numerical results on branching ratios and ratio
     $R(D^{({\ast})})$, and our comments are presented
     in Section \ref{sec03}.
     Section \ref{sec04} devotes to a brief summary.
     Form factors for the $\overline{B}$ ${\to}$ $D^{({\ast})}$
     transition and the helicity decay amplitudes are
     listed in Appendix.

     \section{Theoretical frame for the
         $\overline{B}$ ${\to}$ $D^{({\ast})}$ $+$
         ${\ell}^{-}$ $+$ $\bar{\nu}_{\ell}$ decays}
     \label{sec02}

     \subsection{The helicity amplitudes}
     \label{subsec0201}
     %------------------------------------
     \begin{figure}[h]
     \includegraphics[width=0.25\textwidth,angle=-90]{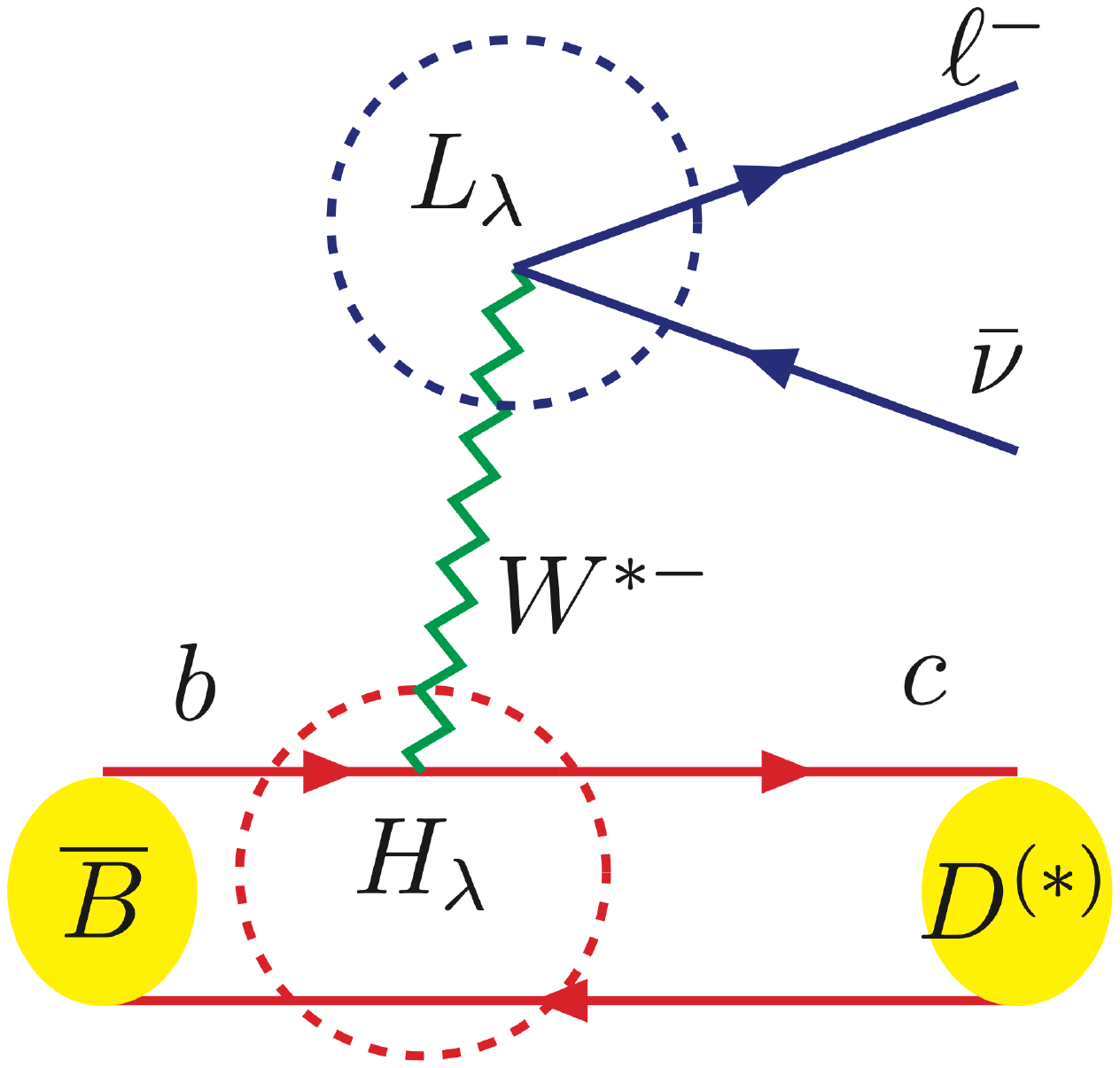}
     \caption{The lowest order Feynman diagram for the
          $\overline{B}$ ${\to}$ $D^{({\ast})}$ $+$ ${\ell}^{-}$ $+$
          $\bar{\nu}_{\ell}$ decays within SM.}
     \label{fig01}
     \end{figure}
     %------------------------------------

     Within SM, the lowest order Feynman diagram for the semileptonic
     $\overline{B}$ ${\to}$ $D^{({\ast})}$ $+$ ${\ell}^{-}$ $+$
     $\bar{\nu}_{\ell}$ decays at the quark level
     is shown in Fig.~\ref{fig01}.
     The corresponding low-energy effective Hamiltonian, which
     is illustrated in Fig.~\ref{fig02}~(a), is written as,
     %------------------------------------
     \begin{equation}
    {\cal H}_{\rm eff} \, = \,
     \frac{ G_{F} }{ \sqrt{2} } \, V_{cb} \,
     \bar{c} \, {\gamma}_{\mu} \, (1-{\gamma}_{5}) \, b \
     \bar{\ell} \, {\gamma}^{\mu} \, (1-{\gamma}_{5}) \, {\nu}_{\ell}
     \, = \,
     \frac{ G_{F} }{ \sqrt{2} } \, V_{cb} \, j_{h,{\mu}} \, j_{\ell}^{\mu}
     \label{eq:Hamiltonian},
     \end{equation}
     %------------------------------------
     with the hadronic and leptonic currents,
     \begin{eqnarray}
     j_{h}^{\mu} & = & \bar{c} \, {\gamma}^{\mu} \, (1-{\gamma}_{5}) \, b
     \label{eq:hadronic-current}, \\
     j_{\ell}^{\mu} & = & \bar{\ell} \, {\gamma}^{\mu} \, (1-{\gamma}_{5}) \, {\nu}_{\ell}
     \label{eq:leptonic-current}.
     \end{eqnarray}

     The decay amplitude is factorized into two parts,
     %------------------------------------
     \begin{equation}
    {\cal A}_{0} \, = \,
    {\langle} \, D^{({\ast})} \, {\ell}^{-} \, \bar{\nu}_{\ell} \,
    {\vert} \, {\cal H}_{\rm eff} \, {\vert} \, \overline{B} \, {\rangle}
     \, = \,
     \frac{ G_{F} }{ \sqrt{2} } \, V_{cb} \, H_{\mu} \, L^{\mu}
     \label{eq:amplitude-leading},
     \end{equation}
     %------------------------------------
     where the hadronic and leptonic current matrix elements are
     respectively defined as,
     %------------------------------------
     \begin{eqnarray}
     H_{\mu} & = &
    {\langle} \, D^{({\ast})} \, {\vert} \, j_{h,{\mu}} \,
    {\vert} \, \overline{B} \, {\rangle}
     \label{eq:hadronic-current-lorentz}, \\
     L_{\mu} & = &
    {\langle} \, {\ell}^{-} \, \bar{\nu}_{\ell} \, {\vert} \, j_{{\ell},{\mu}} \,
    {\vert} \, 0 \, {\rangle}
     \label{eq:leptonic-current-lorentz}.
     \end{eqnarray}
     %------------------------------------
     The leptonic current matrix element $L_{\mu}$ is calculated
     accurately with the perturbative theory.
     So far, the main theoretical challenges and uncertainties come
     from the hadronic current matrix element $H_{\mu}$ because of
     our limited knowledge of dynamics of hadronization.
     Phenomenologically, $H_{\mu}$ is generally expressed by the
     combination of the four-momentum of the initial and final mesons,
     the polarization vectors of the $D^{\ast}$ meson,
     and the $\overline{B}$ ${\to}$ $D^{({\ast})}$
     transition form factors.

     Following the methods described in Ref.~\cite{zpc46.p93},
     and using the orthonormality property and completeness relation
     for the polarization vectors of virtual $W^{\ast}$
     particle,
     %------------------------------------
     \begin{equation}
    {\varepsilon}_{W^{\ast}}({\lambda}) \, {\cdot} \,
    {\varepsilon}_{W^{\ast}}^{\ast}({\lambda}^{\prime})
     \, = \,
    {\varepsilon}_{W^{\ast}}^{\mu}({\lambda}) \,
    {\varepsilon}_{W^{\ast}}^{{\ast}\,{\nu}}({\lambda}^{\prime}) \,
     g_{{\mu}{\nu}}
     \, = \, g_{ {\lambda}, {\lambda}^{\prime} }
     \label{eq:epsilon-orthonormality-property},
     \end{equation}
     %------------------------------------
     \begin{equation}
     \sum\limits_{ {\lambda}, {\lambda}^{\prime} }
    {\varepsilon}_{W^{\ast}}^{\mu}({\lambda}) \,
    {\varepsilon}_{W^{\ast}}^{{\ast}\,{\nu}}({\lambda}^{\prime}) \,
     g_{ {\lambda}, {\lambda}^{\prime} }
     \, = \, g^{ {\mu} {\nu} }
     \label{eq:epsilon-completeness-relation},
     \end{equation}
     %------------------------------------
     with ${\lambda}$, ${\lambda}^{\prime}$ $=$ $+$, $-$, $0$ and $t$
     correspond to the helicity components,
     the product of hadronic and leptonic current matrix elements of
     the decay amplitude in Eq.(\ref{eq:amplitude-leading}) can be
     rewritten as,
     %------------------------------------
     \begin{eqnarray}
     H_{\mu} \, L^{\mu} & = & g^{{\mu}{\nu}} \, H_{\mu} \,  L_{\nu}
     \nonumber \\ & = &
     \sum\limits_{ {\lambda}, \, {\lambda}^{\prime} }
    {\varepsilon}_{W}^{{\ast}{\mu}}({\lambda}) \,
    {\varepsilon}_{W}^{{\nu}}({\lambda}^{\prime}) \,
     g_{{\lambda},{\lambda}^{\prime}} \, H_{\mu} \,  L_{\nu}
     \nonumber \\ & = & \!\!\!
     \sum\limits_{ {\lambda}, \, {\lambda}^{\prime} }
     \big\{ {\varepsilon}_{W}^{{\ast}{\mu}}({\lambda})\, H_{\mu} \big\}\,
     \big\{ {\varepsilon}_{W}^{{\nu}}({\lambda}^{\prime})\,  L_{\nu} \big\}\,
     g_{{\lambda},{\lambda}^{\prime}}
     \nonumber \\ & = &
     \sum\limits_{ {\lambda}, \, {\lambda}^{\prime} }
     H_{\lambda} \, L_{{\lambda}^{\prime}} \,
     g_{{\lambda},{\lambda}^{\prime}}
     \label{eq:amplitude-helicity}
     \end{eqnarray}
     %------------------------------------
     where $H_{\lambda}$ and $L_{\lambda}$ are called as
     hadronic and leptonic helicity amplitudes, respectively,
     %------------------------------------
     \begin{eqnarray}
     H_{\lambda} & = &
    {\varepsilon}_{W}^{{\ast}{\mu}}({\lambda})\, H_{\mu}
     \label{eq:hadronic-helicity-amplitude}, \\
     L_{\lambda} & = &
    {\varepsilon}_{W}^{{\nu}}({\lambda})\,  L_{\nu}
     \label{eq:leptonic-helicity-amplitude}.
     \end{eqnarray}
     %------------------------------------
     The helicity amplitudes $H_{\lambda}$ and $L_{\lambda}$,
     corresponding to the dashed circles in Fig.~\ref{fig01},
     are invariant under the Lorentz transformation and thus
     can be evaluated in different reference frames.
     The helicity amplitudes $H_{\lambda}$ and $L_{\lambda}$
     are usually evaluated in the rest frame of the
     $\overline{B}$ meson and virtual $W^{\ast}$
     particle, respectively.
     The helicity amplitudes $H_{\lambda}$ are listed
     in Appendix \ref{app02} and \ref{app04}.

     \subsection{The QED radiative corrections}
     \label{subsec0202}
\begin{figure}[h]
     \subfigure[]{ \includegraphics[width=0.3\textwidth]{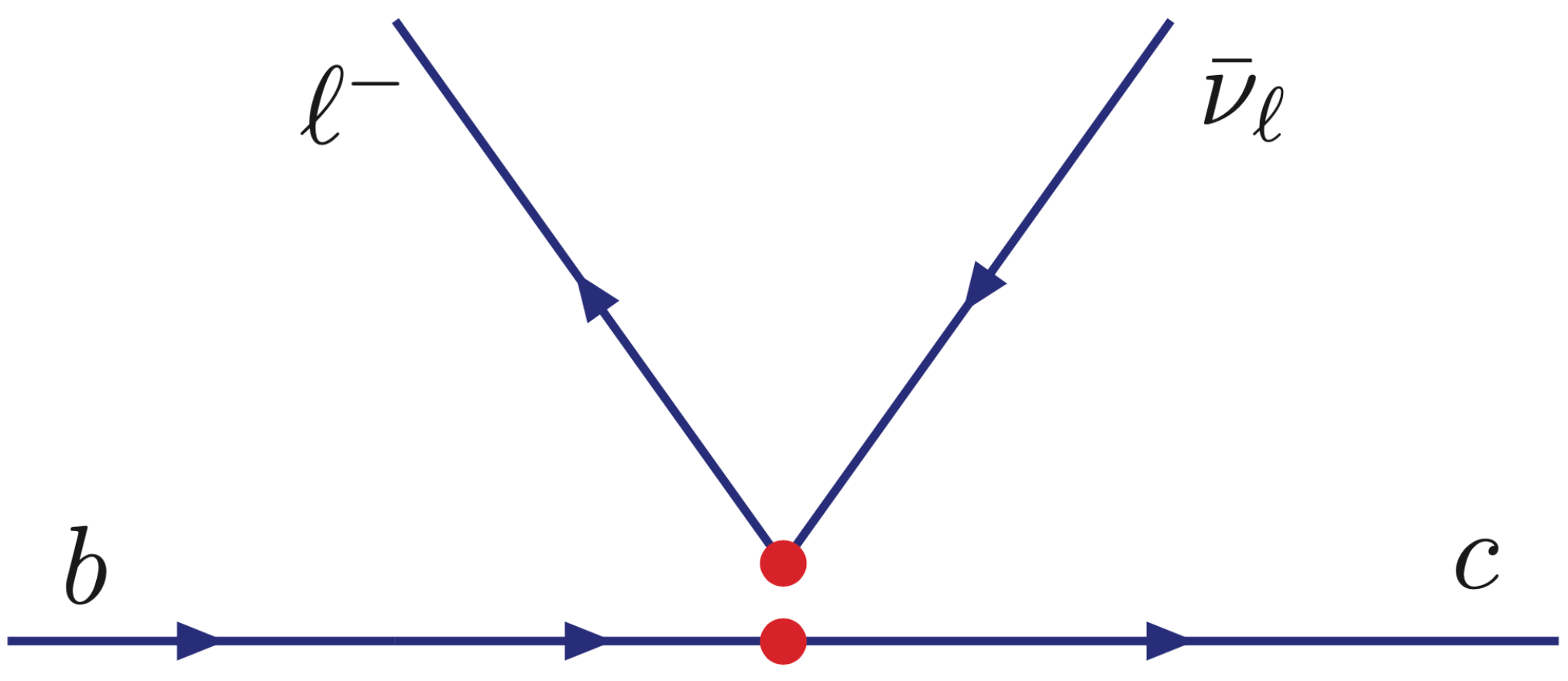} }
     \subfigure[]{ \includegraphics[width=0.3\textwidth]{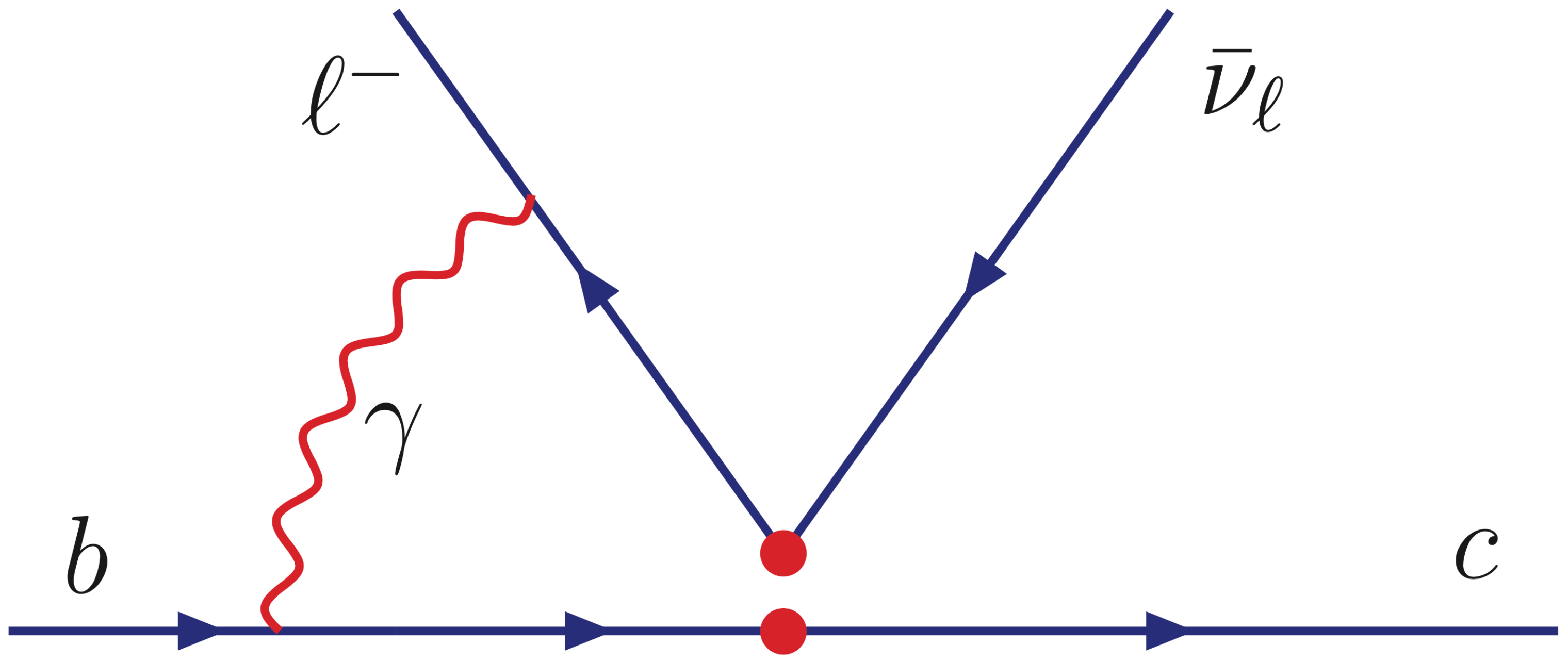} }
     \subfigure[]{ \includegraphics[width=0.3\textwidth]{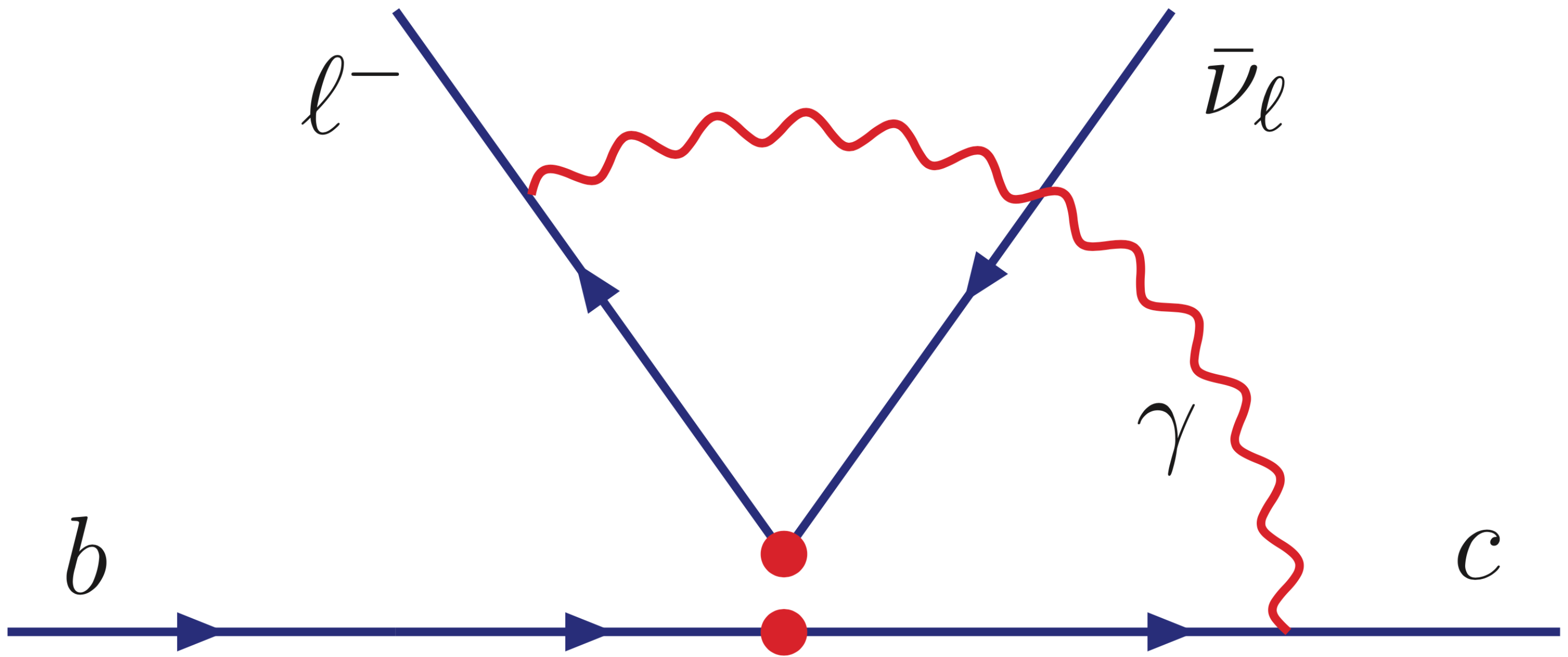} }
     \caption{The Feynman diagrams for the $b$ ${\to}$ $c$ $+$ ${\ell}^{-}$ $+$
          $\bar{\nu}_{\ell}$ decay with the low energy effective Hamiltonian
          in Eq.(\ref{eq:Hamiltonian}), where the dots denote the local
          current interactions, (a) for the leading order contribution,
          (b) and (c) for the QED vertex corrections.}
     \label{fig02}
     \end{figure}

     In order to get more precise theoretical result, the radiative
     corrections beyond the leading order approximation should be
     considered.
     In principle, the contributions from the one-loop self-energy
     corrections to fermion can be absorbed by the renormalized
     field operator and mass.
     According to the arguments of the QCD factorization approach
     in Ref.~\cite{NPB.591.313}, the QCD and QED corrections
     among quarks can be factorized into the $\overline{B}$
     ${\to}$ $D^{({\ast})}$ form factors
     \cite{NPB.591.313,JHEP.2020.11.081,JHEP.2021.10.223}.
     The remaining nonfactorizable contributions come from the
     photon exchanges between quarks and leptons.
     So, the decay amplitudes can be modified as,
     %------------------------------------
     \begin{equation}
    {\cal A} \, = \, {\cal A}_{0} \, {\eta}_{\rm EW}
    \, = \, {\cal A}_{0} \, ( 1+{\alpha}_{\rm em}\,{\eta} )
     \label{eq:amplitude-nlo}.
     \end{equation}
     %------------------------------------
     where ${\cal A}_{0}$ is the leading order amplitude in
     Eq.(\ref{eq:amplitude-leading}),
     corresponds to the Fig.~\ref{fig02} (a).
     Here, we will only consider the nonfactorizable contributions
     from the QED vertex corrections,
     shown in Fig.~\ref{fig02} (b) and (c).
     %------------------------------------
     %------------------------------------

     The calculation of the one-photon-exchange vertex corrections is
     carried out at the quark level with the naive dimensional
     regularization (NDR) and $\overline{\rm MS}$ renormalization
     scheme. Finally, we obtained,
     %------------------------------------
     \begin{eqnarray}
    {\cal M}_{\text{Fig.~\ref{fig02} (b)}} & = &
    {\int}\frac{ d^{D}k_{\gamma} }{ (2{\pi})^{D} }\,
     \bar{c}\, {\gamma}_{\mu}\,(1-{\gamma}_{5})\,
     \frac{i}{\not{p}_{b} \, -\!\!\not{k}_{\gamma}-m_{b}+i\,{\epsilon}}\,
     \big( +i\,Q_{b}\,e\,{\mu}^{\frac{4-D}{2}}\,{\gamma}_{\alpha} \big)\, b\
     \nonumber \\ & &
     \frac{ -i\,g^{ {\alpha}{\beta} } } { k_{\gamma}^{2}+i\,{\epsilon} } \,
     \bar{\ell}\, \big( +i\,Q_{\ell}\,e\,{\mu}^{\frac{4-D}{2}}\,{\gamma}_{\beta} \big)\,
     \frac{i}{\not{p}_{\ell} \, -\!\!\not{k}_{\gamma}-m_{l}+i\,{\epsilon}}\,
    {\gamma}^{\mu}\,(1-{\gamma}_{5})\, {\nu}
     \nonumber \\ & = &
     \frac{ Q_{b}\,Q_{\ell}\,{\alpha}_{\rm em} }{ 4\,{\pi} } \,
     j_{h,{\mu}} \, j_{\ell}^{\mu} \, \Big\{
     \frac{1}{ {\epsilon}_{\rm uv} }+1
   -{\ln}\Big( \frac{ m_{b}^{2} }{ {\mu}_{\overline{\rm MS}}^{2} } \Big)
    +\frac{s_{b}\,{\ln}(s_{b})}{1-s_{b}}
    +\frac{t_{b}\,{\ln}(t_{b})}{1-t_{b}} \Big\}
     \nonumber \\ & - &
     \frac{ Q_{b}\,Q_{\ell}\,{\alpha}_{\rm em} }{ 4\,{\pi} }\,
     j_{h,{\mu}} \, j_{\ell}^{\mu} \, \Big\{
     \frac{ t_{b}+s_{b} }{ t_{b}-s_{b} }\,
    {\ln}\Big( \frac{ t_{b} }{ s_{b} } \Big)
     \Big[ \frac{1}{ {\varepsilon}_{\rm ir} }
   +{\ln}\Big( \frac{ m_{b}^{2} }{ {\mu}_{\overline{\rm MS}}^{2} } \Big) \Big]
   + \frac{1}{2}
     \nonumber \\ &  &
   + \frac{ t_{b}+s_{b} }{ t_{b}-s_{b} }\, \Big[
    2\,{\ln}(t_{b})\,{\ln}\Big(\frac{ t_{b}-s_{b} }{ 1-t_{b} }\Big)
   -2\,{\rm Li}_{2}(t_{b})
   +   {\rm Li}_{2}\Big(\frac{t_{b}}{s_{b}}\Big)
   +i\,{\pi}\,{\ln}\Big(\frac{t_{b}}{s_{b}}\Big)
     \nonumber \\ &  & \qquad\ \quad\,
   -2\,{\ln}(s_{b})\,{\ln}\Big(\frac{ t_{b}-s_{b} }{ 1-s_{b} }\Big)
   +2\,{\rm Li}_{2}(s_{b})
   -   {\rm Li}_{2}\Big(\frac{s_{b}}{t_{b}}\Big) \Big]
     \nonumber \\ &  &
    +\frac{ 1+s_{b} }{ 1-s_{b} }\,{\ln}(s_{b})
    +\frac{ 1+t_{b} }{ 1-t_{b} }\,{\ln}(t_{b}) \Big\}
     \nonumber \\ & = &
     j_{h,{\mu}} \, j_{\ell}^{\mu} \, \Big\{
     \frac{ Q_{b}\,Q_{\ell}\,{\alpha}_{\rm em} }{ 4\,{\pi} } \,
     \Big[ \frac{1}{ {\epsilon}_{\rm uv} }
    -\frac{ t_{b}+s_{b} }{ t_{b}-s_{b} }\,
    {\ln}\Big( \frac{ t_{b} }{ s_{b} } \Big)
     \frac{1}{ {\varepsilon}_{\rm ir} } \Big]
     + {\alpha}_{\rm em}\, {\eta}_{b} \Big\}
     \label{vertex-fig02b},
     \end{eqnarray}
     %------------------------------------
     %------------------------------------
     \begin{eqnarray}
    {\cal M}_{\text{Fig.~\ref{fig02} (c)}} & = &
    {\int}\frac{ d^{D}k_{\gamma} }{ (2{\pi})^{D} }\,
     \bar{c}\, \big( +i\,Q_{c}\,e\,{\mu}^{\frac{4-D}{2}}\,{\gamma}_{\alpha} \big)\,
     \frac{i}{ \not{p}_{c}\, + \!\!\not{k}_{\gamma}-m_{c}+i\,{\epsilon} }\,
    {\gamma}_{\mu}\,(1-{\gamma}_{5})\, b\
     \nonumber \\ & &
     \frac{ -i\,g^{ {\alpha}{\beta} } } { k_{\gamma}^{2}+i\,{\epsilon} } \,
     \bar{\ell}\, \big( +i\,Q_{\ell}\,e\,{\mu}^{\frac{4-D}{2}}\,{\gamma}_{\beta} \big) \,
     \frac{i}{\not{p}_{\ell} \, -\!\!\not{k}_{\gamma}-m_{l}+i\,{\epsilon}} \,
    {\gamma}^{\mu}\,(1-{\gamma}_{5})\, {\nu}
     \nonumber \\ & = &
     -\frac{ Q_{c}\,Q_{\ell}\,{\alpha}_{\rm em} }{ 4\,{\pi} } \,
     j_{h,{\mu}} \, j_{\ell}^{\mu} \, \Big\{
     \frac{4}{ {\epsilon}_{\rm uv} } + 11
   -4\,{\ln}\Big( \frac{ m_{c}^{2} }{ {\mu}_{\overline{\rm MS}}^{2} } \Big)
   + \frac{4\,s_{c}\,{\ln}(s_{c})}{1-s_{c}}
   + \frac{4\,t_{c}\,{\ln}(t_{c})}{1-t_{c}} \Big\}
     \nonumber \\ &  &
    -\frac{ Q_{c}\,Q_{\ell}\,{\alpha}_{\rm em} }{ 4\,{\pi} } \,
     j_{h,{\mu}} \, j_{\ell}^{\mu} \, \Big\{
    -\frac{ t_{c}+s_{c} }{ t_{c}-s_{c} }\,{\ln}\Big( \frac{ t_{c} }{ s_{c} } \Big)
     \Big[ \frac{1}{{\varepsilon}_{\rm ir}}
   +{\ln}\Big( \frac{ m_{c}^{2} }{ {\mu}_{\overline{\rm MS}}^{2} } \Big) \Big] -  2
     \nonumber \\ &  &
   - \frac{ t_{c}+s_{c} }{ t_{c}-s_{c} }\, \Big[
    2\,{\ln}(t_{c})\,{\ln}\Big(\frac{ t_{c}-s_{c} }{ 1-t_{c} }\Big)
   -2\,{\rm Li}_{2}(t_{c})
   +   {\rm Li}_{2}\Big( \frac{ t_{c} }{ s_{c} } \Big)
   +i\,{\pi}\,{\ln}\Big( \frac{ t_{c} }{ s_{c} } \Big)
     \nonumber \\ &  & \qquad\ \quad\,
   -2\,{\ln}(s_{c})\,{\ln}\Big( \frac{t_{c}-s_{c}}{1-s_{c}} \Big)
   +2\,{\rm Li}_{2}(s_{c})
   -   {\rm Li}_{2}\Big(\frac{s_{c}}{t_{c}}\Big)
   -{\ln}\Big( \frac{t_{c}}{s_{c}} \Big) \Big]
     \nonumber \\ &  &
    -\frac{1+s_{c}}{1-s_{c}}\,{\ln}(s_{c})
    -\frac{1+t_{c}}{1-t_{c}}\,{\ln}(t_{c}) \Big\}
     \nonumber \\ & = &
     j_{h,{\mu}} \, j_{\ell}^{\mu} \, \Big\{
     \frac{ Q_{c}\,Q_{\ell}\,{\alpha}_{\rm em} }{ 4\,{\pi} } \,
     \Big[ \frac{ t_{c}+s_{c} }{ t_{c}-s_{c} }\,{\ln}\Big( \frac{ t_{c} }{ s_{c} } \Big)
     \frac{1}{{\varepsilon}_{\rm ir}}
    -\frac{4}{ {\epsilon}_{\rm uv} } \Big]
    +{\alpha}_{\rm em}\, {\eta}_{c} \Big\}
     \label{vertex-fig02c},
     \end{eqnarray}
     %------------------------------------
     where $Q_{b}$ $=$ $-1/3$, $Q_{c}$ $=$ $+2/3$
     and $Q_{\ell}$ $=$ $-1$ are the electric charge quantum
     numbers for the $b$ quark, $c$ quark, and ${\ell}^{-}$ lepton, respectively.
     $j_{h,{\mu}}$ and $j_{\ell}^{\mu}$ are the hadronic
     and leptonic currents in Eq.(\ref{eq:hadronic-current})
     and Eq.(\ref{eq:leptonic-current}), respectively.
     ${\epsilon}_{\rm uv}$ ${\to}$ $0$ for ultraviolet divergences,
     and ${\epsilon}_{\rm ir}$ ${\to}$ $0$ for infrared divergences,
     and
     %------------------------------------
     \begin{eqnarray}
     m_{b}^{2}\, (s_{b}+t_{b}) & = & +2\, p_{b}\, {\cdot} \, p_{\ell}
     \label{vertex-s+t-b}, \\
     m_{c}^{2}\, (s_{c}+t_{c}) & = & -2\, p_{c}\, {\cdot} \, p_{\ell}
     \label{vertex-s+t-c}, \\
     m_{b}^{2}\, s_{b}\, t_{b} \, = \,
     m_{c}^{2}\, s_{c}\, t_{c} & = & m_{\ell}^{2}
     \label{vertex-st-01}, \\
     2\, p_{b}\, {\cdot} \, p_{\ell} -2\, p_{c}\, {\cdot} \, p_{\ell} & = & q^{2}+ m_{\ell}^{2}
     \label{vertex-ql}.
     \end{eqnarray}
     %------------------------------------
     It is easy to see that both the ultraviolet and infrared
     divergences appear in the nonfactorizable amplitudes
     ${\cal M}_{\text{Fig.~\ref{fig02} (b,c)}}$.
     In principle, the ultraviolet divergences could be regularized
     by the redefinition of the fields and parameters, and
     the infrared divergences would vanish by considering the
     real photon emission contributions\footnotemark[1].
     \footnotetext[1]{
     It is pointed out in Refs. \cite{JHEP.2020.11.081,JHEP.2021.10.223}
     that the QED calculation procedure is akin to the QCD case.
     However, the QED effects are more complicated than that of QCD,
     because the mesons are always colour neutral, while the quarks
     are electrically charged.
     Generally, the nonfactorizable QED contributions are no longer
     infrared finite.
     The infrared divergences are dealt with an off-shell regulator
     plus the truncation on ultrasoft photon effects
     \cite{JHEP.2020.11.081,JHEP.2021.10.223}.
     In addition, the long-distance QED contributions from real
     photon emissions and virtual corrections are also considered
     for the semileptonic $B$ decays in Refs.
     \cite{PhysRevLett.120.261804,EPJC.79.744}.
     }
     The relationship among the factor ${\eta}$ in Eq.(\ref{eq:amplitude-nlo}),
     ${\eta}_{b}$ in Eq.(\ref{vertex-fig02b}), and
     ${\eta}_{c}$ in Eq.(\ref{vertex-fig02c}) is
     ${\eta}$ $=$ ${\eta}_{b}$ $+$ ${\eta}_{c}$.

     There are some comments on the factors ${\eta}_{b,c}$.
     (1)
     In Eq.(\ref{vertex-fig02b}) and Eq.(\ref{vertex-fig02c}),
     the term ${\ln}(s_{i}/t_{i})$ $=$ ${\ln}(s_{i}^{2}/s_{i}t_{i})$ ${\propto}$
     ${\ln}(m_{\ell}^{2})$, {\it i.e.}, the factors ${\eta}_{b,c}$
     are the function of the mass of the charged lepton.
     Hence, the effective couplings between the hadronic $j_{h,{\mu}}$ and
     leptonic $j_{\ell}^{\mu}$ currents are no longer universal
     for different lepton at the loop level within SM.
     (2)
     They are the functions of momentum transfer variables $q^{2}$,
     so their contributions may modify the shape lines of the
     differential decay rate distribution.
     (3)
     They are the functions of ${\cos}{\theta}$,
     and will reactivate the contributions of the terms
     $H_{P}$ and $H_{SL}$ absent from
     Eq.(\ref{eq:differential-width-dq2-02}).
     Based on the above comments, it is interesting to comprehensively
     investigate the effects of ${\eta}$ to the partial decay width
     for semileptonic $B$ decays,
     the LFU problem and so on.

     %------------------------------------
     \begin{figure}[h]
     \subfigure[]{ \includegraphics[width=0.3\textwidth]{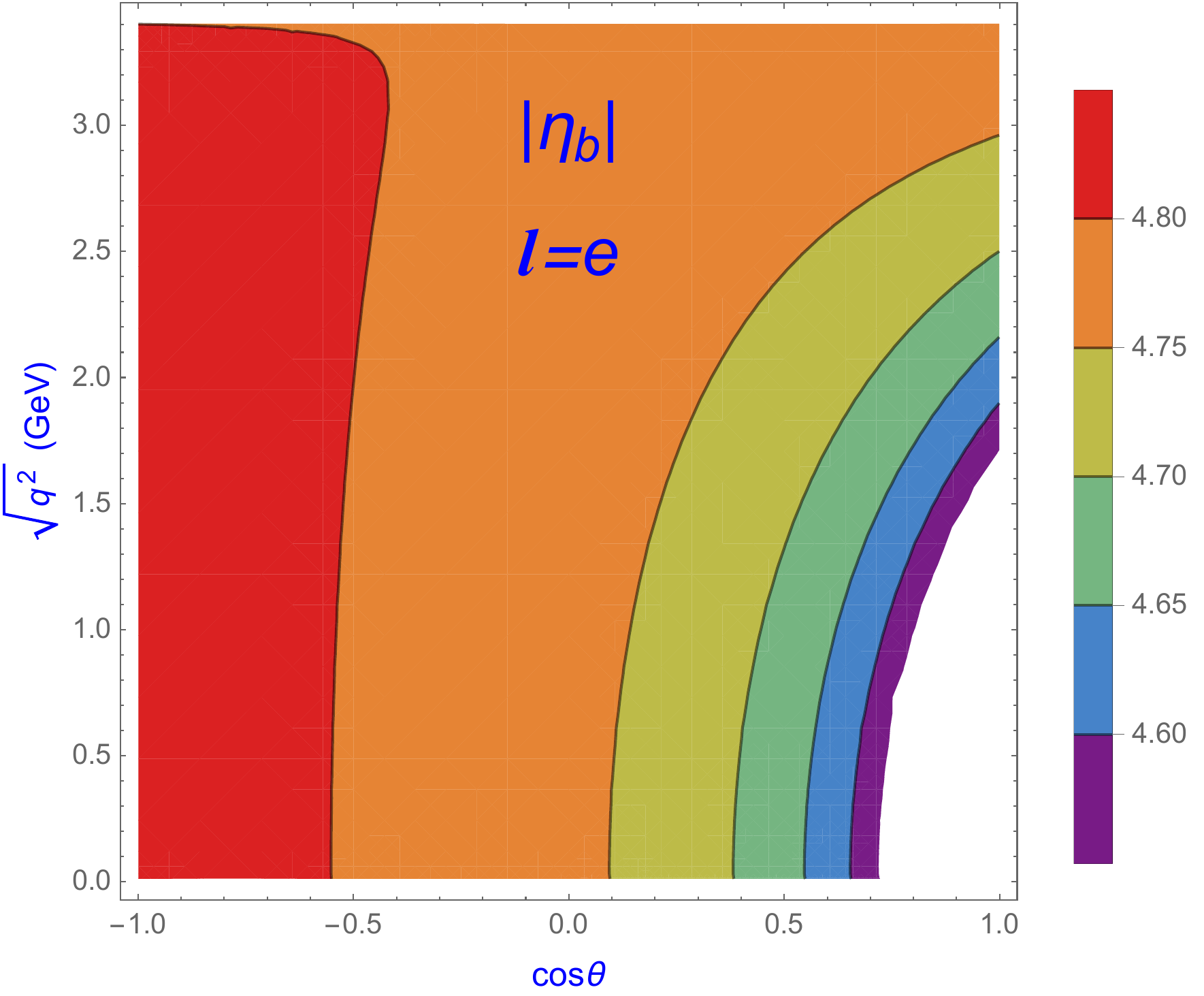} }
     \subfigure[]{ \includegraphics[width=0.3\textwidth]{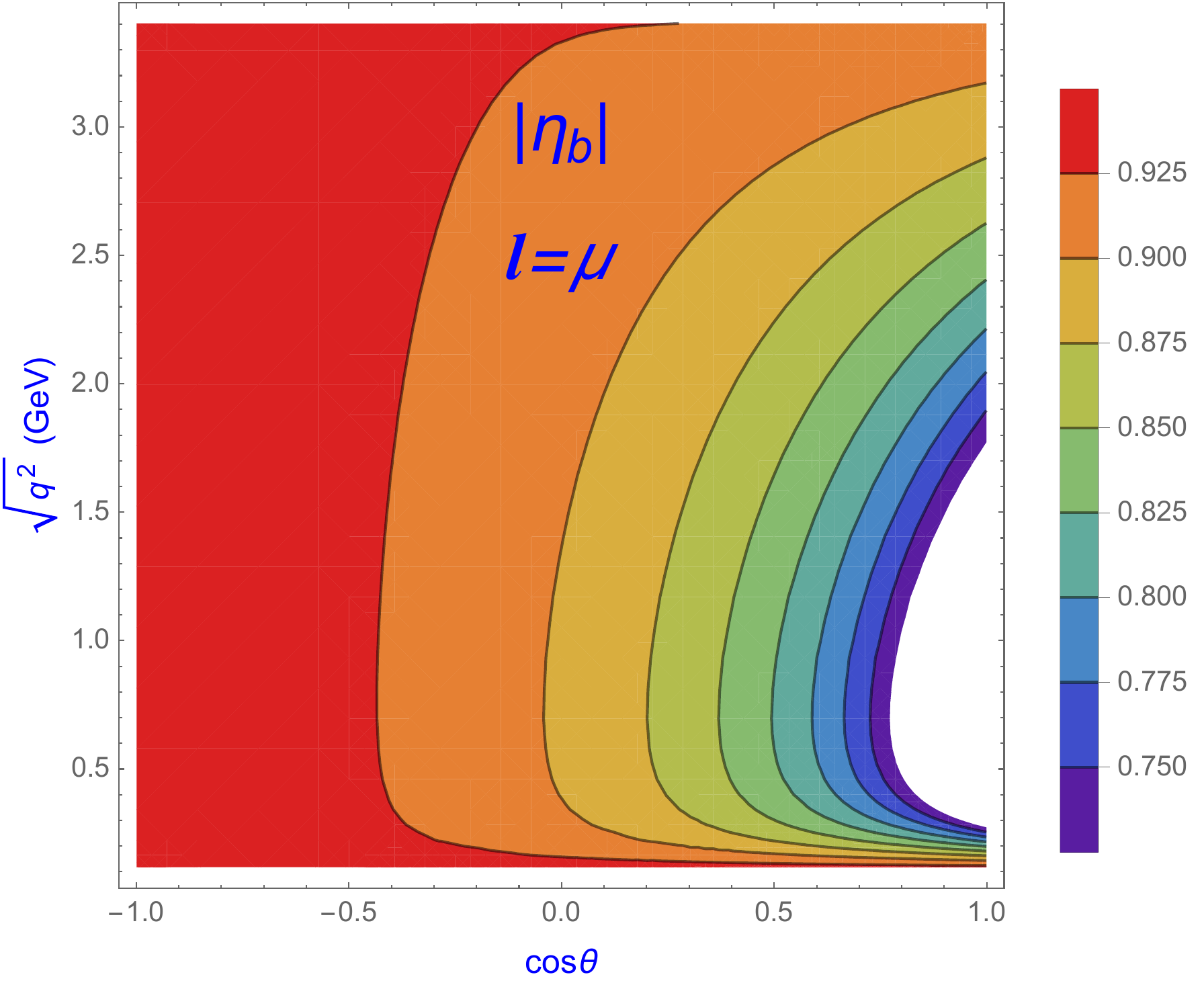} }
     \subfigure[]{ \includegraphics[width=0.3\textwidth]{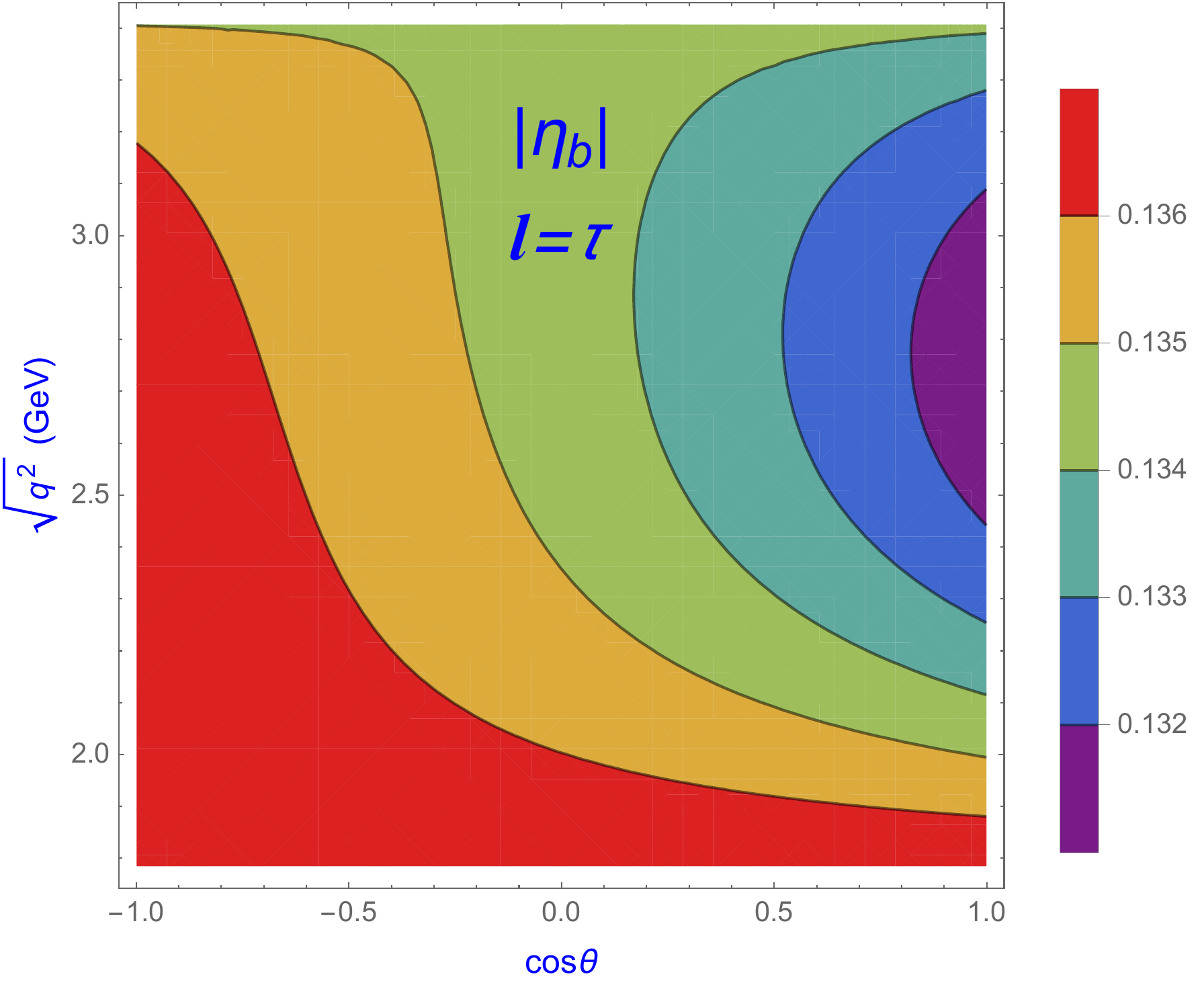} } \\
     \subfigure[]{ \includegraphics[width=0.3\textwidth]{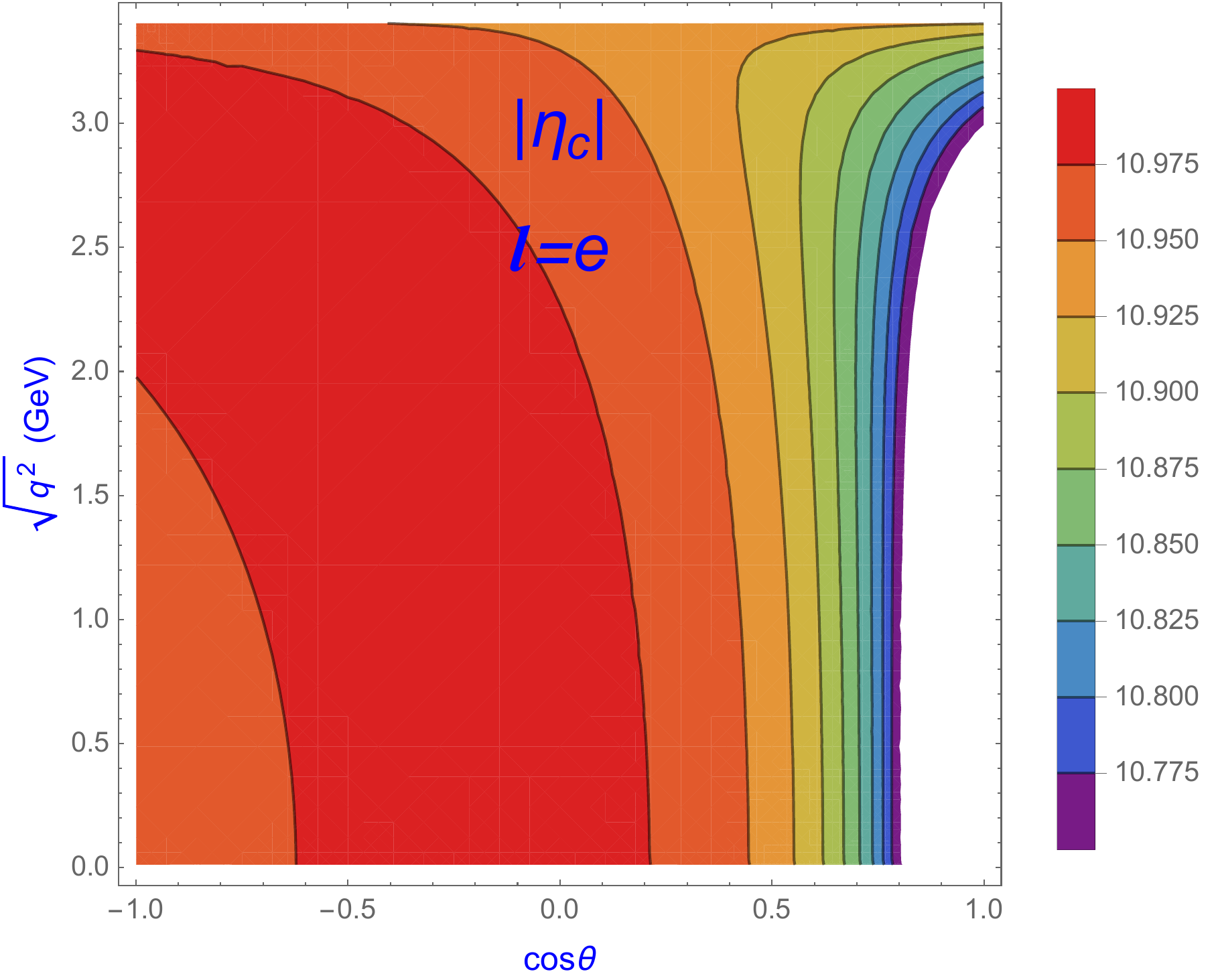} }
     \subfigure[]{ \includegraphics[width=0.3\textwidth]{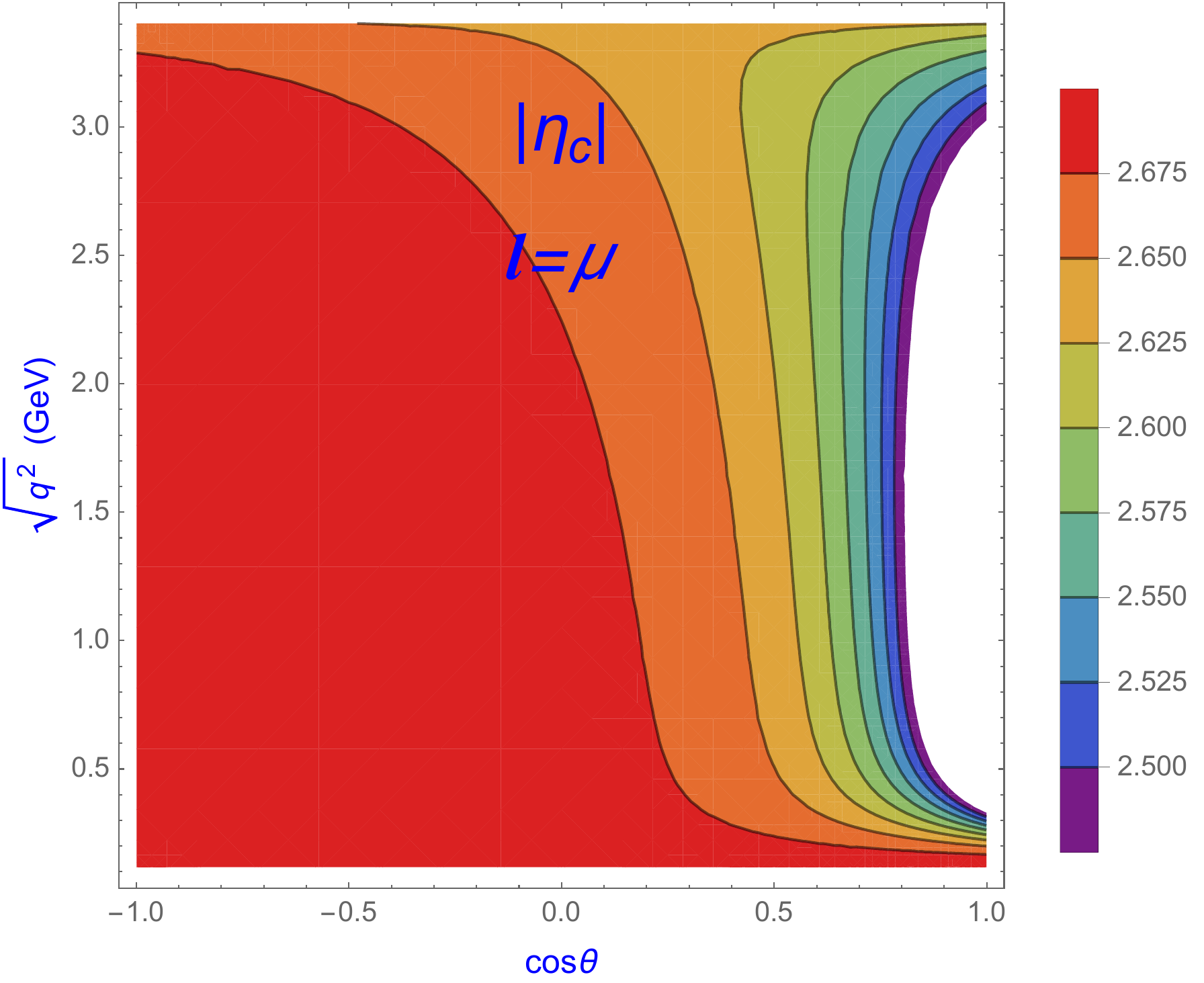} }
     \subfigure[]{ \includegraphics[width=0.3\textwidth]{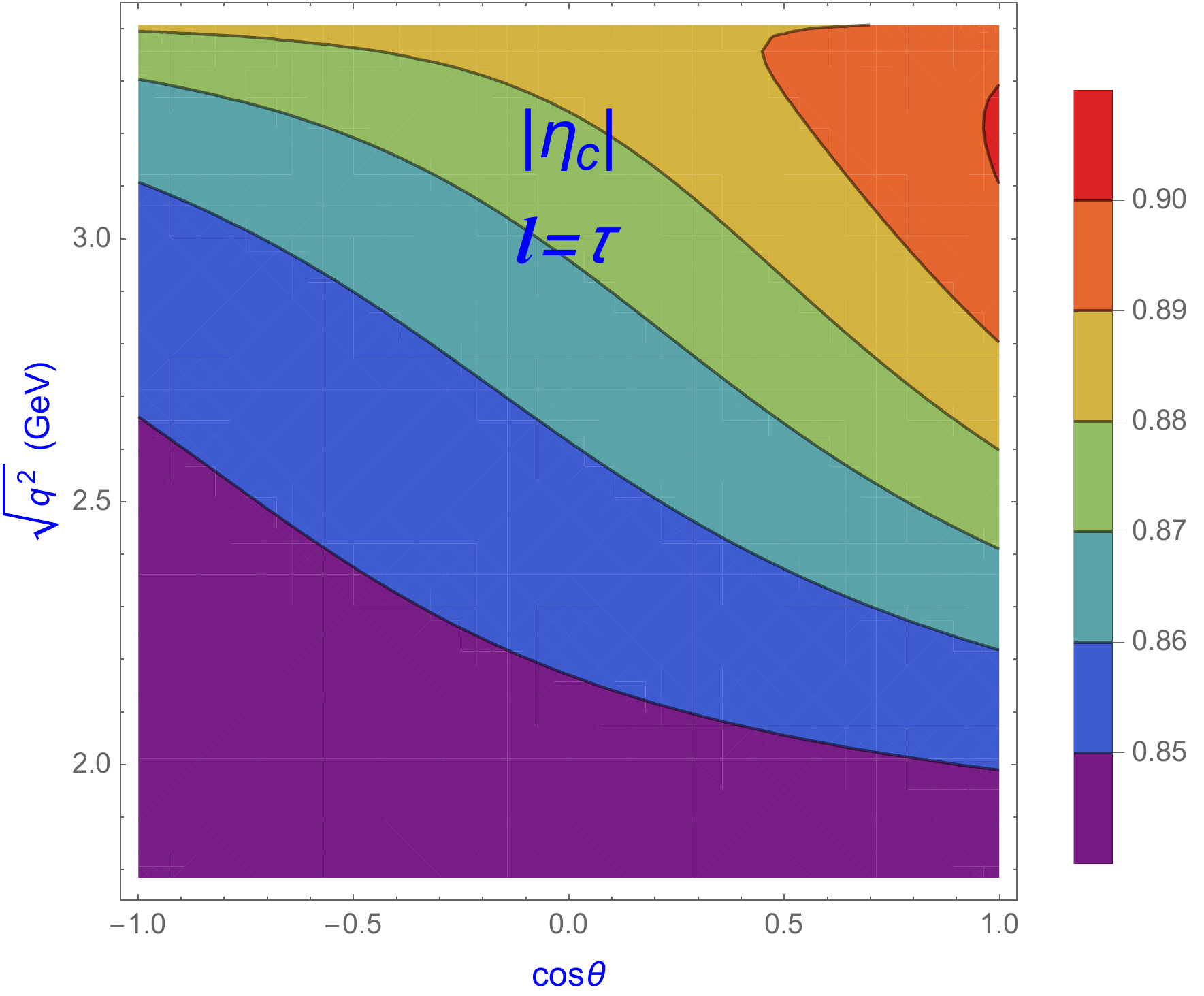} } \\
     \subfigure[]{ \includegraphics[width=0.3\textwidth]{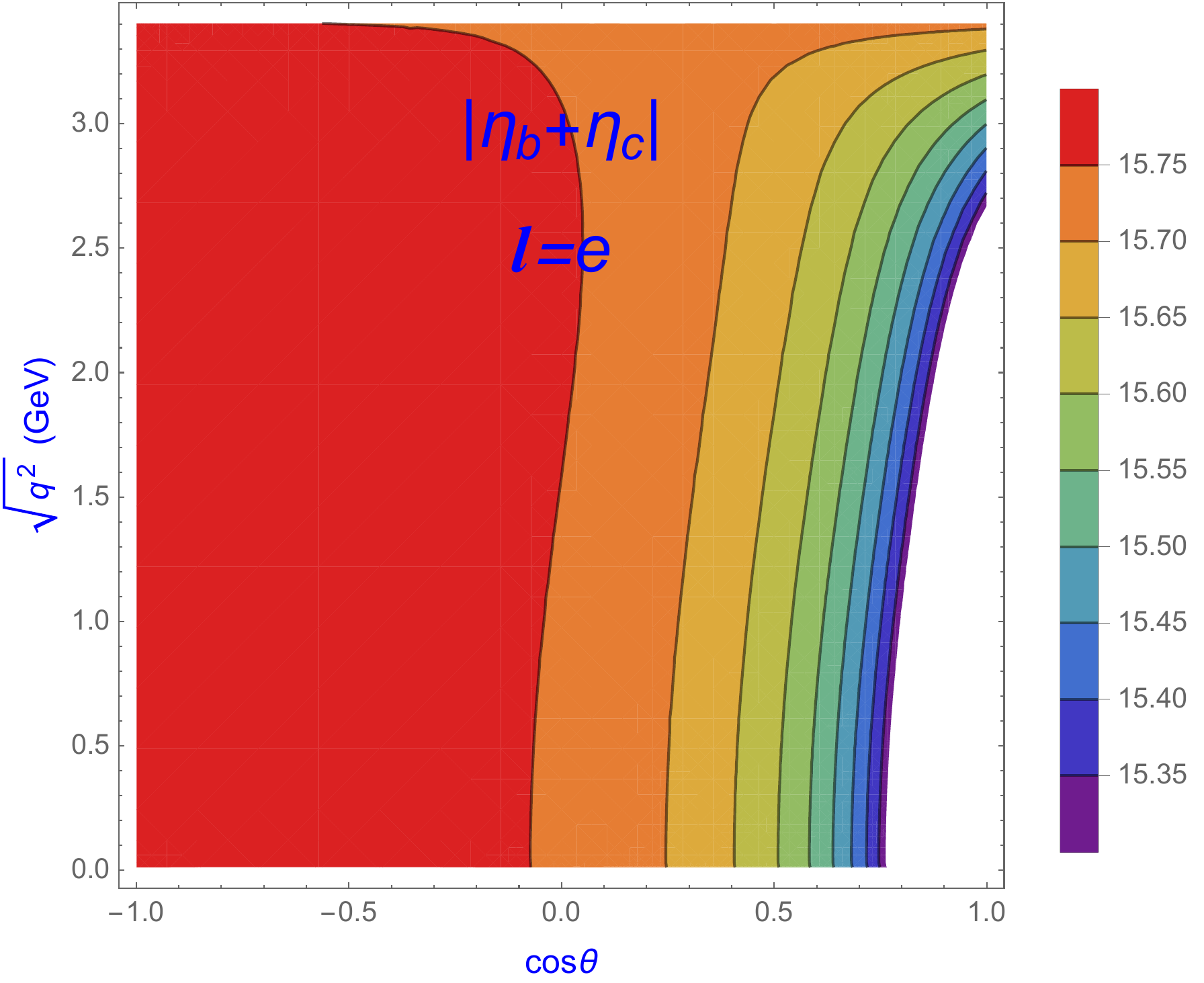} }
     \subfigure[]{ \includegraphics[width=0.3\textwidth]{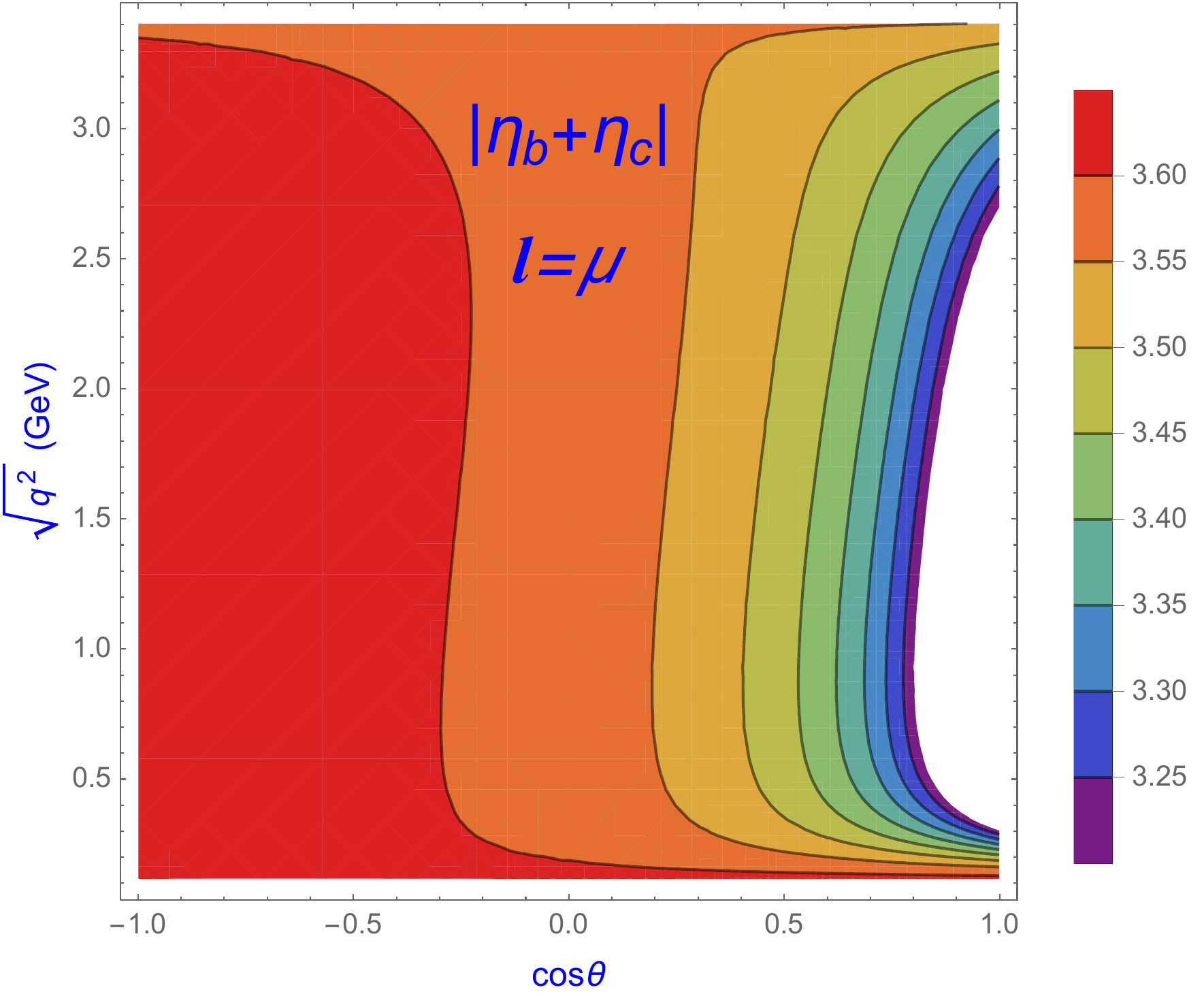} }
     \subfigure[]{ \includegraphics[width=0.3\textwidth]{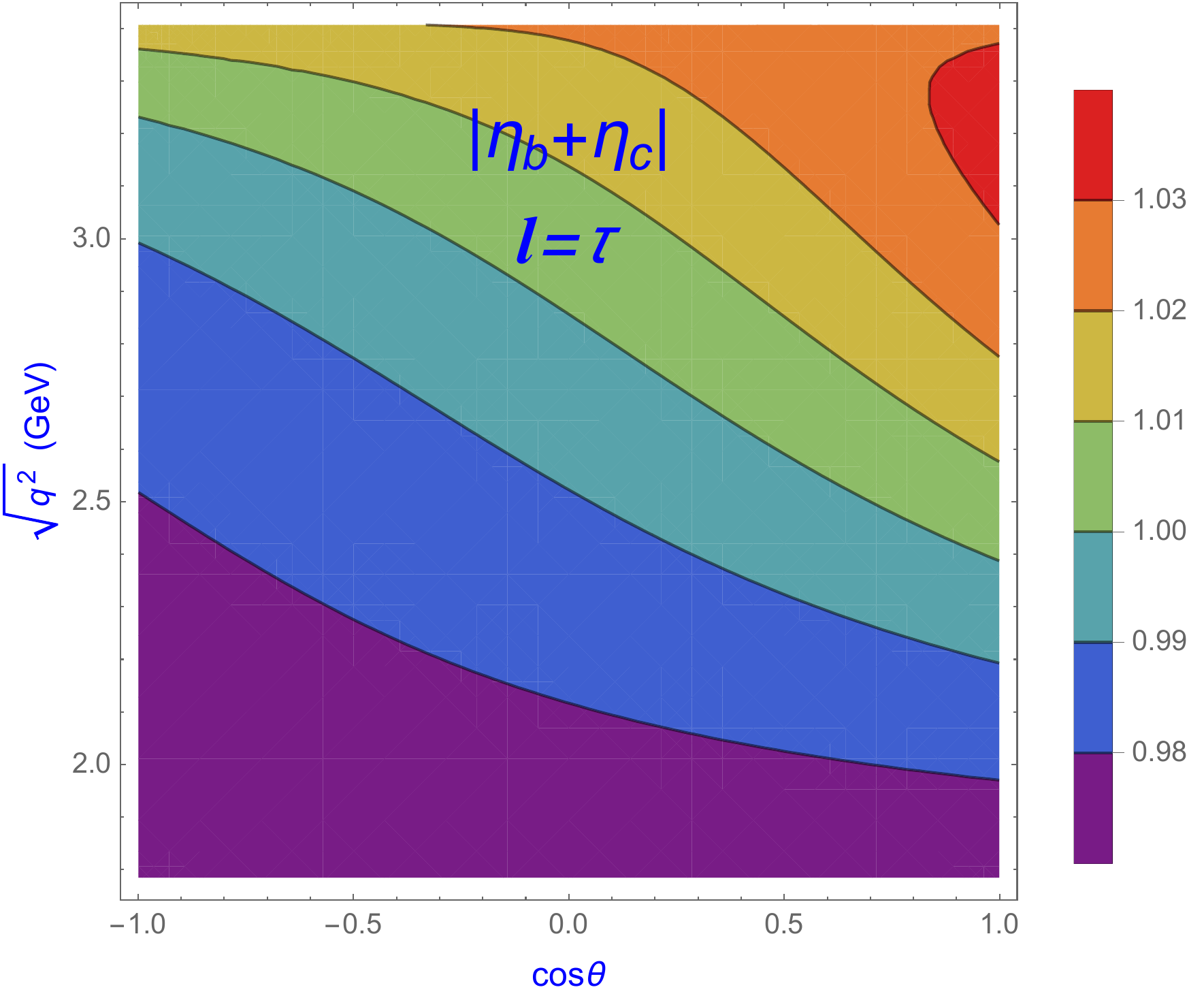} }
     \caption{The contour plot of the ${\eta}_{b,c}$ versus variables of
     $\sqrt{q^{2}}$ (vertical axis) and ${\cos}{\theta}$ (horizontal axis)
     with ${\ell}$ $=$ $e$, ${\mu}$, and ${\tau}$.}
     \label{fig03}
     \end{figure}
     %------------------------------------
     To have a quantitative impression on the factors ${\eta}_{b,c}$,
     the contours plot of the ${\eta}_{b,c}$ versus variables of
     both $\sqrt{q^{2}}$ and ${\cos}{\theta}$ are shown in
     Fig.~\ref{fig03} for three different leptons ${\ell}$ $=$ $e$, ${\mu}$,
     and ${\tau}$, where ${\mu}_{\overline{\rm MS}}$ $=$ $m_{b}$,
     $m_{b}$ ${\approx}$ $m_{B}$ and $m_{c}$ ${\approx}$ $m_{D}$
     is used in the numerical calculation\footnotemark[2].
     \footnotetext[2]{The QED correction factors ${\eta}_{b,c}$
      are proportional to the fine-structure constant ${\alpha}_{\rm em}$.
      The small ${\alpha}_{\rm em}$ will heavily weaken
      the renormalization scale dependence.
      Our numerical results show that the uncertainties on
      branching ratios are not more than 5\%.
     So we will fix the renormalization scale
     ${\mu}_{\overline{\rm MS}}$ $=$ $m_{b}$
     in the following calculation and analysis.}
     It is seen that
     (1)
     The contributions of ${\eta}_{c}$ are generally much larger
     than those of ${\eta}_{b}$.
     The interferences between Fig.~\ref{fig02} (b) and (c) are
     constructive.
     (2)
     For either ${\eta}_{b}$ or ${\eta}_{c}$ or their sum,
     the nonfactorizable QED contributions are sensitive to
     the lepton mass. The smaller the lepton mass,
     the greater the amendments to the effective couplings.
     This fact seems to imply that the effective couplings of
     the current-current $j_{h,{\mu}}$-$j_{\ell}^{\mu}$
     operators of the semileptonic decay
     Hamiltonian are naturally process-dependent by including the
     higher order radiative corrections within SM, even without
     introducing the non-universal couplings of NP.
     Similar phenomena have already been evidenced by
     nonleptonic $B$ decays with the QCD factorization approach
     in Refs.~\cite{NPB.591.313,NPB.606.245,PLB.488.46,PLB.509.263,
     PhysRevD.64.014036,EPJC.36.365,PhysRevD.69.054009,PhysRevD.77.074013}.
     The factor ${\eta}_{\rm EW}$ should not be universal
     for all semileptonic charmed $B$ decays.
     For semi-electronic (semi-muonic) decays, the value of
     ${\vert}{\alpha}_{\rm em}\,{\eta}{\vert}$ is about seventeen
     (four) times larger than that of ${\eta}_{\rm EW}$ $-$ $1$,
     For semi-tauonic decays, the value of
     ${\vert}{\alpha}_{\rm em}\,{\eta}{\vert}$ is approximately
     equal to that of ${\eta}_{\rm EW}$ $-$ $1$, which indirectly
     demonstrates that the proposed QED nonfactorizable scheme and
     calculation is feasible and adaptable.
     For the convenience of the following discussion, we will introduce the
     symbol of $\tilde{\eta}_{\rm EW}$ $=$ $1$ $+$ ${\alpha}_{\rm em}\,{\eta}$
     for distinguishing the ${\eta}_{\rm EW}$.
     (3)
     For the semi-electronic and semi-muonic decays, and
     for most of $q^{2}$ regions, % except for $q^{2}_{\rm min}$,
     the contributions to both ${\eta}_{b}$ and ${\eta}_{c}$ from the
     ${\cos}{\theta}$ $<$ $0$ regions where the charmed meson
     and the lepton ${\ell}$ escape from each other in the
     opposite direction, are slightly larger than the ones
     from the ${\cos}{\theta}$ $>$ $0$ regions.
     And ${\eta}_{b,c}$ vanishes in some areas in
     the ${\cos}{\theta}$ $>$ $0$ regions.
     While for the semi-tauonic decays, ${\eta}_{b,c}$ changes
     with ${\cos}{\theta}$ relatively gently, because the tauon
     is massive, $m_{\tau}$ ${\approx}$ $m_{D}$.
     (4)
     For a specific semileptonic decay, ${\eta}_{b,c}$ changes
     slowly with $q^{2}$ for a particular ${\cos}{\theta}$.

     \section{Numerical results and discussion}
     \label{sec03}

\begin{figure}[h]
     \includegraphics[width=0.3\textwidth]{ 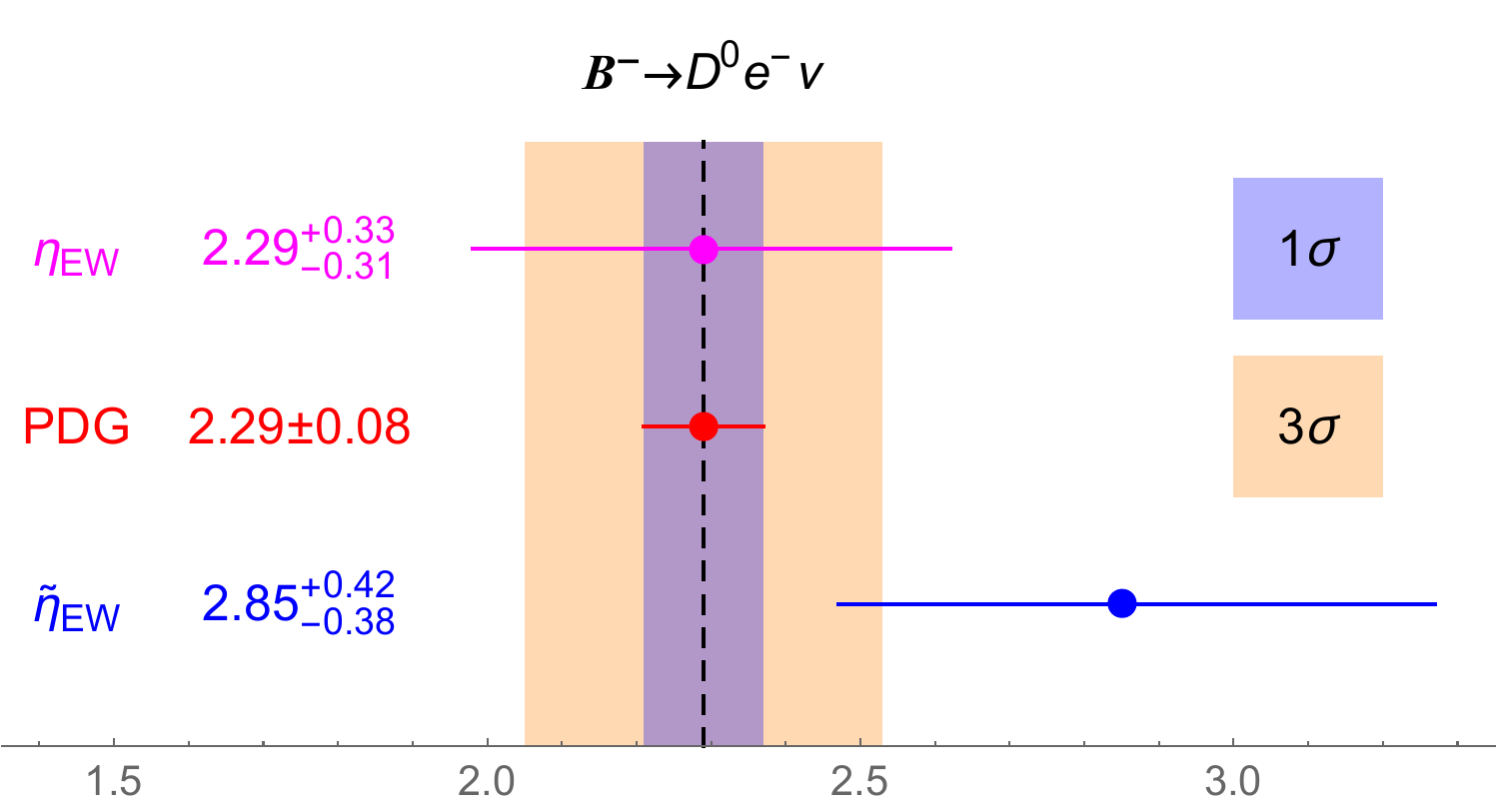} \quad
     \includegraphics[width=0.3\textwidth]{ 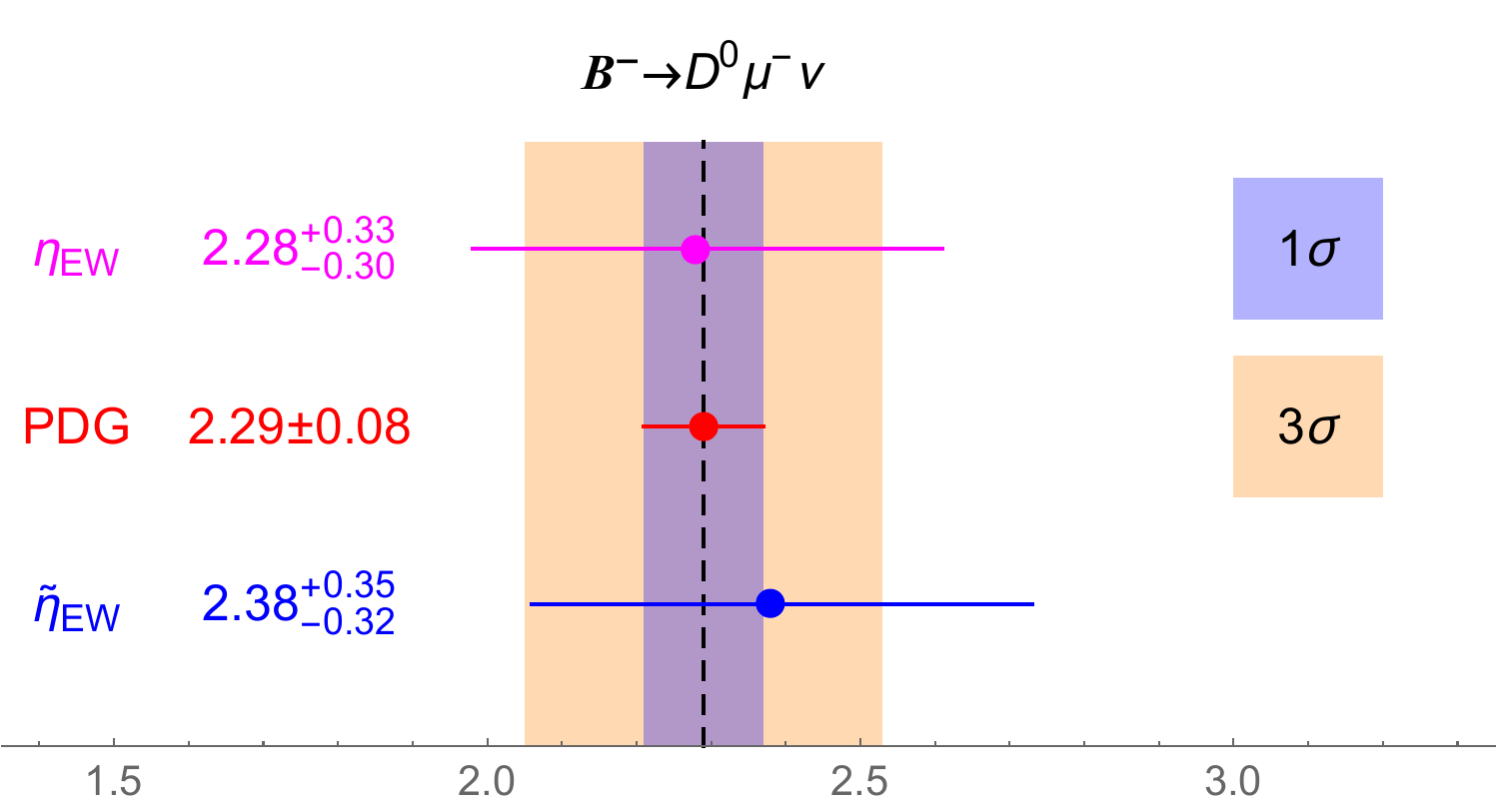} \quad
     \includegraphics[width=0.3\textwidth]{ 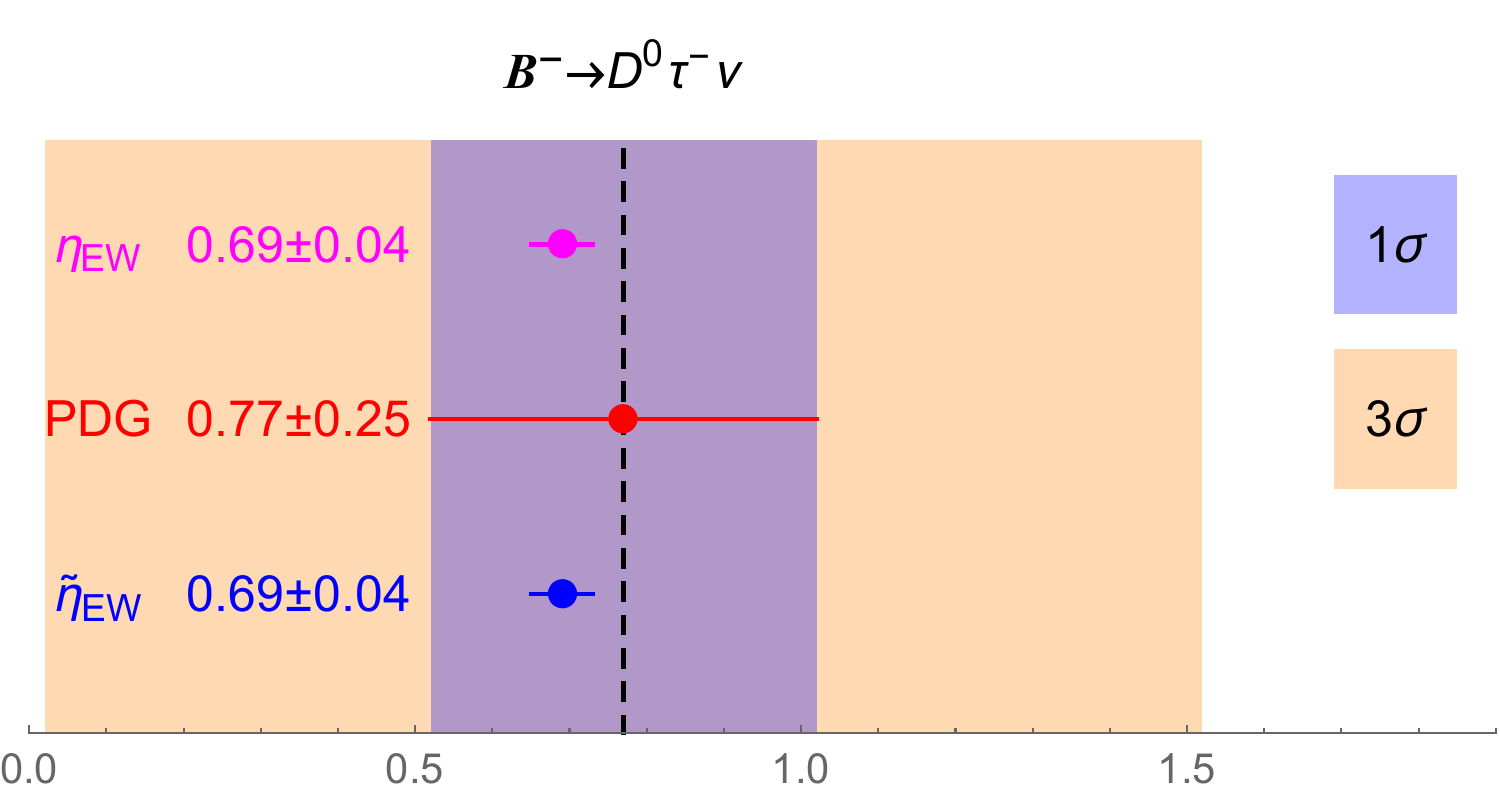} \\
     \includegraphics[width=0.3\textwidth]{ 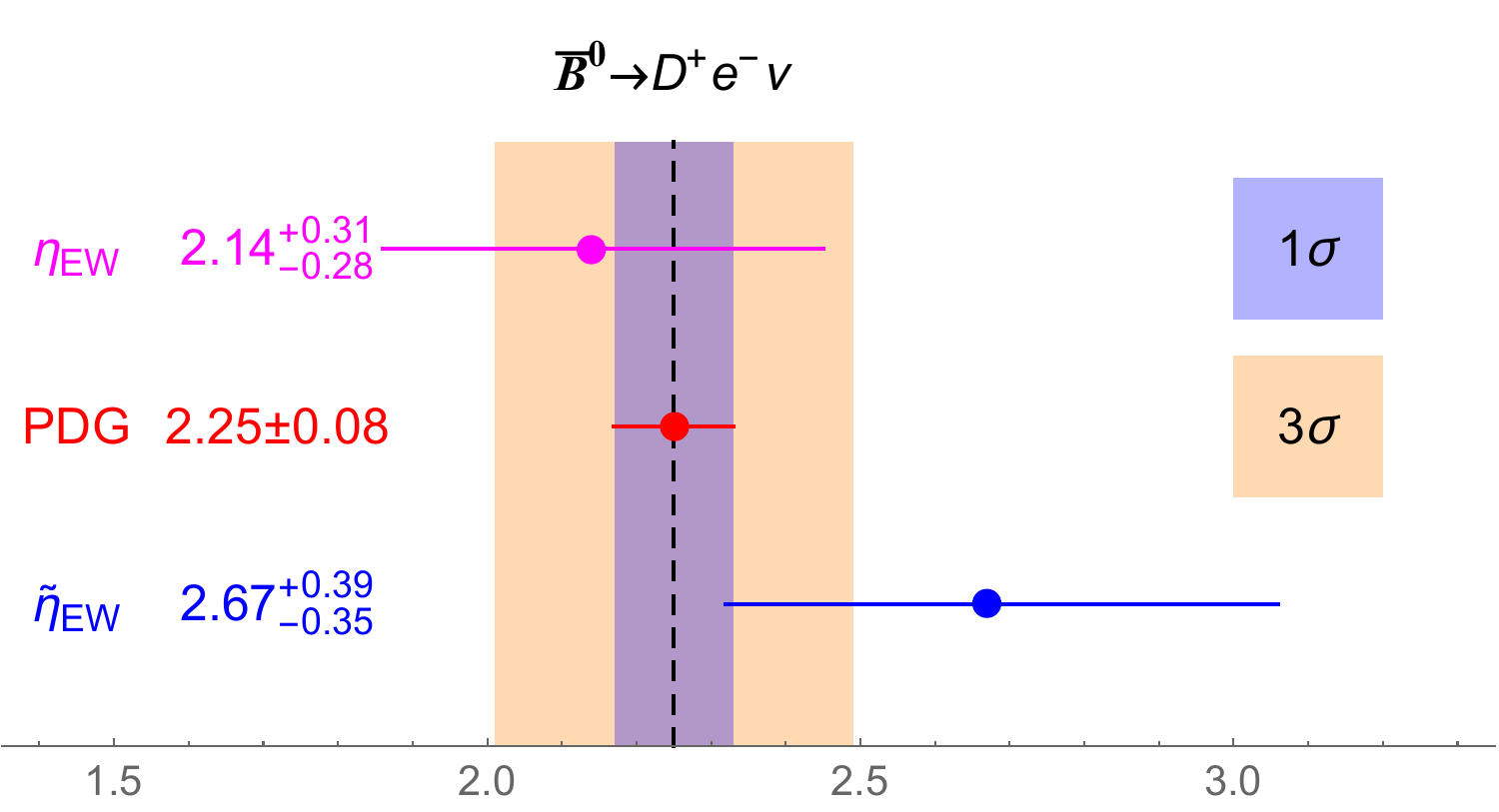} \quad
     \includegraphics[width=0.3\textwidth]{ 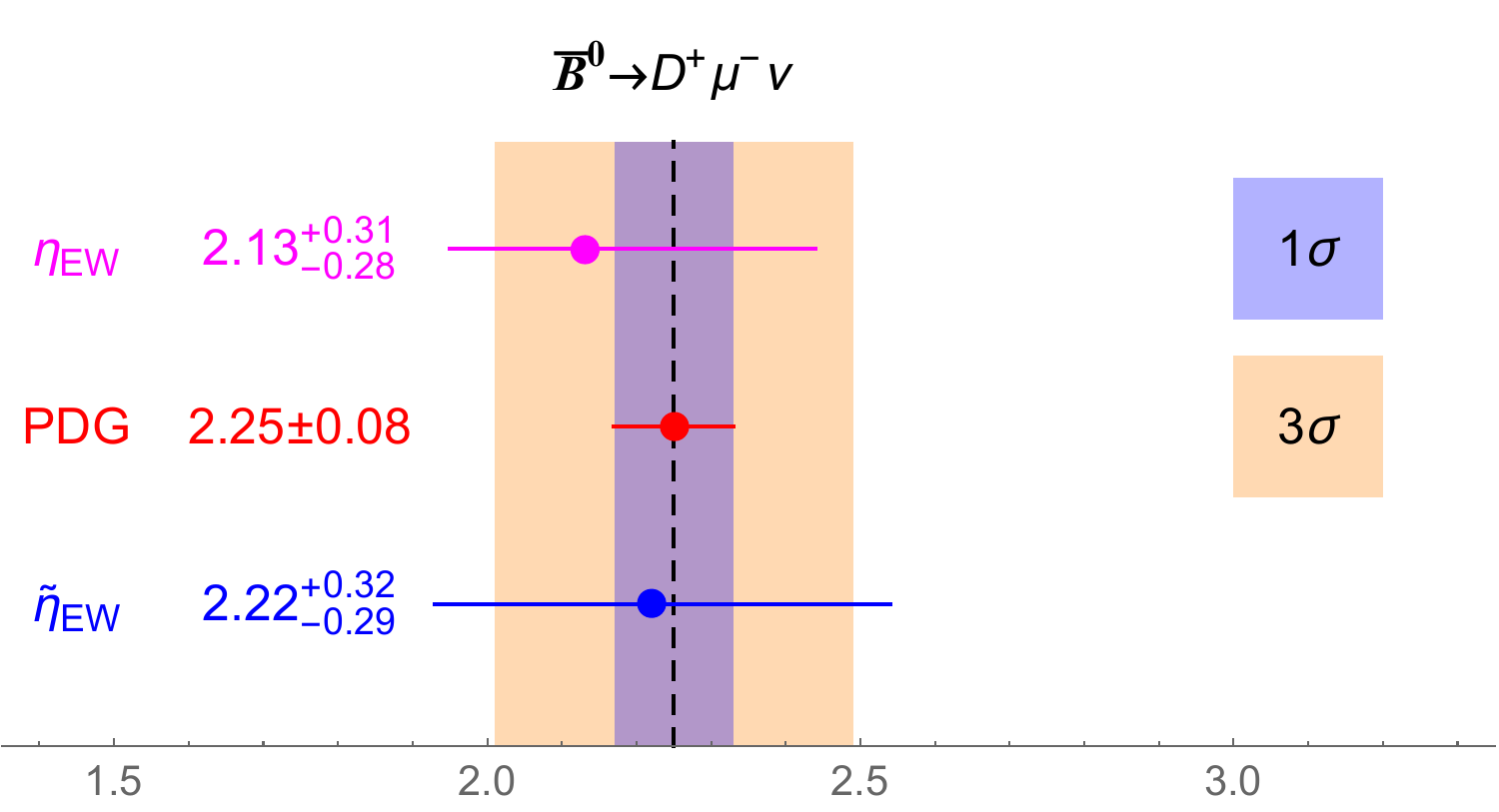} \quad
     \includegraphics[width=0.3\textwidth]{ 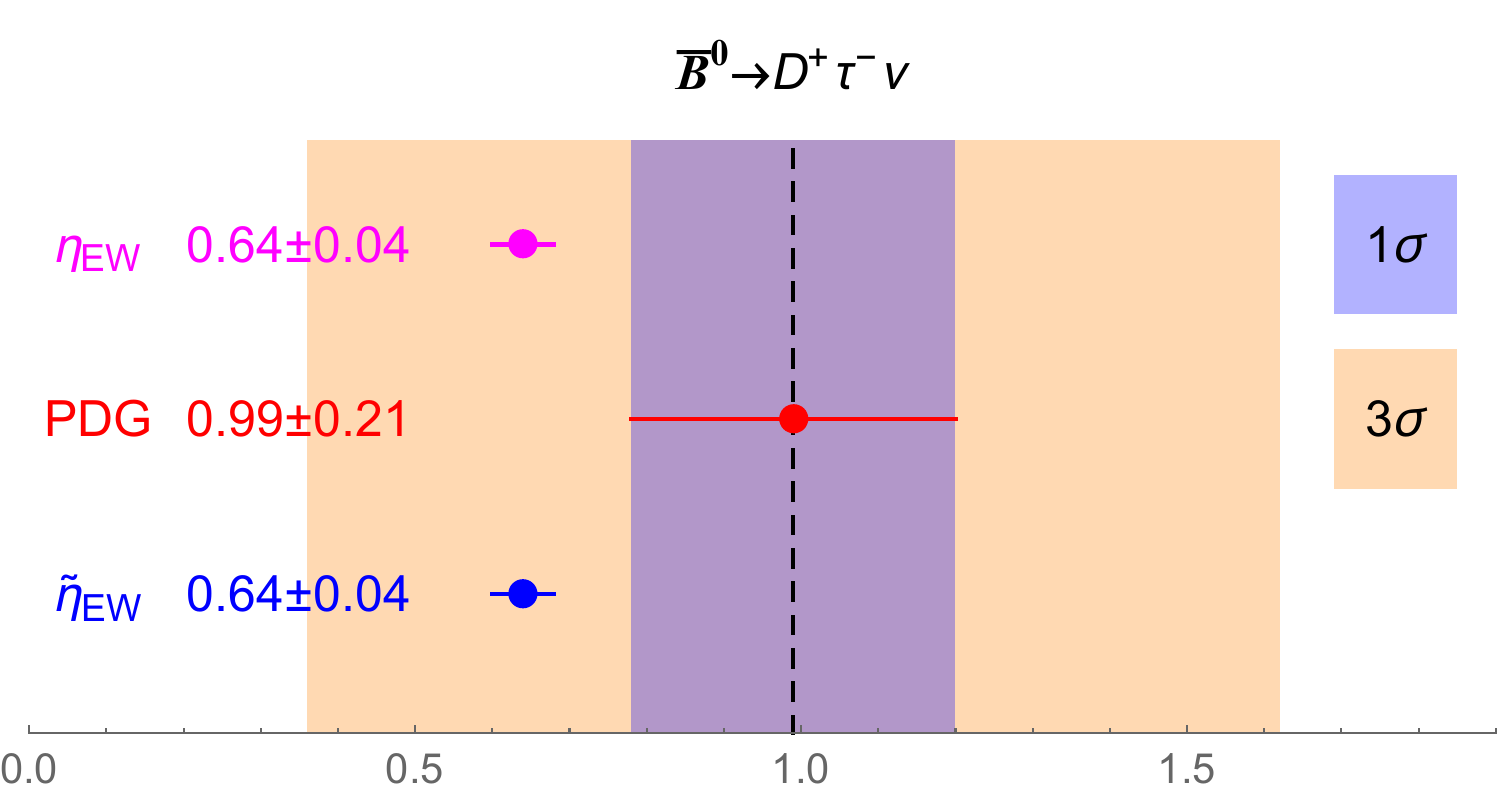} \\
     \includegraphics[width=0.3\textwidth]{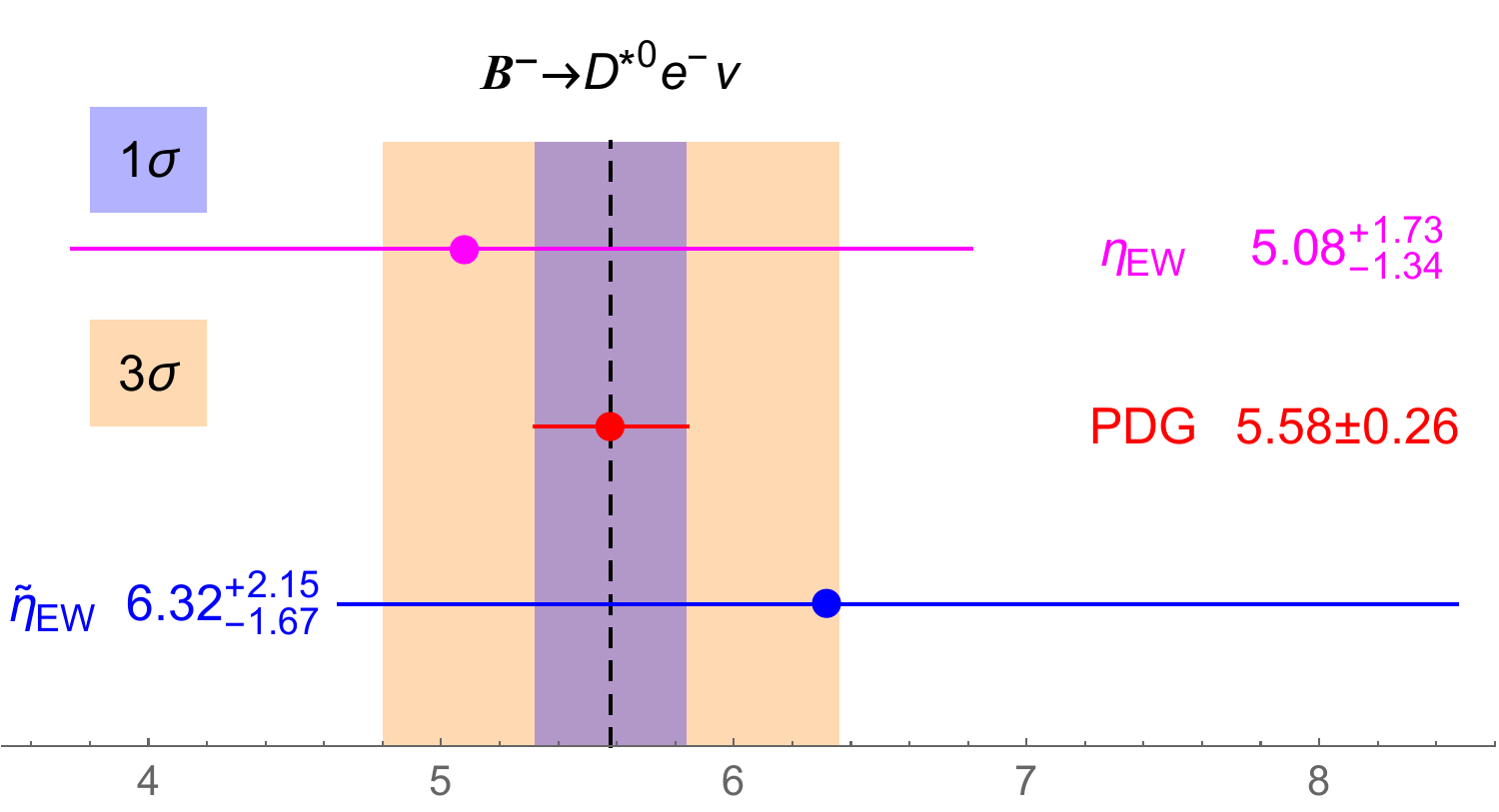} \quad
     \includegraphics[width=0.3\textwidth]{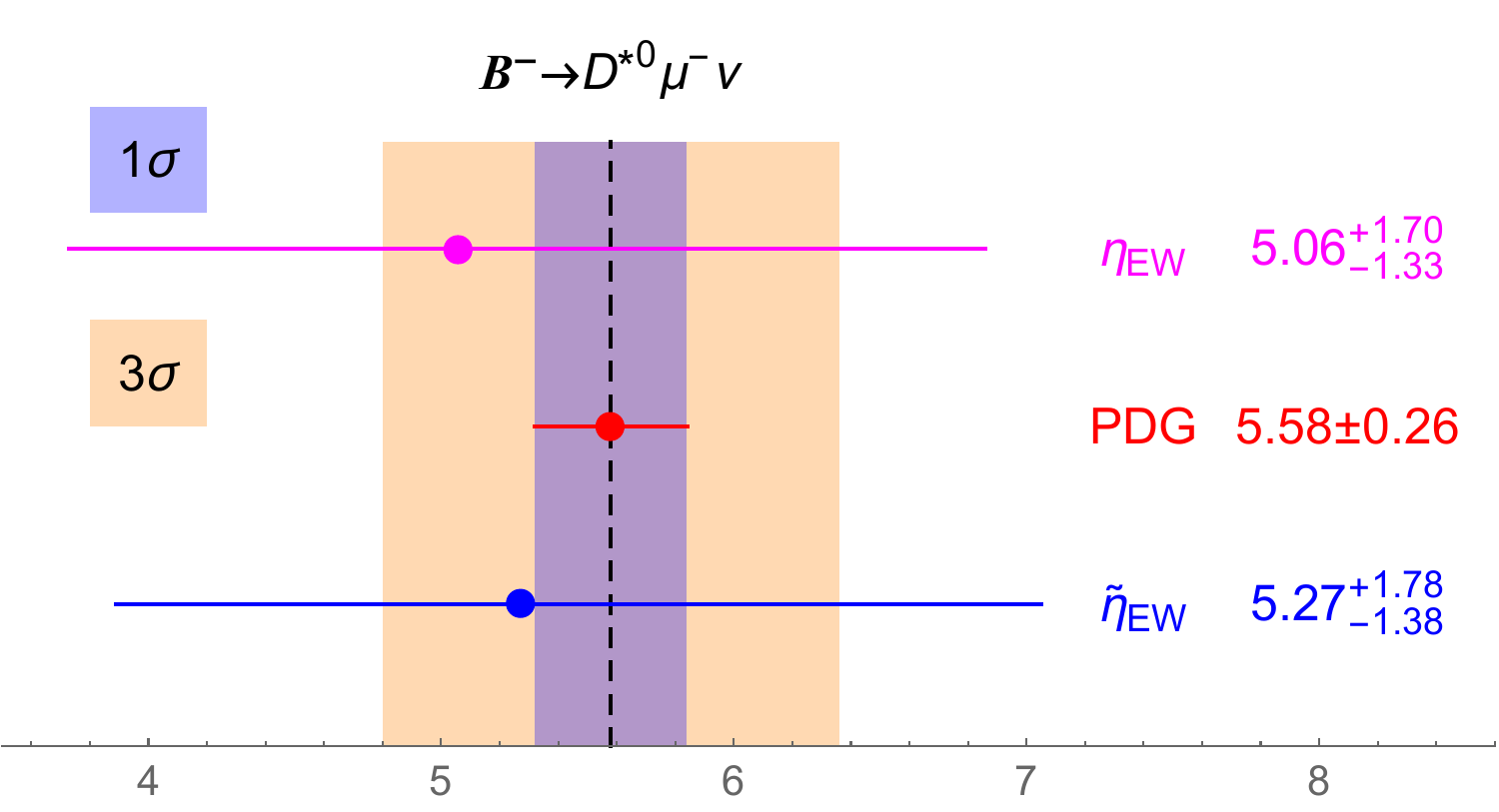} \quad
     \includegraphics[width=0.3\textwidth]{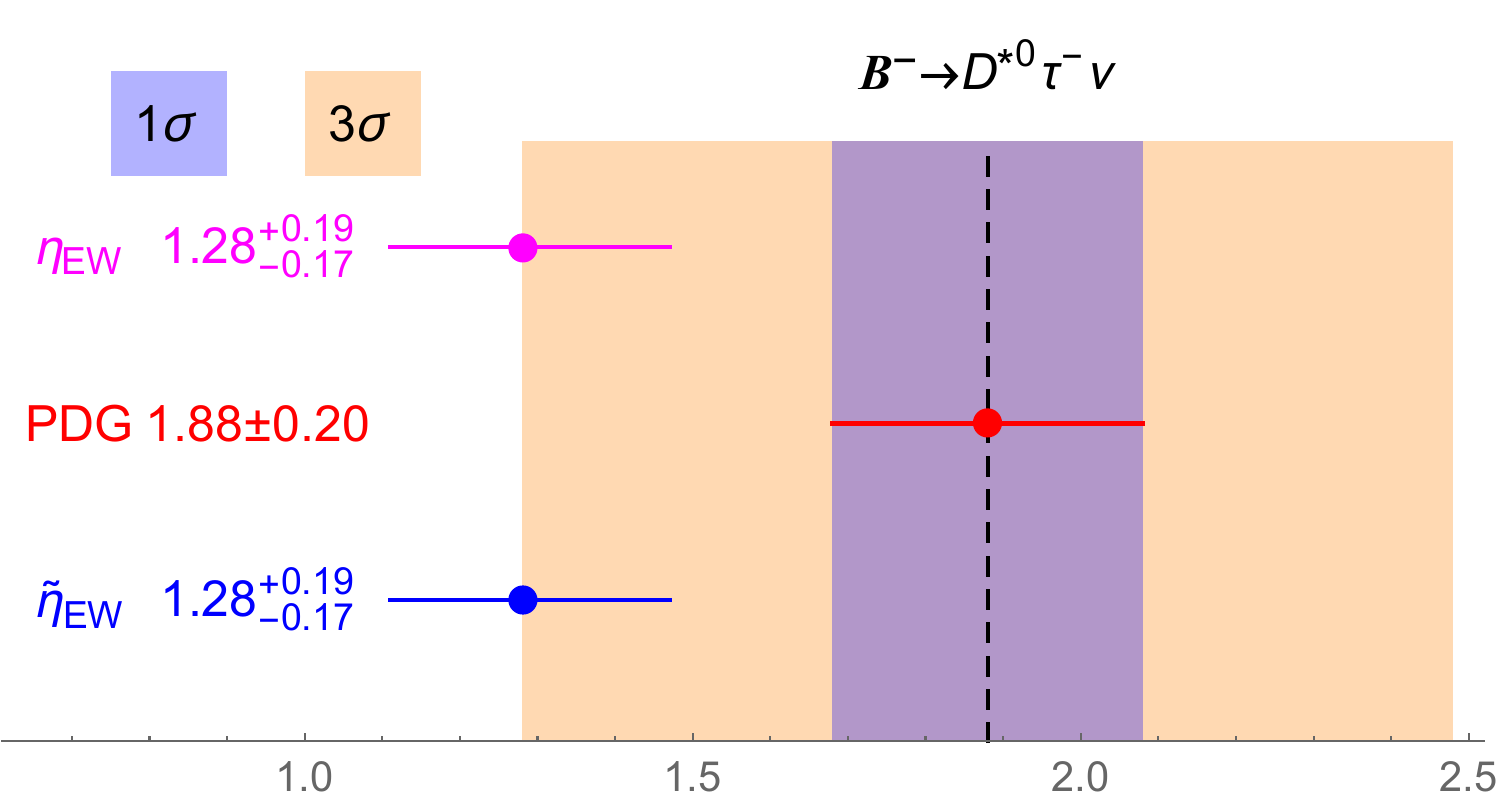} \\
     \includegraphics[width=0.3\textwidth]{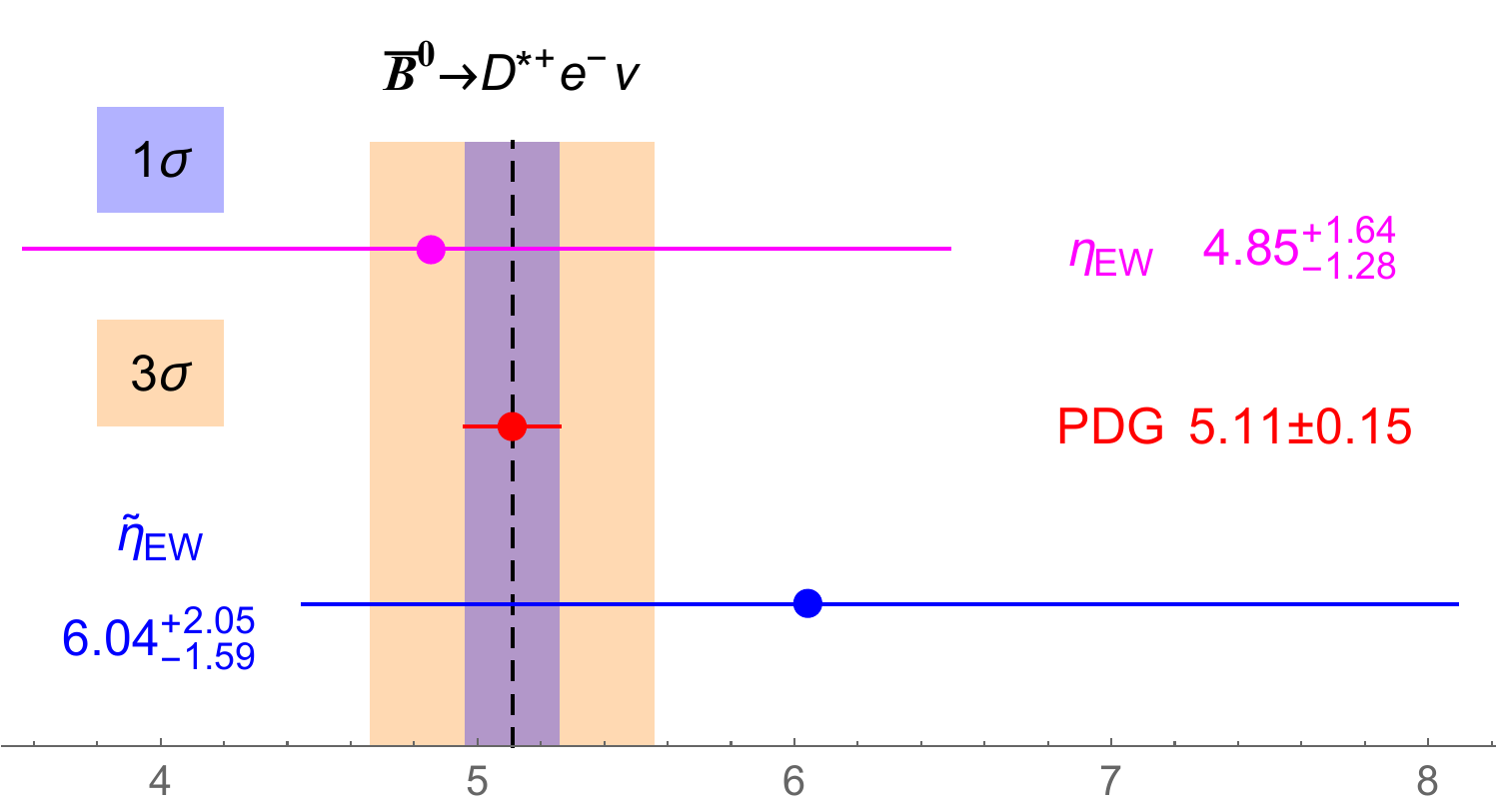} \quad
     \includegraphics[width=0.3\textwidth]{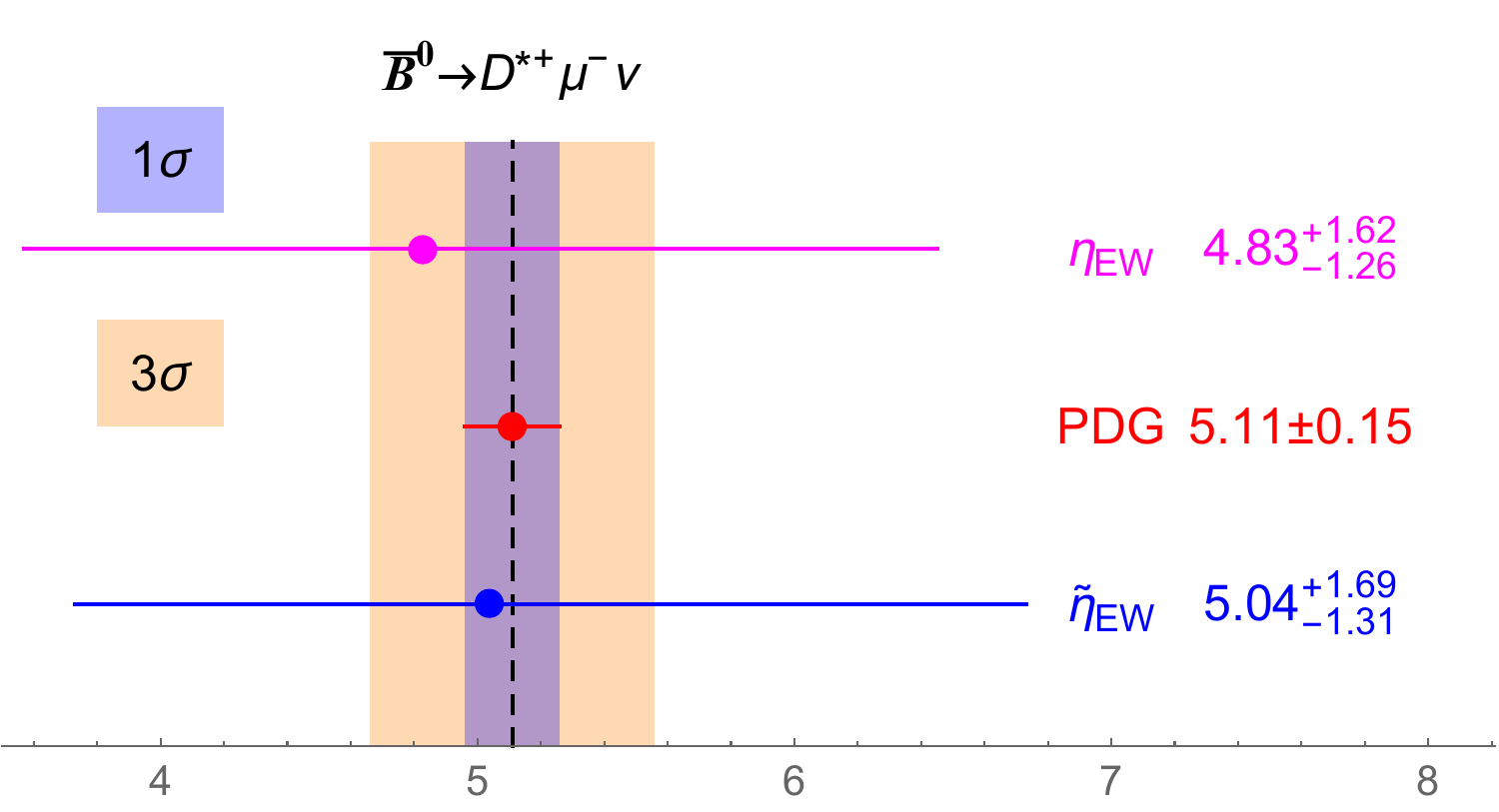} \quad
     \includegraphics[width=0.3\textwidth]{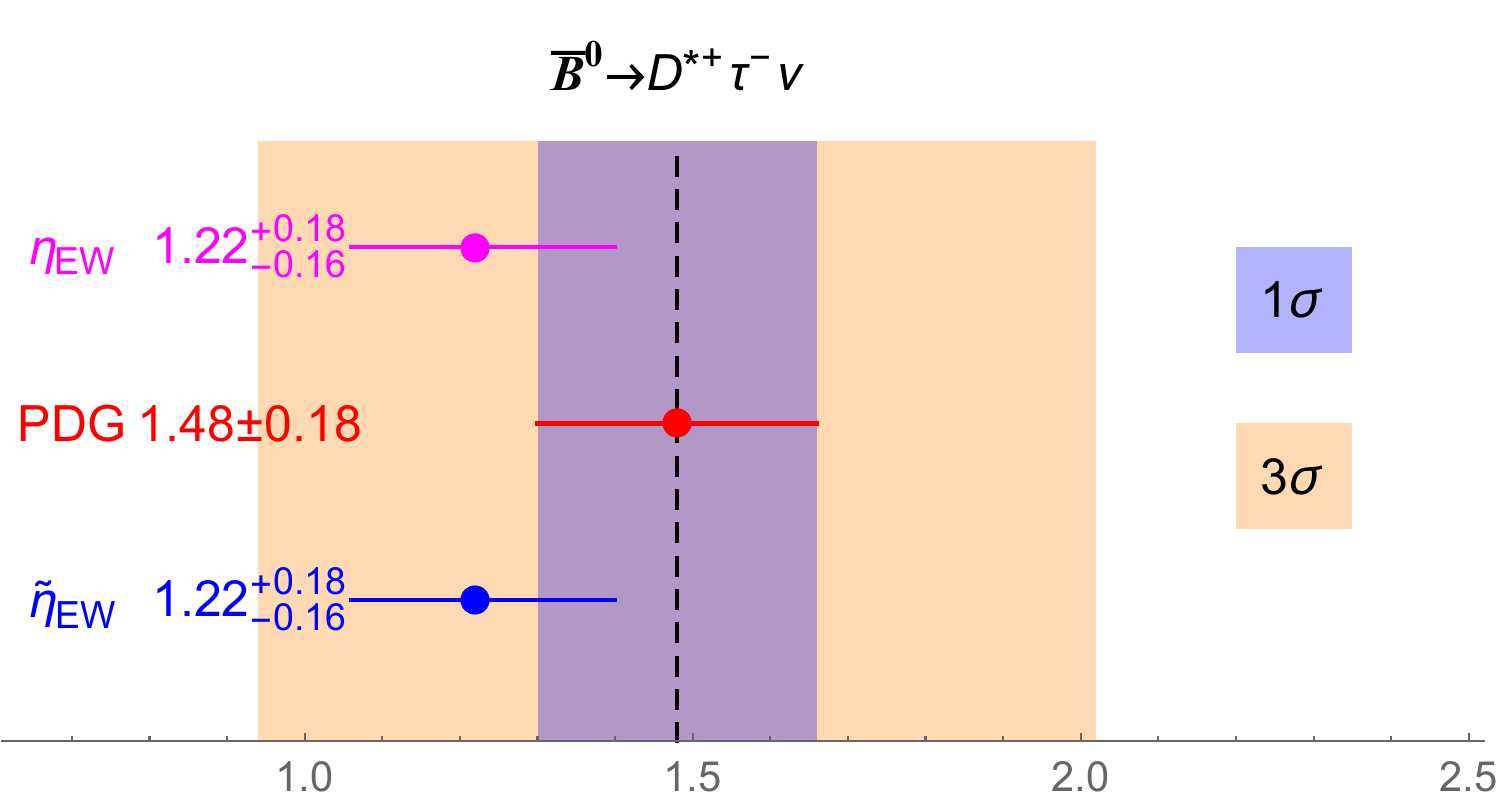} \vspace{-5mm}
     \caption{Branching ratios (in the unit of percentage) for
       the semileptonic $\overline{B}$
       ${\to}$ $D^{({\ast})}$ $+$ ${\ell}^{-}$ $+$
       $\bar{\nu}_{\ell}$ decays, where the dots and lines
       correspond to the central values and uncertainties,
       respectively.}
     \label{fig:br}
     \end{figure}

\begin{figure}[h]
     \subfigure[]{ \includegraphics[width=0.4\textwidth]{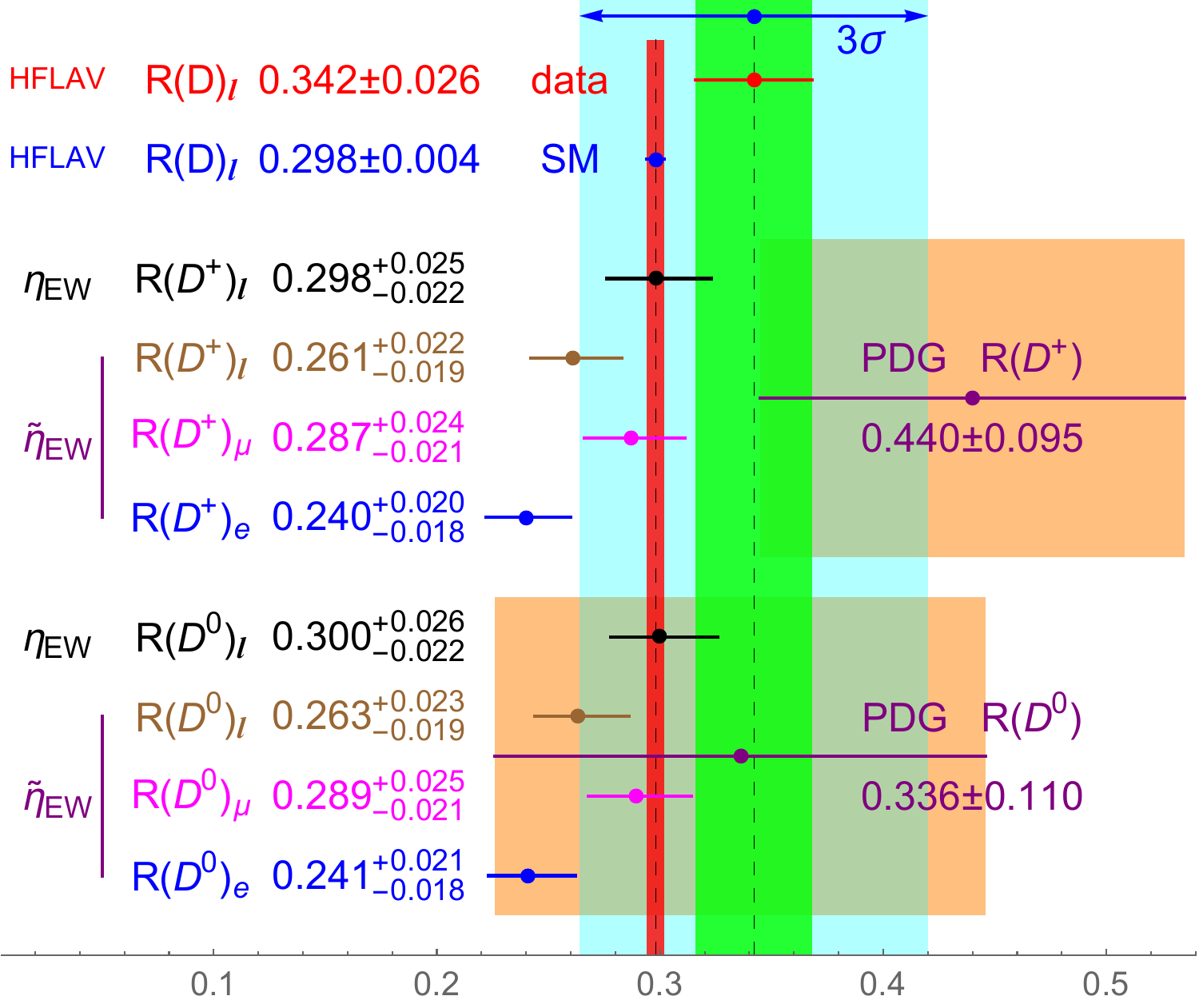}  }
     \subfigure[]{ \includegraphics[width=0.4\textwidth]{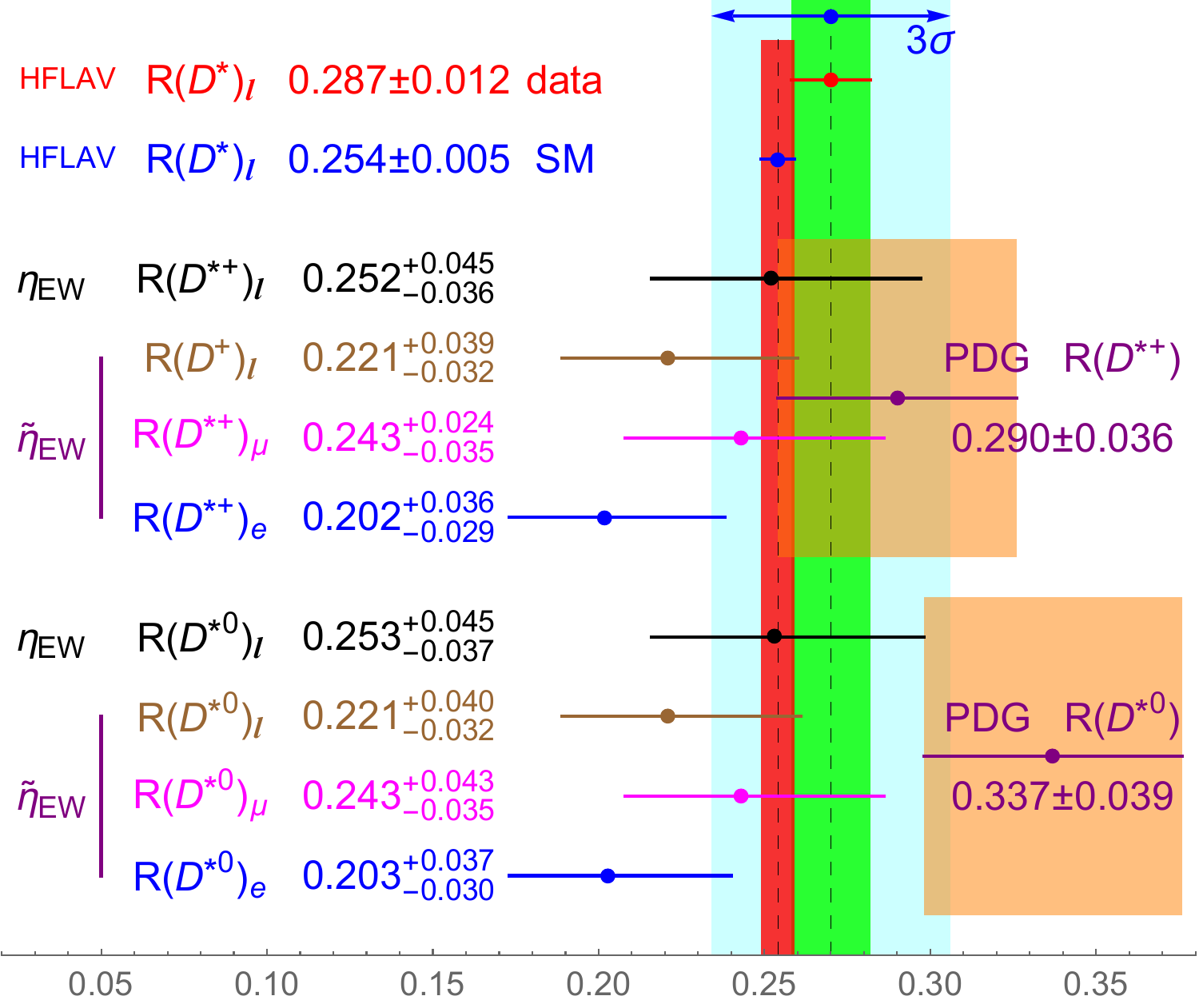} }
     \caption{The distributions of $R(D)$ (a) and $R(D^{\ast})$ (b),
       where the dots and lines correspond to the central values and
       errors, respectively.
       The value of the HFLAV data and SM expectation can
       refer to Ref. \cite{hflav}.}
     \label{fig:rdrdv01}
     \end{figure}

\begin{figure}[h]
     \includegraphics[width=0.5\textwidth]{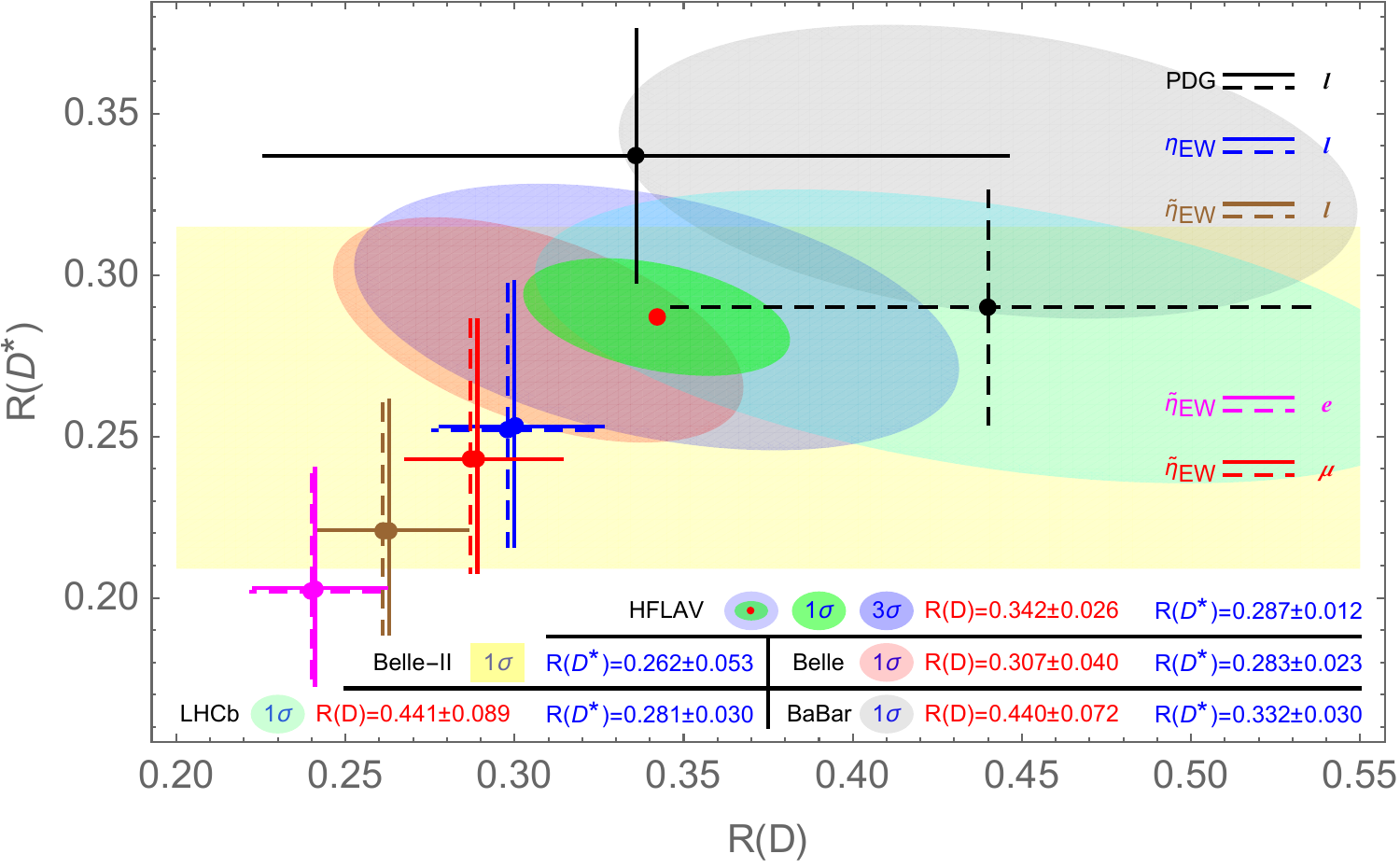} \vspace{-5mm}
     \caption{The $R(D)$-$R(D^{\ast})$ correlation distributions,
          where the dots and lines correspond to the central values
          and errors, respectively.
          The solid (dashed) lines correspond to the ratios
          obtained from the semileptonic $\overline{B}^{0}$ ($B^{-}$)
          decays, respectively. The symbols ${\ell}$ denote the
          subscript of ratios $R(D^{(\ast)})_{\ell}$ with
          ${\ell}$ $=$ $e$ and ${\mu}$.
          The experimental data of PDG, HFLAV, BaBar, Belle, Belle-II,
          and LHCb groups can refer to Refs. \cite{PhysRevD.110.030001,
          hflav,PhysRevD.88.072012,PhysRevLett.124.161803,
          PhysRevD.108.012018,arXiv:2401.02840}. }
     \label{fig:rdrdv02}
     \end{figure}

     With the formula Eq.(\ref{eq:differential-width-dq2-dcos})
     and the inputs listed in Table \ref{tab:input},
     we can obtained the partial decay width, and hence
     branching ratios for the semileptonic $\overline{B}$
     ${\to}$ $D^{({\ast})}$ $+$ ${\ell}^{-}$ $+$
     $\bar{\nu}_{\ell}$ decays.
     Branching ratios are collected in Table \ref{tab:branchingratio}
     and Fig. \ref{fig:br}.
     The dominant theoretical uncertainties come from form factors.
     In addition, it is seen from Eq.(\ref{eq:differential-width-dq2-dcos})
     that the uncertainties arising from the CKM element
     $V_{cb}$ is about $3\%$.
     To reduce the theoretical uncertainties originating from inputs, the ratios
     of branching ratios $R(D)$ in Eq.(\ref{rd-definition}) and
     $R(D^{\ast})$ in Eq.(\ref{rdstar-definition}) are suggested
     by the intelligent particle physicists, and their
     numerical results are presented in Table \ref{tab:ratio},
     and Fig. \ref{fig:rdrdv01} and \ref{fig:rdrdv02}.

     %------------------------------------
     \begin{table}[h]
     \caption{Values of input parameters given by PDG \cite{PhysRevD.110.030001},
              where their central values are regarded as the default inputs
              unless otherwise specified.}
     \label{tab:input}
     \begin{ruledtabular}
     \begin{tabular}{lll}
       $m_{D^{\pm}}$ $=$ $1869.66 \, {\pm} \, 0.05$ MeV,
     & $m_{B^{\pm}}$ $=$ $5279.41 \, {\pm} \, 0.07$ MeV,
     & ${\tau}_{B^{\pm}}$ $=$ $1638 \, {\pm} \, 4$ fs,   \\ \hline
       $m_{D^{0}}$ $=$ $1864.84 \, {\pm} \, 0.05$ MeV,
     & $m_{B^{0}}$ $=$ $5279.72 \, {\pm} \, 0.08$ MeV,
     & ${\tau}_{B^{0}}$ $=$ $1517 \, {\pm} \, 4$ fs,   \\ \hline
       $m_{D^{{\ast}{\pm}}}$ $=$ $2010.26 \, {\pm} \, 0.05$ MeV,
     & $m_{D^{{\ast}0}}$ $=$ $2006.85 \, {\pm} \, 0.05$ MeV,
     & $m_{e}$ $=$ $0.511$ MeV, \\ \hline
       ${\vert} \, V_{cb} \, {\vert}$  $=$ $(39.8 \, {\pm} \, 0.6) \, {\times} \, 10^{-3}$,
     & $m_{\tau}$ $=$ $1776.93 \, {\pm} \, 0.09$ MeV,
     & $m_{\mu}$ $=$ $105.658$ MeV.
     \end{tabular}
     \end{ruledtabular}
     \end{table}
     %------------------------------------
     %------------------------------------
     \begin{table}[h]
     \caption{Branching ratios of the semileptonic $\overline{B}$
        ${\to}$ $D^{({\ast})}$ $+$ ${\ell}^{-}$ $+$ $\bar{\nu}_{\ell}$
        decays in the unit of percentage, where the theoretical
        uncertainties come only from form factors, and
        $\tilde{\eta}_{\rm EW}$ $=$ $1$ $+$ ${\alpha}_{\rm em}\,{\eta}$.}
     \label{tab:branchingratio}
     \begin{ruledtabular}
     \begin{tabular}{lccc}
       \multicolumn{1}{c}{decay mode}
     & ${\eta}_{\rm EW}$ $=$ $1.0066$
     & $\tilde{\eta}_{\rm EW}$ (this work)
     & PDG data \cite{PhysRevD.110.030001} \\ \hline
       $B^{-}$ ${\to}$ $D^{0}$ $+$ $e^{-}$ $+$ $\bar{\nu}$
     & $2.29^{+0.33}_{-0.31}$
     & $2.85^{+0.42}_{-0.38}$
     & $2.29 \, {\pm} \, 0.08$
       \\ \hline
       $B^{-}$ ${\to}$ $D^{0}$ $+$ ${\mu}^{-}$ $+$ $\bar{\nu}$
     & $2.28^{+0.33}_{-0.30}$
     & $2.38^{+0.35}_{-0.32}$
     & $2.29 \, {\pm} \, 0.08$
       \\ \hline
       $B^{-}$ ${\to}$ $D^{0}$ $+$ ${\tau}^{-}$ $+$ $\bar{\nu}$
     & $0.69\, {\pm} \, 0.04$
     & $0.69\, {\pm} \, 0.04$
     & $0.77 \, {\pm} \, 0.25$
       \\ \hline
       $\overline{B}^{0}$ ${\to}$ $D^{+}$ $+$ $e^{-}$ $+$ $\bar{\nu}$
     & $2.14^{+0.31}_{-0.28}$
     & $2.67^{+0.39}_{-0.35}$
     & $2.25 \, {\pm} \, 0.08$
       \\ \hline
       $\overline{B}^{0}$ ${\to}$ $D^{+}$ $+$ ${\mu}^{-}$ $+$ $\bar{\nu}$
     & $2.13^{+0.31}_{-0.28}$
     & $2.22^{+0.32}_{-0.29}$
     & $2.25 \, {\pm} \, 0.08$
       \\ \hline
       $\overline{B}^{0}$ ${\to}$ $D^{+}$ $+$ ${\tau}^{-}$ $+$ $\bar{\nu}$
     & $0.64\, {\pm} \, 0.04$
     & $0.64\, {\pm} \, 0.04$
     & $0.99 \, {\pm} \, 0.21$
       \\ \hline \hline
       $B^{-}$ ${\to}$ $D^{{\ast}0}$ $+$ $e^{-}$ $+$ $\bar{\nu}$
     & $5.08^{+1.73}_{-1.34}$
     & $6.32^{+2.15}_{-1.67}$
     & $5.58 \, {\pm} \, 0.26$
       \\ \hline
       $B^{-}$ ${\to}$ $D^{{\ast}0}$ $+$ ${\mu}^{-}$ $+$ $\bar{\nu}$
     & $5.06^{+1.70}_{-1.33}$
     & $5.27^{+1.78}_{-1.38}$
     & $5.58 \, {\pm} \, 0.26$
       \\ \hline
       $B^{-}$ ${\to}$ $D^{{\ast}0}$ $+$ ${\tau}^{-}$ $+$ $\bar{\nu}$
     & $1.28^{+0.19}_{-0.17}$
     & $1.28^{+0.19}_{-0.17}$
     & $1.88 \, {\pm} \, 0.20$
       \\ \hline
       $\overline{B}^{0}$ ${\to}$ $D^{{\ast}+}$ $+$ $e^{-}$ $+$ $\bar{\nu}$
     & $4.85^{+1.64}_{-1.28}$
     & $6.04^{+2.05}_{-1.59}$
     & $5.11 \, {\pm} \, 0.15$
       \\ \hline
       $\overline{B}^{0}$ ${\to}$ $D^{{\ast}+}$ $+$ ${\mu}^{-}$ $+$ $\bar{\nu}$
     & $4.83^{+1.62}_{-1.26}$
     & $5.04^{+1.69}_{-1.31}$
     & $5.11 \, {\pm} \, 0.15$
       \\ \hline
       $\overline{B}^{0}$ ${\to}$ $D^{{\ast}+}$ $+$ ${\tau}^{-}$ $+$ $\bar{\nu}$
     & $1.22^{+0.18}_{-0.16}$
     & $1.22^{+0.18}_{-0.16}$
     & $1.48 \, {\pm} \, 0.18$
     \end{tabular}
     \end{ruledtabular}
     \end{table}
     %------------------------------------
     %------------------------------------
     %------------------------------------
     %------------------------------------
     \begin{table}[h]
     \caption{Ratios of branching ratios for semileptonic
        $\overline{B}$ ${\to}$ $D^{({\ast})}$ $+$ ${\ell}^{-}$ $+$
        $\bar{\nu}_{\ell}$ decays, where the theoretical uncertainties
        come only from form factors.
        The definition of $R(D)$ and $R(D^{\ast})$ are given in
        Eq.(\ref{rd-definition}) and  Eq.(\ref{rdstar-definition}), respectively .
        The subscript ${\ell}$ of ratio $R(D)_{\ell}$ denote the
        final charged lepton, with ${\ell}$ $=$ $e$ and ${\mu}$.
        The PDG ratios are obtained
        with data in Table \ref{tab:branchingratio}.}
     \label{tab:ratio}
     \begin{ruledtabular}
     \begin{tabular}{ccccc}
       ratios
     & ${\eta}_{\rm EW}$ $=$ $1.0066$
     & $\tilde{\eta}_{\rm EW}$ (this work)
     & PDG data \cite{PhysRevD.110.030001}
     & HFLAV \cite{hflav} \\ \hline
       $R(D^{0})_{e}$
     & $ 0.300^{+0.026}_{-0.022} $
     & $ 0.241^{+0.021}_{-0.018} $
     & \multirow{3}{*}{$0.336 \, {\pm} \, 0.110$}
     & \multirow{6}{*}{$0.342 \, {\pm} \, 0.026$}
       \\ \cline{1-3}
      $R(D^{0})_{\mu}$
     & $ 0.301^{+0.026}_{-0.022} $
     & $ 0.289^{+0.025}_{-0.021} $
       \\ \cline{1-3}
       $R(D^{0})_{\ell}$
     & $ 0.300^{+0.026}_{-0.022} $
     & $ 0.263^{+0.023}_{-0.019} $
       \\ \cline{1-4}
       $R(D^{+})_{e}$
     & $ 0.298^{+0.025}_{-0.022} $
     & $ 0.240^{+0.020}_{-0.018} $
     & \multirow{3}{*}{$0.440 \, {\pm} \, 0.095$}
       \\ \cline{1-3}
       $R(D^{+})_{\mu}$
     & $ 0.299^{+0.025}_{-0.022} $
     & $ 0.287^{+0.024}_{-0.021} $
       \\ \cline{1-3}
       $R(D^{+})_{\ell}$
     & $ 0.298^{+0.025}_{-0.022} $
     & $ 0.261^{+0.022}_{-0.019} $
       \\ \hline \hline
       $R(D^{{\ast}0})_{e}$
     & $ 0.252^{+0.046}_{-0.037} $
     & $ 0.203^{+0.037}_{-0.030} $
     & \multirow{3}{*}{$0.337 \, {\pm} \, 0.039$}
     & \multirow{6}{*}{$0.287 \, {\pm} \, 0.012$}
       \\ \cline{1-3}
       $R(D^{{\ast}0})_{\mu}$
     & $ 0.253^{+0.045}_{-0.036} $
     & $ 0.243^{+0.043}_{-0.035} $
       \\ \cline{1-3}
       $R(D^{{\ast}0})_{\ell}$
     & $ 0.253^{+0.045}_{-0.037} $
     & $ 0.221^{+0.040}_{-0.032} $
       \\ \cline{1-4}
       $R(D^{{\ast}+})_{e}$
     & $ 0.251^{+0.045}_{-0.037} $
     & $ 0.202^{+0.036}_{-0.029} $
     & \multirow{3}{*}{$0.290 \, {\pm} \, 0.036$}
       \\ \cline{1-3}
       $R(D^{{\ast}+})_{\mu}$
     & $ 0.252^{+0.045}_{-0.036} $
     & $ 0.243^{+0.043}_{-0.035} $
       \\ \cline{1-3}
       $R(D^{{\ast}+})_{\ell}$
     & $ 0.252^{+0.045}_{-0.036} $
     & $ 0.221^{+0.039}_{-0.032} $
     \end{tabular}
     \end{ruledtabular}
     \end{table}
     %------------------------------------
     %------------------------------------
     %------------------------------------
     %------------------------------------
     %------------------------------------

     There are some comments on the numerical results.

     (1)
     It is seen from Table \ref{tab:branchingratio} and Fig.~\ref{fig:br} that
     when the contributions from the QED vertex corrections are
     considered, branching ratios for the semi-electronic and
     semi-muonic decays are relatively larger than those obtained
     with the universal factor ${\eta}_{\rm EW}$.
     Branching ratios for the semi-muonic decays for the
     $\tilde{\eta}_{\rm EW}$ case are in better agreement with the
     experimental data and Belle-II recent results\footnotemark[3],
     \footnotetext[3]{Only the statistical uncertainties are given.}
     ${\cal B}(B^{-}{\to}D^{{\ast}0}{\ell}^{-}\bar{\nu})$
     $=$ $(5.50^{+0.28}_{-0.27})\,\%$ and
     ${\cal B}(\overline{B}^{0}{\to}D^{{\ast}+}{\ell}^{-}\bar{\nu})$
     $=$ $(5.27^{+0.25}_{-0.24})\,\%$ \cite{arXiv:2401.02840}.
     Branching ratios for the semi-tauonic decays with $\tilde{\eta}_{\rm EW}$
     and ${\eta}_{\rm EW}$ are well consistent with each other.
     This is the reason why the ratios $R(D)$ and $R(D^{\ast})$
     obtained with $\tilde{\eta}_{\rm EW}$ are generally smaller than
     those with ${\eta}_{\rm EW}$ in Table \ref{tab:ratio}.
     In addition, the difference of branching ratios between the
     semi-electronic and semi-muonic decays is negligible for
     the ${\eta}_{\rm EW}$ case, but significant for the
     $\tilde{\eta}_{\rm EW}$ case.
     The effects of the $\tilde{\eta}_{\rm EW}$ on branching ratios for
     the semi-electronic and semi-muonic decays are comparable with
     those of uncertainties from form factors,
     which is also manifested in Ref. \cite{JHEP.2021.10.223}.

     (2)
     Theoretically, branching ratios are highly subject to the shape
     lines of form factors.
     It is seen from Fig. \ref{fig-ffp} and \ref{fig-ffv} that
     the form factors are relatively exactly determined near the
     $q^{2}_{\rm max}$ regions, but uncertain at the lower $q^{2}$
     regions for the moment.
     For the semi-electronic and semi-muonic decays, the form factors
     across almost the entire $q^{2}$ regions are involved.
     For the semi-tauonic decays, only the form factors within
     the higher $q^{2}$ partial regions are involved due to the
     massive tauon.
     In addition, experimentally, both the electron and muon are
     more easily identified than the tauon, because
     the tauon is too short-lived to clearly leave its footprints
     in the detectors.
     This is why the theoretical uncertainties of branching ratios for
     the semi-electronic and semi-muonic decays are much larger
     than experimental ones in Table \ref{tab:branchingratio} and
     Fig.~\ref{fig:br},
     while the reverse is true for the semi-tauonic decays.
     Moreover, most of branching ratios for the semi-electronic and
     semi-muonic decays agree well with data within one error, except
     for the $\overline{B}$ ${\to}$ $D$ $+$ $e^{-}$ $+$ $\bar{\nu}_{e}$
     decays.
     The experimental data on branching ratios for the semi-tauonic
     decays are generally larger than expected by SM.
     It is eagerly expected to improve the theoretical and
     experimental precision in the future.

     (3)
     It is widely known that the smaller the lepton mass, the larger
     the phase spaces of the final particles. Hence, according to
     the mass of lepton, the magnitude on branching ratio for the
     semileptonic $B$ decays containing a particular charmed meson
     in the final states, in the descending order, is
     ${\cal B}(\overline{B}{\to}D^{(\ast)}e^{-}\bar{\nu}_{e})$ $>$
     ${\cal B}(\overline{B}{\to}D^{(\ast)}{\mu}^{-}\bar{\nu}_{\mu})$ $>$
     ${\cal B}(\overline{B}{\to}D^{(\ast)}{\tau}^{-}\bar{\nu}_{\tau})$
     with either $\tilde{\eta}_{\rm EW}$ or ${\eta}_{\rm EW}$, which
     has been well verified in part by the data on the semi-muonic and
     semi-tauonic $B$ decays, and results in the general
     relationship $R(D^{(\ast)})_{e}$ $<$ $R(D^{(\ast)})_{\mu}$
     in Table \ref{tab:ratio}
     waiting to be checked by the more precise measurements.
     In the future, a more careful experimental investigation on the
     ratios $R(D^{(\ast)})_{e}$ $<$ $R(D^{(\ast)})_{\mu}$   %% $R_{e/{\mu}}$ $=$ $R(D^{(\ast)})_{\mu}/R(D^{(\ast)})_{e}$
     is expected to testify the QED nonfactoriziable effects.

     %------------------------------------
     \begin{table}[h]
     \caption{Contributions of different helicity amplitudes
       for the $B^{-}$ ${\to}$ $D^{{\ast}0}$ $+$ ${\ell}$
       $+$ ${\nu}$ decays (in the unit of percentage),
       where ${\Gamma}_{B}$ is the total width of the $B^{-}$ meson,
       ${\Gamma}$ is the partial width of the concerned semileptonic
       decay, and ${\Gamma}_{i}$ (with $i$ $=$ $U$, $L$, $P$, $S$)
       corresponds to the $H_{i}$
       in Eq.(\ref{eq:unpolarized-transverse-component}),
       Eq.(\ref{eq:parity-odd-component}),
       Eq.(\ref{eq:longitudinal-component}), and
       Eq.(\ref{eq:scalar-component}).
       $f_{\perp}$ $=$ ${\Gamma}_{U}/{\Gamma}$,
       $f_{L}$ $=$ ${\Gamma}_{L}/{\Gamma}$,
       and $f_{S}$ $=$ ${\Gamma}_{S}/{\Gamma}$.}
     \label{tab:helicity-contribution}
     \begin{ruledtabular}
     \begin{tabular}{c|c|ccccc|ccc|c}
       \multicolumn{2}{c|}{case}
     & ${\Gamma}_{U}/{\Gamma}_{B}$
     & ${\Gamma}_{L}/{\Gamma}_{B}$
     & ${\Gamma}_{P}/{\Gamma}_{B}$
     & ${\Gamma}_{S}/{\Gamma}_{B}$
     & ${\Gamma}/{\Gamma}_{B}$
     & $f_{\perp}$
     & $f_{L}$
     & $f_{S}$
     & $f_{L}^{\rm exp.}$ \cite{PhysRevD.108.012002}  \\ \hline
       \multirow{2}{*}{${\ell}$ $=$ $e$}
     & ${\eta}_{\rm EW}$       & $2.494$ & $2.587$ &   ---   & ${\sim}$ $0$ & $5.081$ & $49.1$ & $50.9$ & ${\sim}$ $0$
     & \multirow{2}{*}{$50.5{\pm}2.8$} \\ \cline{2-10}
     & $\tilde{\eta}_{\rm EW}$ & $3.100$ & $3.218$ & $0.0025$ & ${\sim}$ $0$ & $6.320$ & $49.1$ & $50.9$ & ${\sim}$ $0$ \\ \hline
       \multirow{2}{*}{${\ell}$ $=$ ${\mu}$}
     & ${\eta}_{\rm EW}$       & $2.485$ & $2.551$ &   ---   & $0.023$ & $5.059$ & $49.1$ & $50.4$ & $0.5$
     & \multirow{2}{*}{$52.2{\pm}2.6$}\\ \cline{2-10}
     & $\tilde{\eta}_{\rm EW}$ & $2.588$ & $2.658$ & $0.0019$ & $0.024$ & $5.272$ & $49.1$ & $50.4$ & $0.5$ \\ \hline
       \multirow{2}{*}{${\ell}$ $=$ ${\tau}$}
     & ${\eta}_{\rm EW}$       & $0.710$ & $0.463$ &   ---        & $0.108$ & $1.281$ & $55.4$ & $36.1$ & $8.4$ \\ \cline{2-10}
     & $\tilde{\eta}_{\rm EW}$ & $0.711$ & $0.464$ & ${\sim}$ $0$ & $0.108$ & $1.283$ & $55.4$ & $36.2$ & $8.4$
     \end{tabular}
     \end{ruledtabular}
     \end{table}
     %------------------------------------
     %------------------------------------
     \begin{figure}[h]
     \includegraphics[width=0.3\textwidth]{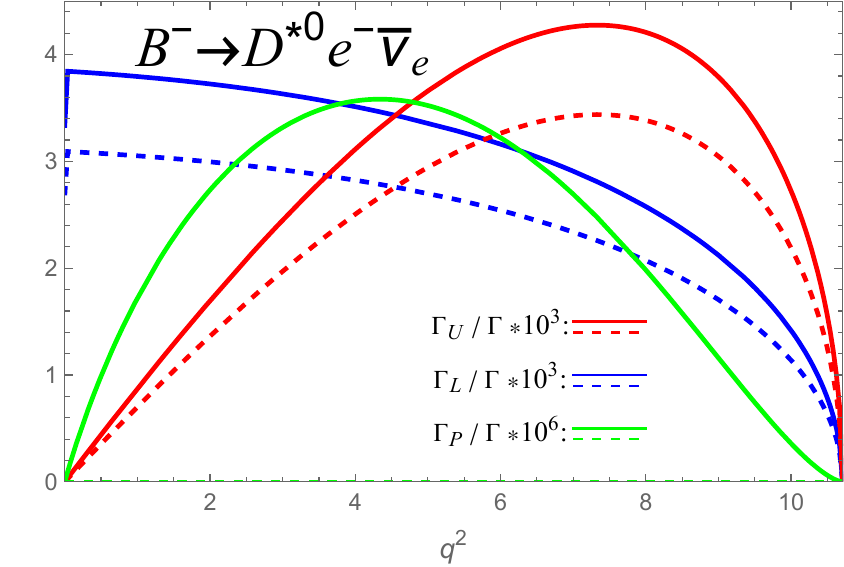} \quad
     \includegraphics[width=0.3\textwidth]{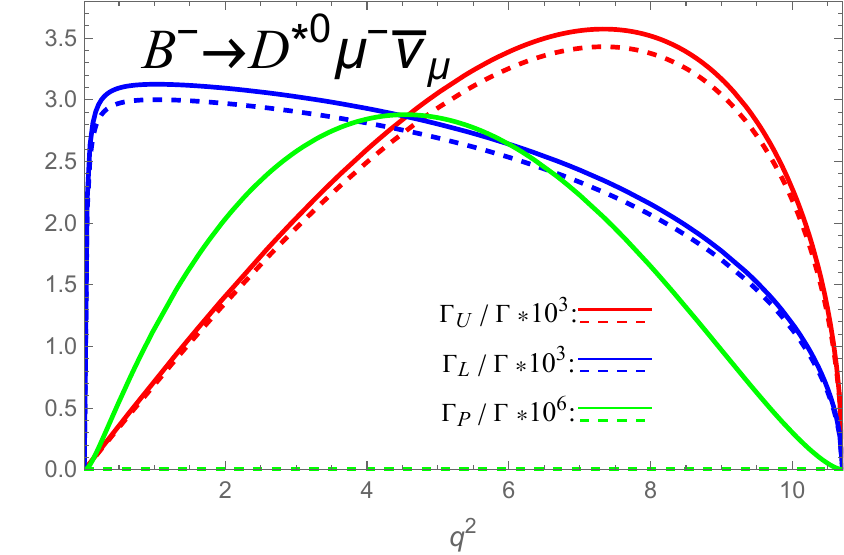} \quad
     \includegraphics[width=0.3\textwidth]{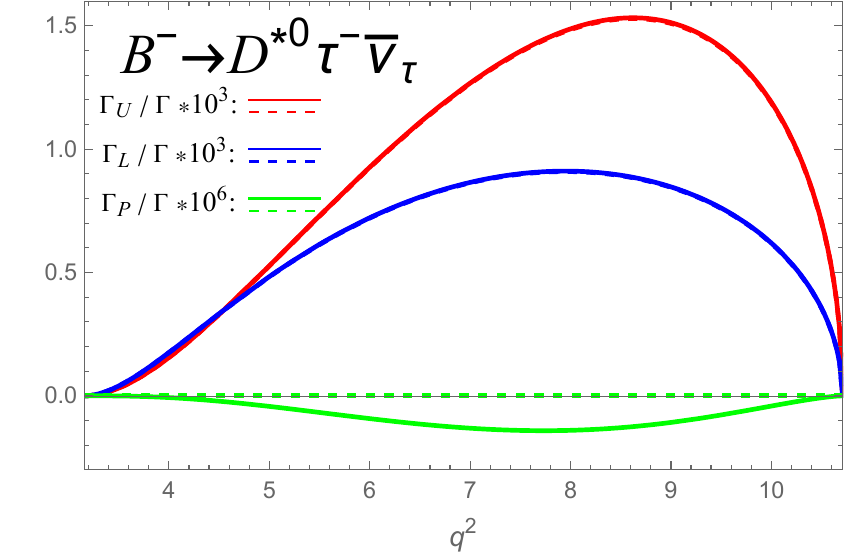} \\ \vspace{5mm}
     \includegraphics[width=0.3\textwidth]{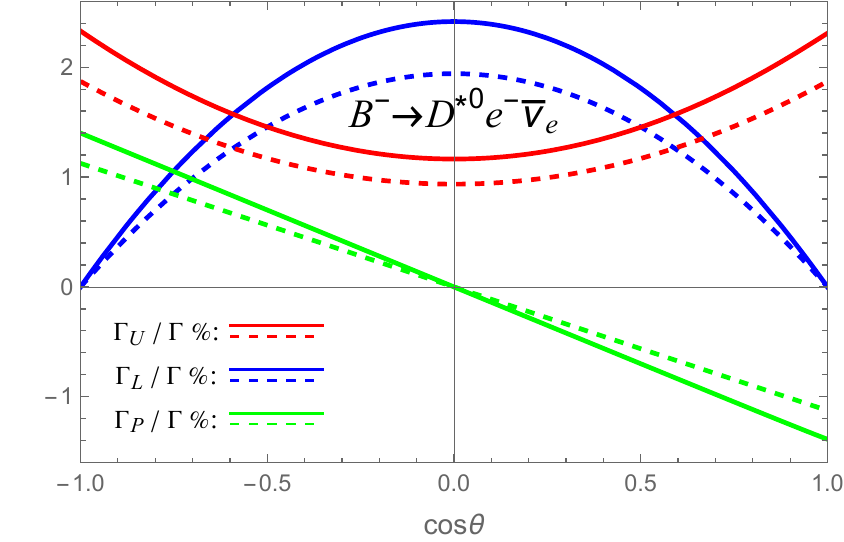} \quad
     \includegraphics[width=0.3\textwidth]{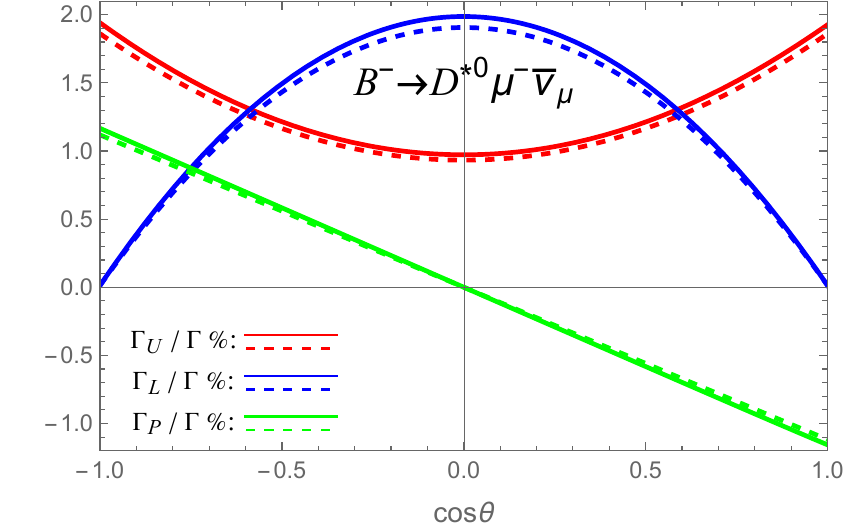} \quad
     \includegraphics[width=0.3\textwidth]{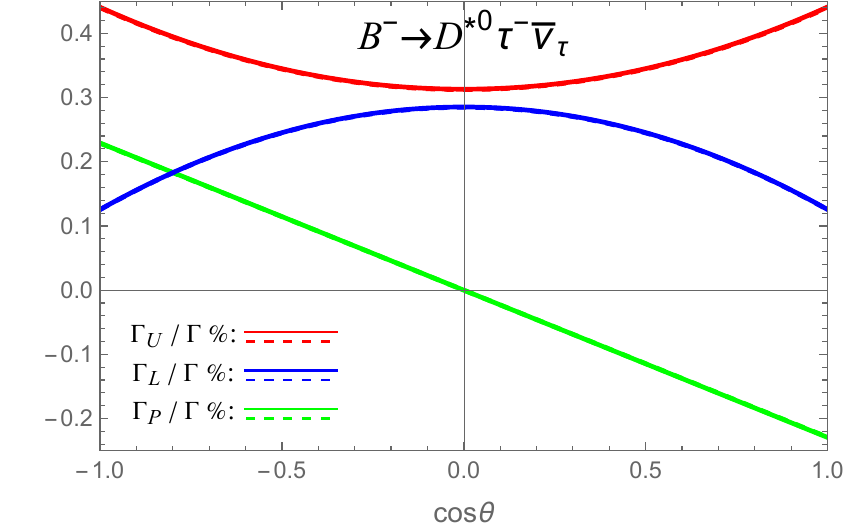} \vspace{-5mm}
     \caption{Contributions of different helicity amplitudes
       for the $B^{-}$ ${\to}$ $D^{{\ast}0}$ $+$ ${\ell}$
       $+$ ${\nu}$ decays, where the solid (dashed) lines correspond
       to the $\tilde{\eta}_{\rm EW}$ (${\eta}_{\rm EW}$) cases.}
     \label{fig:helicity-contribution}
     \end{figure}
     %------------------------------------

     (4)
     A fascinating phenomenon is that branching ratios
     ${\cal B}(\overline{B}{\to}D{\ell}^{-}\bar{\nu})$ $<$
     ${\cal B}(\overline{B}{\to}D^{\ast}{\ell}^{-}\bar{\nu})$
     for a particular lepton pair ${\ell}\bar{\nu}$ in the
     final states.
     The reason is that the spin angular momentum of the
     initial $\overline{B}$ meson is zero.
     According to the law of conservation of angular momentum,
     the orbital angular momentum between the pseudoscalar $D$ meson
     and the ${\ell}\bar{\nu}$ pair should be $L$ $=$ $0$.
     Only the $S$-wave contributes to the helicity amplitudes
     for the $\overline{B}$ ${\to}$ $D$ $+$ ${\ell}^{-}$ $+$
     $\bar{\nu}$ decay, corresponding to the scalar $H_{t}$ and
     longitudinal $H_{0}$ components, see Eq.(\ref{eq:hp-hm-B2D}),
     Eq.(\ref{eq:hz-B2D}) and Eq.(\ref{eq:ht-B2D}).
     However, the orbital angular momentum between the vector
     $D^{\ast}$ meson and the ${\ell}\bar{\nu}$ pair could be
     $L$ $=$ $0$, $1$ and $2$. Besides the scalar $H_{t}$ and
     longitudinal $H_{0}$ components, there are also the nonzero
     transverse $H_{\pm}$ components contributing to the
     helicity amplitudes for the $\overline{B}$ ${\to}$
     $D^{\ast}$ $+$ ${\ell}^{-}$ $+$ $\bar{\nu}$ decay,
     see Eq.(\ref{eq:hp-hm-B2Dv}), Eq.(\ref{eq:hz-B2Dv})
     and Eq.(\ref{eq:ht-B2Dv}).
     What's more, it is seen from Table \ref{tab:branchingratio} that
     there exists a certain relationship
     ${\cal B}(\overline{B}{\to}D^{\ast}{\ell}^{-}\bar{\nu})$
     ${\ge}$ $2\,{\cal B}(\overline{B}{\to}D{\ell}^{-}\bar{\nu})$ with
     ${\ell}$ $=$ $e$ and ${\mu}$ for both the $\tilde{\eta}_{\rm EW}$
     and ${\eta}_{\rm EW}$ cases, which on one hand have been
     evidenced by experimental data, and on the other hand indicate the
     significant role of the contributions of transverse helicity
     amplitudes.
     Here, we take the $B^{-}$ ${\to}$ $D^{{\ast}0}$ $+$ ${\ell}$
     $+$ ${\nu}$ decays as an example to illustrate the contributions
     of different helicity amplitudes in Table \ref{tab:helicity-contribution}
     and Fig. \ref{fig:helicity-contribution}.
     These results seem to show the following facts.
     (a)
     the QED nonfactorizable corrections will enhance both
     transverse and longitudinal amplitudes according to the
     mass of the charged lepton. The smaller the lepton mass,
     the larger the enhancements including both the longitudinal and
     transverse contributions versus whether the $q^{2}$ distributions
     or the ${\cos}{\theta}$ distributions.
     The enhancements for the semi-tauonic decays are
     unnoticeable and negligible.
     (b)
     Although the transverse and longitudinal amplitudes differ
     with each other,
     the polarization fractions have almost nothing to do with the
     factor ${\eta}_{\rm EW}$ and $\tilde{\eta}_{\rm EW}$.
     The longitudinal fractions for the semi-electronic and
     semi-muonic decays agree well with the Belle
     measurements \cite{PhysRevD.108.012002}.
     (c)
     The transverse (longitudinal) fractions increase (decrease)
     with the decreases of $q^{2}$ on one hand, and simultaneously
     with the decreases of the phase spaces of final states
     which is very similar to the situation of the nonleptonic
     $B$ decays, such as the $B$ ${\to}$ ${\psi}(1S,2S)$ $+$
     $K^{{\ast}0}$ decays \cite{PhysRevD.110.030001}.
     The transverse fractions are comparable with the
     longitudinal ones for the semi-electronic and semi-muonic
     decays, while the transverse fractions exceed
     the longitudinal ones for the semi-tauonic decays.
     It will be very promising to measure and study the relatively
     large transverse fractions with the progress of experimental
     analytical techniques in the future.
     (d)
     The scalar helicity contributions are directly proportional to
     the mass square of the lepton, which is seen from
     Eq.(\ref{eq:differential-width-dq2-dcos}).
     Because the tauon is massive, the importance of the scalar
     fractions becomes evident for the semi-tauonic decays,
     compared with the semi-electronic and semi-muonic decays
     where the scalar fractions are considerably less stark.

     (5)
     It is seen from Eq.(\ref{eq:differential-width-dq2-dcos}) and
     Fig. \ref{fig:helicity-contribution} that the contributions
     from the parity-odd $H_{P}$ term in the ${\cos}{\theta}$
     $<$ $0$ regions have the same magnitudes but with the
     opposite signs as those in the ${\cos}{\theta}$ $>$ $0$
     regions, which results in the absence of the $H_{P}$ term
     in Eq.(\ref{eq:differential-width-dq2-02}) for the universal
     ${\eta}_{\rm EW}$ case.
     The QED corrections make the $H_{P}$ contributions reappear.
     If only the ${\cos}{\theta}$ ${\in}$ $[-1,0]$
     (or ${\cos}{\theta}$ ${\in}$ $[0,+1]$) regions rather than
     the entire ${\cos}{\theta}$ ${\in}$ $[-1,+1]$ regions are
     considered, the contributions from the parity-odd $H_{P}$ term
     to branching ratios for the semileptonic $B$ decays
     are the same order of magnitude as those of the transverse
     $H_{U}$ and/or longitudinal $H_{L}$ contributions.
     Hence, the physical observables for the same direction and
     opposite direction distributions over ${\cos}{\theta}$,
     which was called the forward-backward asymmetry in
     Ref. \cite{PhysRevD.108.012002}, are suggested for the
     $\overline{B}$ ${\to}$ $D^{\ast}$ $+$ ${\ell}^{-}$ $+$
     $\bar{\nu}$ decays,
     %------------------------------------
     \begin{equation}
    {\cal A}_{\rm FB} \, = \,
     \frac{\displaystyle {\int}_{-1}^{0}\frac{d\,{\Gamma}(\overline{B}{\to}D^{\ast}{\ell}^{-}\bar{\nu})}
                                             {d\,{\cos}{\theta}}\, d\,{\cos}{\theta}
                        -{\int}_{0}^{+1}\frac{d\,{\Gamma}(\overline{B}{\to}D^{\ast}{\ell}^{-}\bar{\nu})}
                                             {d\,{\cos}{\theta}}\, d\,{\cos}{\theta} }
          {\displaystyle {\int}_{-1}^{0}\frac{d\,{\Gamma}(\overline{B}{\to}D^{\ast}{\ell}^{-}\bar{\nu})}
                                             {d\,{\cos}{\theta}}\, d\,{\cos}{\theta}
                        +{\int}_{0}^{+1}\frac{d\,{\Gamma}(\overline{B}{\to}D^{\ast}{\ell}^{-}\bar{\nu})}
                                             {d\,{\cos}{\theta}}\, d\,{\cos}{\theta} }
     \label{eq:forward-backward-asymmetry},
     \end{equation}
     %------------------------------------
     which not only provide another constraints to determine the $B$ ${\to}$
     $D^{\ast}$ form factors, but also can be used to investigate the
     parity-odd contributions\footnotemark[4], the parity-violating
     asymmetries and the QED radiative corrections as well.
     \footnotetext[4]{For the semi-electronic and semi-muonic decays,
        the terms proportional to $m_{\ell}^{2}$ in
        Eq.(\ref{eq:differential-width-dq2-dcos}) are often neglected.
        So actually, the forward-backward asymmetry is twice as
        much as the parity-odd fraction ${\Gamma}_{P}/{\Gamma}$ within
        the ${\cos}{\theta}$ $<$ $0$ regions.
        For the semi-tauonic decays, the movement direction of tauon
        can hardly be reconstructed correctly at experiments due to at
        least two neutrinos in the final states.
        So the forward-backward asymmetry will have no practical
        significance for the semi-tauonic decays.}
     The numerical results on the forward-backward asymmetries are
     presented in Table \ref{tab:forward-backward}. It is shown that
     (a)
     theoretical predictions on ${\cal A}_{\rm FB}^{e}$ for the
     semi-electronic decays are about $0.22$ for both the
     ${\eta}_{\rm EW}$ and $\tilde{\eta}_{\rm EW}$ cases,
     and agree well with the Belle measurements inside one error
     margin scope \cite{PhysRevD.108.012002}.
     (b)
     Theoretical predictions on ${\cal A}_{\rm FB}^{\mu}$ for
     the semi-muonic decays are slightly smaller than
     ${\cal A}_{\rm FB}^{e}$, while the Belle measurements
     show ${\cal A}_{\rm FB}^{\mu}$ ${\simeq}$ $0.28$
     beyond expectations for the $\overline{B}^{0}$ ${\to}$
     $D^{{\ast}+}$ $+$ ${\mu}^{-}$ $+$ $\bar{\nu}$ decay.
     (c)
     The QED radiative corrections to ${\cal A}_{\rm FB}$ are
     completely overwhelmed by uncertainties from from factors.

     %------------------------------------
     \begin{table}[h]
     \caption{The forward-backward asymmetries ${\cal A}_{\rm FB}$ for
       the $\overline{B}$ ${\to}$ $D^{\ast}$ $+$ ${\ell}^{-}$ $+$
       $\bar{\nu}$ decays, where the theoretical uncertainties come
       only from form factors.}
     \label{tab:forward-backward}
     \begin{ruledtabular}
     \begin{tabular}{cccc}
       decay mode
     & ${\eta}_{\rm EW}$
     & $\tilde{\eta}_{\rm EW}$
     & data \cite{PhysRevD.108.012002}  \\ \hline
       $B^{-}$ ${\to}$ $D^{{\ast}0}$ $+$ $e^{-}$ $+$ $\bar{\nu}$
     & $0.221^{+0.015}_{-0.013}$
     & $0.222^{+0.015}_{-0.013}$
     & $0.234\, {\pm} \, 0.027$
       \\ \hline
       $B^{-}$ ${\to}$ $D^{{\ast}0}$ $+$ ${\mu}^{-}$ $+$ $\bar{\nu}$
     & $0.216^{+0.016}_{-0.014}$
     & $0.217^{+0.016}_{-0.014}$
     & $0.243\, {\pm} \, 0.027$
       \\ \hline
       $\overline{B}^{0}$ ${\to}$ $D^{{\ast}+}$ $+$ $e^{-}$ $+$ $\bar{\nu}$
     & $0.219^{+0.015}_{-0.013}$
     & $0.220^{+0.015}_{-0.013}$
     & $0.218\, {\pm} \, 0.031$
       \\ \hline
       $\overline{B}^{0}$ ${\to}$ $D^{{\ast}+}$ $+$ ${\mu}^{-}$ $+$ $\bar{\nu}$
     & $0.214^{+0.015}_{-0.013}$
     & $0.215^{+0.015}_{-0.013}$
     & $0.281\, {\pm} \, 0.033$
     \end{tabular}
     \end{ruledtabular}
     \end{table}
     %------------------------------------

     (6)
     Generally, theoretical results on the ratios of both $R(D)$
     and $R(D^{\ast})$ are smaller than the PDG and HFLAV
     averages (see Fig.~\ref{fig:rdrdv01}), especially the
     differences between the SM predictions and HFLAV averages
     including correlations among the related measurements are
     larger than three errors in the $R(D)$-$R(D^{\ast})$ joint
     distributions (see Fig.~\ref{fig:rdrdv02}), which are usually
     called the LFU problem, and have attracted much attention.
     It is shown that
     (a)
     the isospin breaking effects on the ratios $R(D^{(\ast)})_{\ell}$
     for a specific lepton are extremely small for theoretical predictions
     but clearly distinguishable for experimental data.
     (b)
     The QED nonfactoriziable effects on the differences between
     $R(D^{(\ast)})_{e}$ and $R(D^{(\ast)})_{\mu}$ are remarkable
     for theory but undetectable for the running experiments.
     (c)
     The QED nonfactoriziable contributions will further enlarge
     these discrepancies between the SM theory and HFLAV averages on the
     $R(D)$-$R(D^{\ast})$ joint distributions, and hence
     increase the tension sensitivity.
     Recently, the latest result on $R(D^{\ast})$ from Belle-II
     group $R(D^{\ast})$ $=$ $0.262^{+0.041+0.035}_{-0.039-0.032}$
     \cite{arXiv:2401.02840} are more consistent with the SM
     prediction, although the experimental uncertainties are still
     very large.
     For the semileptonic charmed $B$ decays,
     much more efforts are needed to further improve theoretical
     and experimental precision.

     \section{SUMMARY}
     \label{sec04}

     In the time of high-luminosity, high-precision and huge-data
     for particle physics, the $B$ meson physics undoubtedly plays
     as an important and potential arena to scrutinize SM and
     search for NP.
     The more than three error differences on the $R(D)$-$R(D^{\ast})$
     correlation distributions between SM predictions and HFLAV
     averages have attracted much attention, and make the study on
     the semileptonic $\overline{B}$ ${\to}$ $D^{({\ast})}$
     $+$ ${\ell}^{-}$ $+$ $\bar{\nu}_{\ell}$ decays highly interesting.
     Phenomenologically, the major research interests and efforts
     concentrate on the $B$ ${\to}$ $D^{({\ast})}$ transition form
     factors and the non-universal couplings arising from NP.
     In this paper, by considering the nonfactorizable one-photon
     exchange between the charged lepton and quarks, the QED
     radiative contributions to the semileptonic charmed $B$
     decays are investigated within SM. It is found that
     (a)
     $\tilde{\eta}_{\rm EW}$ demonstrates a non-trivial function of
     the lepton mass, the momentum transfer square $q^{2}$, and
     angular distribution ${\cos}{\theta}$, which naturally make
     the effective coupling process-dependent without the
     introduction of NP.
     (b)
     The QED contributions will enhance branching ratios
     according to the mass of the charged lepton.
     The smaller the mass of the charged lepton,
     the larger the enhancements.
     The enhancements for the semi-tauonic decays are negligible.
     These factors lead to the reduction of ratios $R(D^{(\ast)})$
     and the relationship $R(D^{(\ast)})_{e}$ $<$ $R(D^{(\ast)})_{\mu}$.
     (c)
     The dominated theoretical uncertainties on branching ratios
     come from the $B$ ${\to}$ $D^{(\ast)}$ transition form factors.
     The accurate determination of line shapes of form factors versus
     $q^{2}$ is the key to the FLU problem.
     More theoretical and experimental efforts are needed
     for the semileptonic $B$ decays.

     \section*{Acknowledgments}
     The work is supported by the National Natural Science Foundation
     of China (Grant Nos. 11705047, 12275068, 12275067),
     National Key R\&D Program of China (Grant No. 2023YFA1606000),
     and Natural Science Foundation of Henan Province
     (Grant Nos. 222300420479, 242300420250).
     We thank
     Prof. Shuangshi Fang (IHEP@CAS),
     Prof. Huijing Li, Prof. Qingping Ji,
     Dr. Suping Jin (Henan Normal University),
     Dr. Kaikai He (Hangzhou Normal University),
     Prof. Xinqiang Li (Central China Normal University),
     Alejandro Vaquero,
     Teppei Kitahara, and
     Gael Finauri
     for their kind assistance and valuable discussion.

     \begin{appendix}

     \section{The definition and parametrization of the
              $B$ ${\to}$ $D$ form factors}
     \label{app01}
     In this paper, we will take the conventions of
     Refs.~\cite{PhysRevD.92.034506,PhysRevD.92.054510,PhysRevD.101.074513}
     for the $B$ ${\to}$ $D$ transition form factors.
     %------------------------------------
     \begin{equation}
     {\langle} \, D \, {\vert} \, V^{\mu} \, {\vert} \, B \, {\rangle}
     \, = \,
     f_{+}(q^{2}) \, \big[ ( p_{B}^{\mu} + p_{D}^{\mu} ) -
     \frac{ m_{B}^{2} - m_{D}^{2} }{ q^{2} } \, q^{\mu} \big]
   + f_{0}(q^{2}) \, \frac{ m_{B}^{2} - m_{D}^{2} }{ q^{2} } \, q^{\mu}
     \label{eq:ff-B2D},
     \end{equation}
     %------------------------------------
     where $q^{\mu}$ $=$ $p_{B}^{\mu}$ $-$ $p_{D}^{\mu}$.
     The constraint condition for form factors,
     $f_{+}(0)$ $=$ $f_{0}(0)$,
     is generally required to cancel the singularity
     at the pole $q^{2}$ $=$ $0$.

\begin{figure}[h]
     \includegraphics[width=0.3\textwidth]{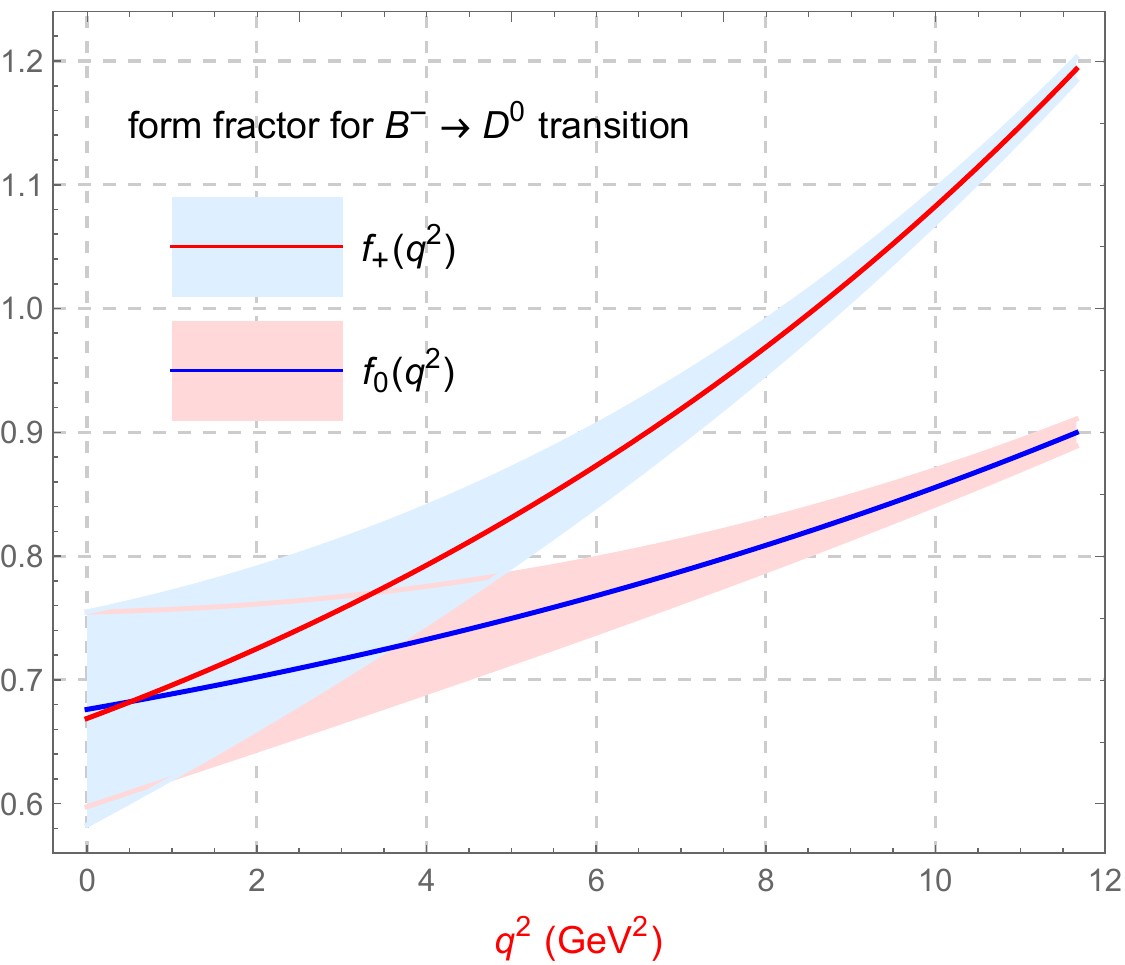} \vspace{-5mm}
     \caption{The shape lines of form factors for the $B^{-}$
          ${\to}$ $D^{0}$ transition, where the solid lines
          are computed with the central values of coefficients
          $a_{i,n}$, and the bands corresponds to uncertainties
          arising from $a_{i,n}$.}
     \label{fig-ffp}
     \end{figure}

     Form factors $f_{+}$ and $f_{0}$ are Lorentz scalar functions.
     The $z$ expansion of both
     Boyd-Grinstein-Lebed (BGL) \cite{PhysRevLett.74.4603}
     and Bourrely-Lellouch-Caprini (BLC) \cite{PhysRevD.79.013008}
     parametrization are generally used to carry out the
     interpolation/extrapolation of form factors over the
     kinematic range of $q^{2}$.
     Following the conventions of Ref.~\cite{PhysRevD.92.034506},
     the BGL form factors are parametrized as,
     %------------------------------------
     \begin{equation}
     f_{i}(z) \, = \, \frac{1}{ P_{i}(z) \, {\phi}_{i}(z) }\,
     \sum\limits_{n} a_{i,n}\, z^{n}
     \label{eq:FF-B2D-BGL},
     \end{equation}
     %------------------------------------
     with
     %------------------------------------
     \begin{equation}
     z \, = \,
     \frac{ \sqrt{ t_{+} - q^{2} } - \sqrt{ t_{+} - t_{-} } }
          { \sqrt{ t_{+} - q^{2} } + \sqrt{ t_{+} - t_{-} } }
       \, = \,
     \frac{ \sqrt{ {\omega}+1 } - \sqrt{ 2 } }
          { \sqrt{ {\omega}+1 } + \sqrt{ 2 } }
     \label{eq:FF-B2D-BGL-z},
     \end{equation}
     %------------------------------------
     %------------------------------------
     \begin{eqnarray}
     t_{\pm} & = & ( m_{B} {\pm} m_{D} )^{2}
     \label{eq:FF-B2D-BGL-tp-tm}, \\
    {\omega} & = &
     \frac{ m_{B}^{2} + m_{D}^{2} - q^{2} }
          { 2 \, m_{B} \, m_{D} }
     \label{eq:FF-B2D-BGL-omega}.
     \end{eqnarray}
     %------------------------------------
     The function $P_{i}(z)$, also called Blaschke factors,
     takes into account the resonances in the $b\bar{c}$ system
     below the $BD$ threshold in the channel variable $q^{2}$.
     Here, just as Ref.~\cite{PhysRevD.92.034506}, any
     poles are not introduced, {\it i.e.},
     $P_{i}(z)$ $=$ $1$.
     The weighting functions ${\phi}_{i}(z)$ corresponding to
     form factor $f_{i}$ are written as,
     %------------------------------------
     \begin{eqnarray}
    {\phi}_{+}(z) & = &
     \frac{ 1.1213 \, (1+z)^{2} \, \sqrt{ 1-z } }
          { \{ (1+r) \, (1-z) + 2 \, \sqrt{r} \, (1+z) \}^{5} }
     \label{eq:FF-B2D-BGL-phip}, \\
    {\phi}_{0}(z) & = &
     \frac{ 0.5299 \, (1-z^{2}) \, \sqrt{ 1-z } }
          { \{ (1+r) \, (1-z) + 2 \, \sqrt{r} \, (1+z) \}^{4} }
     \label{eq:FF-B2D-BGL-phiz},
     \end{eqnarray}
     %------------------------------------
     with $r$ $=$ $m_{D}/m_{B}$.
     The coefficients $a_{i,n}$ in Eq.(\ref{eq:FF-B2D-BGL})
     are given in Table XI of Ref.~\cite{PhysRevD.92.034506}.
     The shape lines of form factors are shown in Fig.~\ref{fig-ffp}.
     %------------------------------------
     %------------------------------------

     \section{The hadronic helicity amplitudes for $\bar{B}$
              ${\to}$ $D$ $+$ ${\ell}^{-}$ $+$
              $\bar{\nu}_{\ell}$ decays}
     \label{app02}
     Both the initial $B$ meson and recoiled $D$ meson
     are pseudoscalar particles, and have zero spin.
     Based on the angular momentum conservation, only the
     helicity projections of the spin of the virtual
     $W^{\ast}$ particle, ${\lambda}$ $=$ $0$ and $t$,
     contribute to the decay amplitudes for the
     $\bar{B}$ ${\to}$ $D$ $+$ ${\ell}^{-}$ $+$
     $\bar{\nu}_{\ell}$ decays.

     With the form factors in Eq.(\ref{eq:ff-B2D}),
     the helicity amplitudes $H_{\lambda}$
     in Eqs.(\ref{eq:unpolarized-transverse-component}--\ref{eq:scalar-longitudinal-interference-component})
     are written as,
     %------------------------------------
     \begin{eqnarray}
     H_{\pm} & = & 0
     \label{eq:hp-hm-B2D}, \\
     H_{0} & = & \frac{ 2\, m_{B} \, {\vert} {\bf p} {\vert} }{ \sqrt{q^{2}} } \, f_{+}(q^{2})
     \label{eq:hz-B2D}, \\
     H_{t} & = & \frac{ m_{B}^{2} - m_{D}^{2} }{ \sqrt{q^{2}} } \, f_{0}(q^{2})
     \label{eq:ht-B2D}.
     \end{eqnarray}
     %------------------------------------

     \section{The definition and parametrization of the
              $B$ ${\to}$ $D^{\ast}$ form factors}
     \label{app03}
     In this paper, we will take the conventions of
     Ref.~\cite{PhysRevD.109.074503}
     for the $B$ ${\to}$ $D^{\ast}$ transition form factors.
     %------------------------------------
     \begin{eqnarray}
     {\langle} \, D^{\ast} \, {\vert} \, V_{\mu} \, {\vert} \, B \, {\rangle}
     & = &  - i \, {\varepsilon}_{ {\mu} \, {\nu} \, {\alpha} \, {\beta} } \,
    {\varepsilon}_{D}^{{\ast}\,{\nu}} \,
     p_{B}^{\alpha} \, p_{D^{\ast}}^{\beta} \, g(z)
     \label{eq:prd.109-p.14-eq.25}, \\
    {\langle} \, D^{\ast} \, {\vert} \, A_{\mu} \, {\vert} \, B \, {\rangle}
     & = &
    {\varepsilon}_{D^{\ast}}^{{\ast}\,{\nu}} \, f(z)
   + ( {\varepsilon}_{D}^{{\ast}} \, {\cdot} \, p_{B} ) \big\{
     ( p_{B}^{\mu} + p_{D^{\ast}}^{\mu} ) \, a_{+}(z)
   + ( p_{B}^{\mu} - p_{D^{\ast}}^{\mu} ) \, a_{-}(z) \big\}
     \label{eq:prd.109-p.14-eq.26},
     \end{eqnarray}
     %------------------------------------
     and the combinations of form factors are
     %------------------------------------
     \begin{eqnarray}
     2\, m_{D^{\ast}} \, {\cal F}_{1} & = &
     ( m_{B} ^{2} - m_{D^{\ast}}^{2} - q^{2} ) \, f(z)
    + 4 \, m_{B}^{2} \, {\vert} {\bf p} {\vert}^{2} \, a_{+}(z)
     \label{eq:prd.109-p.14-eq.27}, \\
     m_{D^{\ast}} \, {\cal F}_{2}
     & = & f(z)
           + ( m_{B}^{2} - m_{D^{\ast}}^{2} ) \, a_{+}(z)
           + q^{2} \, a_{-}(z)
     \label{eq:prd.109-p.14-eq.28}.
     \end{eqnarray}
     %------------------------------------

     The BGL form factors are parametrized as
     \cite{EPJC.82.1141,PhysRevD.109.074503},
     %------------------------------------
     \begin{equation}
     f_{i}(z) \, = \, \frac{1}{ P_{i}(z) \, {\phi}_{i}(z) }\,
     \sum\limits_{n} a_{i,n}\, z^{n},
     \quad (f_{i} \, = \, f,\,g,\,{\cal F}_{1},\,{\cal F}_{2} )
     \label{eq:prd.109-p.16-eq.39},
     \end{equation}
     %------------------------------------
     The expansion coefficients $a_{i,n}$ in Eq.(\ref{eq:prd.109-p.16-eq.39})
     are given in Table VII of Ref.~\cite{PhysRevD.109.074503}.

     The Blaschke factors are
     %------------------------------------
     \begin{eqnarray}
     P_{J^{P}}(z) & = & \prod\limits_{k}^{n_{\rm pole}}
     \frac{ z- z_{{\rm pole},k} }{ 1 - z\, z_{{\rm pole},k} }
     \label{eq:prd.109-p.16-eq.42}, \\
     P_{1^{+}}(z) & = & P_{f}(z) \, = \, P_{{\cal F}_{1}}(z)
     \label{eq:prd.109-p.16-eq.41-1p}, \\
     P_{1^{-}}(z) & = & P_{g}(z)
     \label{eq:prd.109-p.16-eq.41-1m}, \\
     P_{0^{-}}(z) & = & P_{{\cal F}_{2}}(z)
     \label{eq:prd.109-p.16-eq.41-0m},
     \end{eqnarray}
     %------------------------------------
     where
     %------------------------------------
     \begin{equation}
     z_{{\rm pole},k} \, = \,
     \frac{ \sqrt{ t_{+} - m_{{\rm pole},k}^{2} } - \sqrt{ t_{+} - t_{-} } }
          { \sqrt{ t_{+} - m_{{\rm pole},k}^{2} } + \sqrt{ t_{+} - t_{-} } }
     \label{eq:prd.109-p.16-eq.43},
     \end{equation}
     %------------------------------------
     corresponding to the $k$-th resonance.
     The concerned resonance mass $m_{{\rm pole}}$ corresponding to
     different quantum number $J^{P}$ are listed in Table V
     of Ref.~\cite{PhysRevD.109.074503}.

\begin{figure}[h]
     \includegraphics[width=0.3\textwidth]{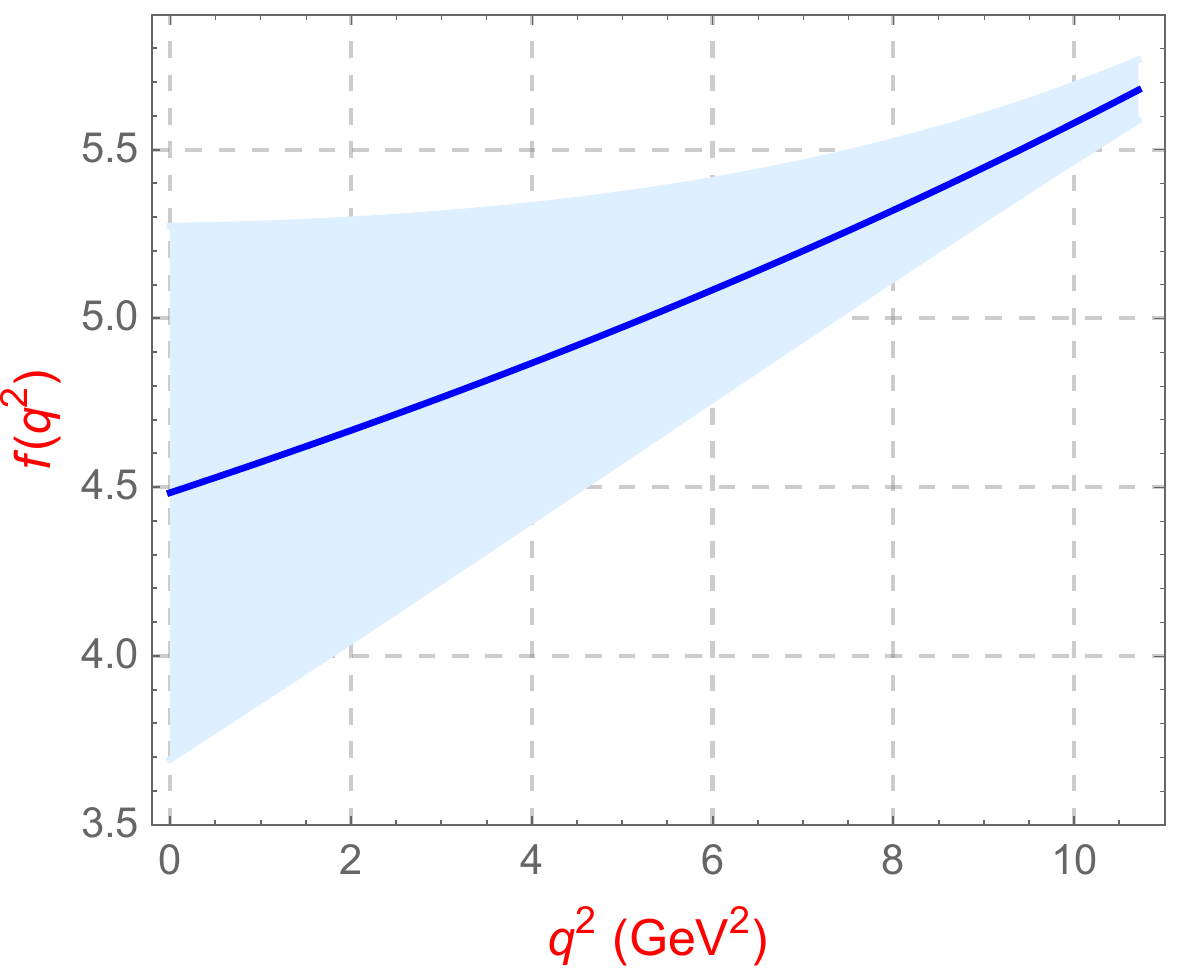} \quad
     \includegraphics[width=0.3\textwidth]{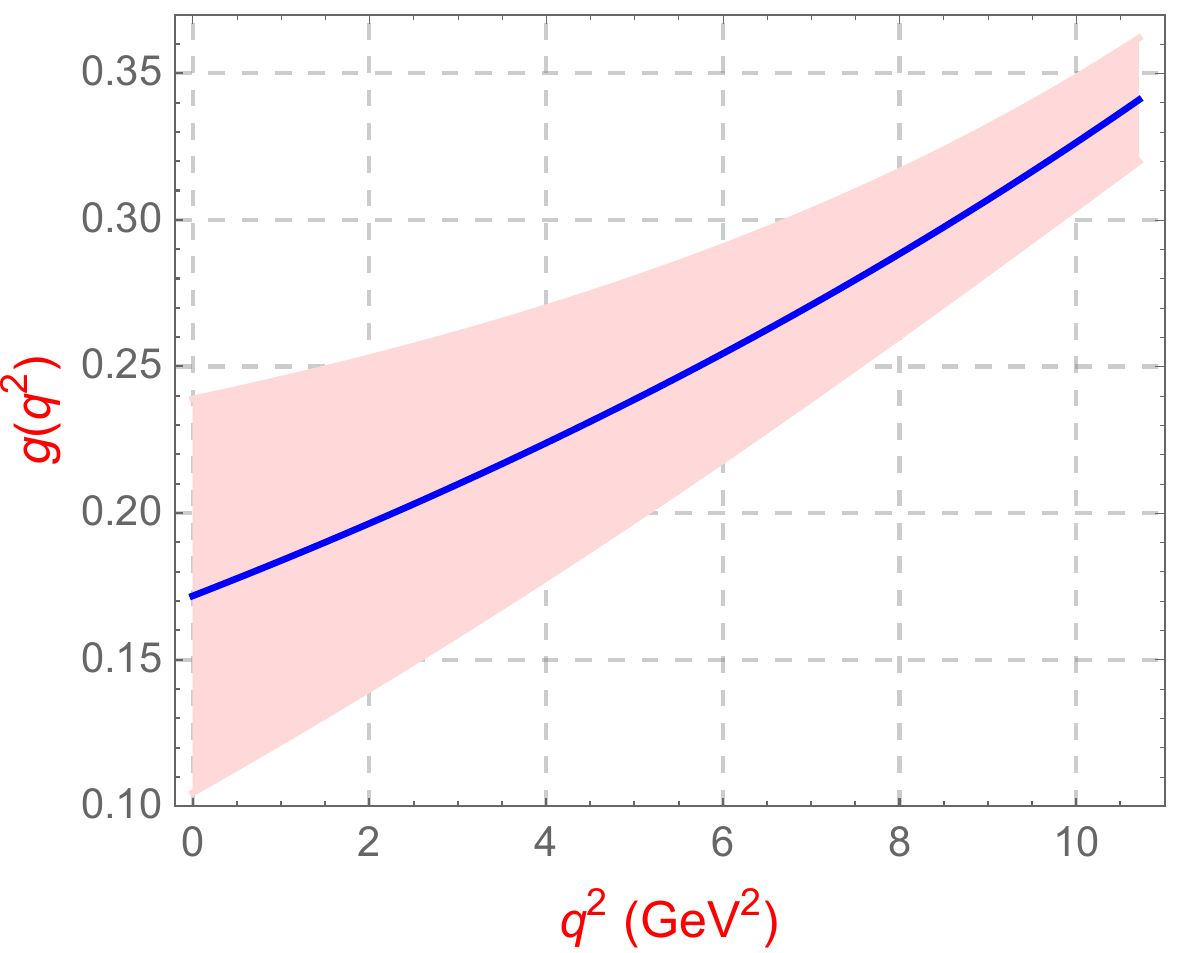} \\ ~ \\
     \includegraphics[width=0.3\textwidth]{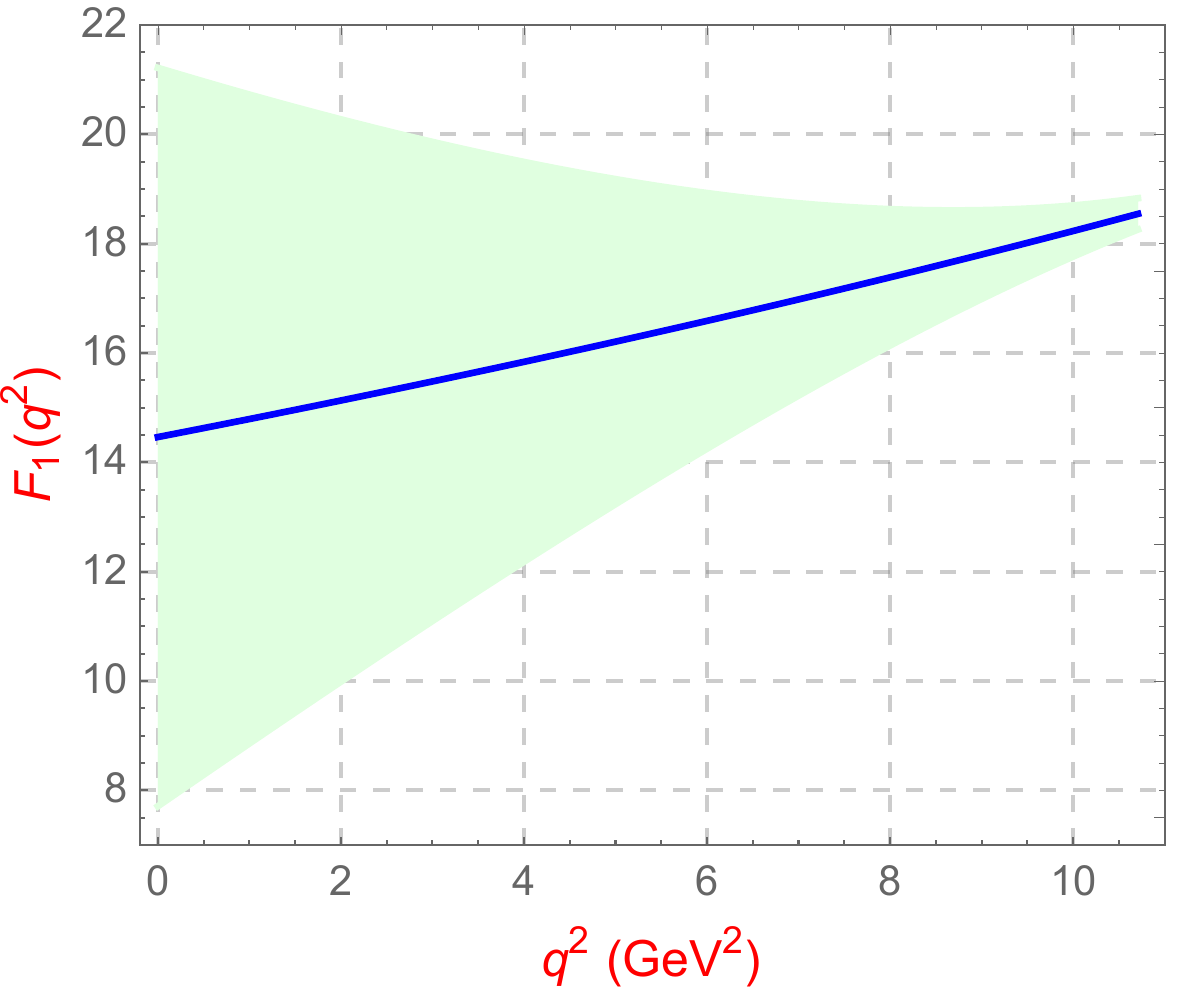} \quad
     \includegraphics[width=0.3\textwidth]{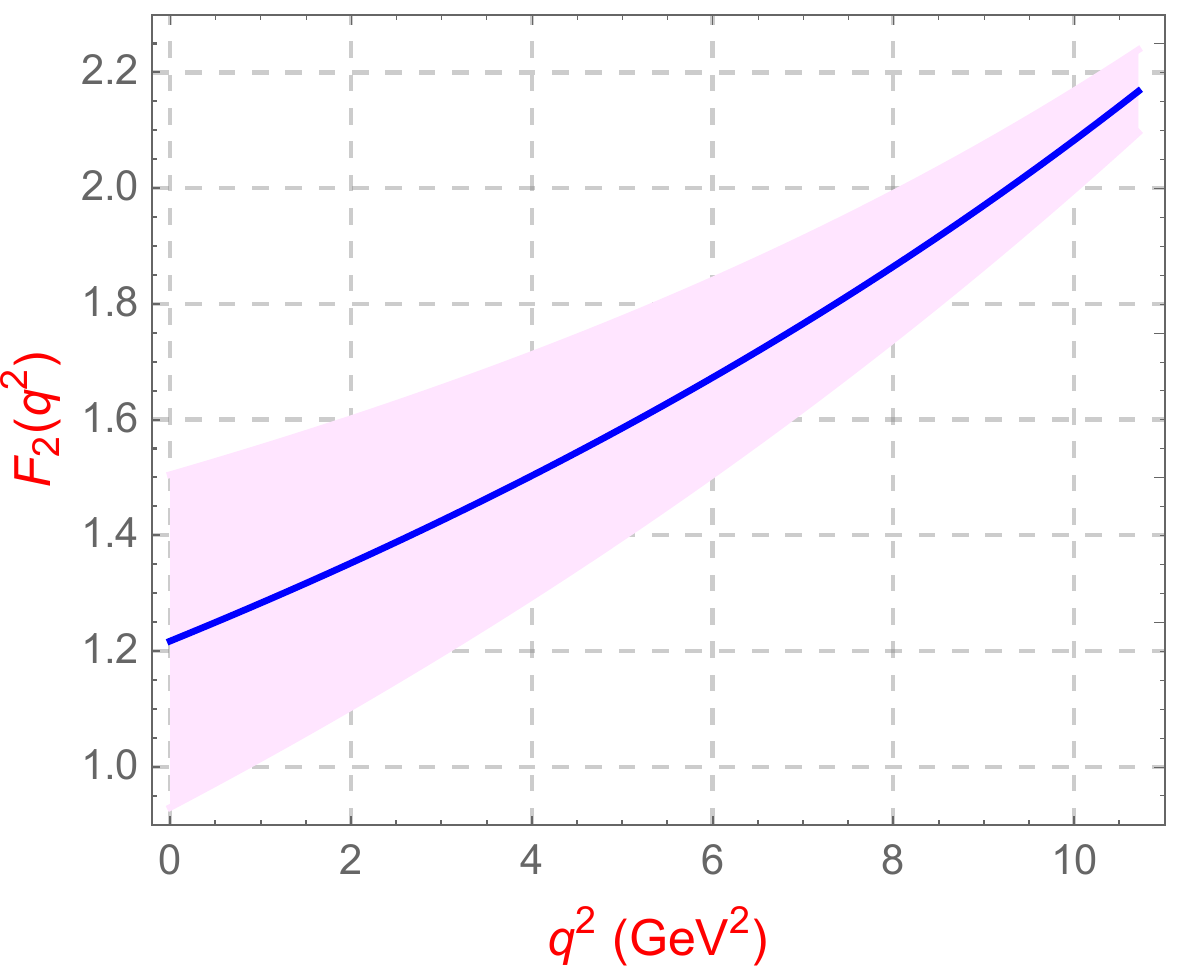} \vspace{-5mm}
     \caption{The shape lines of form factors for the $B^{-}$
          ${\to}$ $D^{{\ast}0}$ transition. Other legends are
          the same as those of Fig. \ref{fig-ffp}.}
     \label{fig-ffv}
     \end{figure}

     The weighting functions ${\phi}_{i}(z)$ corresponding to
     form factor $f_{i}$ are written as,
     %------------------------------------
     \begin{eqnarray}
    {\phi}_{f}(z) & = &
     \frac{ 4 \, r }{ m_{B}^{2} } \, \sqrt{ \frac{ n_{I} }{ 3 \, {\pi} \, {\chi}_{1^{+}}^{T}(0) } }\,
     \frac{ (1+z) \, (1-z)^{3/2} }
          { \{ (1+r) \, (1-z) + 2 \, \sqrt{r} \, (1+z) \}^{4} }
     \label{eq:prd.109-p.17-eq.46}, \\
    {\phi}_{g}(z) & = &
     16 \, r^{2} \, \sqrt{ \frac{ n_{I} }{ 3 \, {\pi} \, {\chi}_{1^{-}}^{T}(0) } }\,
     \frac{ (1+z)^{2} \, (1-z)^{-1/2} }
          { \{ (1+r) \, (1-z) + 2 \, \sqrt{r} \, (1+z) \}^{4} }
     \label{eq:prd.109-p.17-eq.45}, \\
    {\phi}_{{\cal F}_{1}}(z) & = &
     \frac{ 4 \, r }{ m_{B}^{3} } \, \sqrt{ \frac{ n_{I} }{ 6 \, {\pi} \, {\chi}_{1^{+}}^{T}(0) } }\,
     \frac{ (1+z) \, (1-z)^{5/2} }
          { \{ (1+r) \, (1-z) + 2 \, \sqrt{r} \, (1+z) \}^{5} }
     \label{eq:prd.109-p.17-eq.47}, \\
    {\phi}_{{\cal F}_{2}}(z) & = &
     16 \, r^{2} \, \sqrt{ \frac{ n_{I} }{ 2 \, {\pi} \, {\chi}_{1^{+}}^{L}(0) } }\,
     \frac{ (1+z)^{2} \, (1-z)^{-1/2} }
          { \{ (1+r) \, (1-z) + 2 \, \sqrt{r} \, (1+z) \}^{4} }
     \label{eq:prd.109-p.17-eq.48},
     \end{eqnarray}
     %------------------------------------
     where $r$ $=$ $m_{D^{\ast}}/m_{B}$ and $n_{I}$ $=$ $2.6$ is
     the number of the spectator quarks with a correction due to
     $SU(3)$ breaking.
     The shape lines of form factors are shown in Fig.~\ref{fig-ffv}.
     %------------------------------------
     %------------------------------------

     \section{The hadronic helicity amplitudes for $\bar{B}$
              ${\to}$ $D^{\ast}$ $+$ ${\ell}^{-}$ $+$
              $\bar{\nu}_{\ell}$ decays}
     \label{app04}
     With the form factors in Appendix \ref{app03},
     the helicity amplitudes $H_{\lambda}$
     in Eqs.(\ref{eq:unpolarized-transverse-component}--\ref{eq:scalar-longitudinal-interference-component})
     are written as Ref.~\cite{PhysRevD.109.074503},
     %------------------------------------
     \begin{eqnarray}
     H_{\pm} & = & f(z) \, {\mp} \,  m_{B}\, {\vert} {\bf p} {\vert} \, g(z)
     \label{eq:hp-hm-B2Dv}, \\
     H_{0} & = & \frac{ 1 }{ \sqrt{q^{2}} } \, {\cal F}_{1}(z)
     \label{eq:hz-B2Dv}, \\
     H_{t} & = & \frac{ m_{B} \, {\vert} {\bf p} {\vert} }{ \sqrt{q^{2}} } \, {\cal F}_{2}(z)
     \label{eq:ht-B2Dv}.
     \end{eqnarray}
     %------------------------------------

     \end{appendix}

     %------------------------------------
     
     %------------------------------------

     \end{document}